\documentclass[manuscript]{aastex}

\usepackage{natbib}

\usepackage{epsfig,times,lscape,rotating}
\usepackage{color}
\usepackage{amsmath}
\usepackage{amssymb}
\usepackage{txfonts}

\shorttitle{Global disk dynamo}
\shortauthors{Flock et al.}

\begin{document}

%\title{The Significance Of Large Scale Azimuthal Modes 
%In Turbulent Accretion Disks}

\title{Large Scale Azimuthal Structures Of Turbulence 
In Accretion Disks\\ 
\it{Dynamo triggered variability of accretion.}
}
%\subtitle{Hallo}

\author{M. Flock\altaffilmark{1}, N. Dzyurkevich\altaffilmark{1}, H.
Klahr\altaffilmark{1},
N. Turner\altaffilmark{1,2},Th. Henning\altaffilmark{1}}
\affil{$^1$Max Planck Institute for Astronomy, K\"onigstuhl 17,
 69117 Heidelberg, Germany}
\affil{$^2$Jet Propulsion Laboratory, California Institute of Technology,
Pasadena,
CA 91109, USA}

\begin{abstract}
We investigate the significance of large scale azimuthal, magnetic and
velocity modes for the MRI turbulence in accretion disks.
We perform 3D global ideal MHD simulations of 
global stratified proto-planetary disk models. 
Our domains span azimuthal angles of $\pi/4$, $\pi/2$, $\pi$ and $2\pi$.
We observe up to $100\%$ stronger magnetic fields and 
stronger turbulence for the restricted azimuthal
domain models $\pi/2$ and $\pi/4$ compared to the full $2\pi$ model. 
We show that for those models, the Maxwell Stress is larger due to
strong axisymmetric magnetic fields, generated by the $\alpha \Omega$
dynamo.
Large radial extended axisymmetric toroidal fields trigger 
temporal magnification of accretion stress. % For the restricted domains 
%the magnetic field strength increases between $50\%$ and $100\%$.} % larger magnetic fields 
All models display a positive dynamo-$\alpha$ 
%with a positive sign for
%$\alpha_{\phi\phi}^{dyn.}$ 
in the northern hemisphere (upper disk). 
The parity is distinct in each model and changes on timescales of 40 local
orbits. 
In model $2\pi$, the toroidal field is mostly antisymmetric in respect to the midplane.
The eddies of the MRI turbulence are highly anisotropic. 
The major wavelengths of the turbulent velocity and magnetic fields are between
one and two disk scale heights. At the midplane, we find magnetic 
tilt angles around $8-9^\circ$ increasing up to $12-13^\circ$ in the
corona.
We conclude that an azimuthal extent of $\pi$ is sufficient to reproduce 
most turbulent properties in 3D global stratified simulations of magnetised
accretion disks. % as they are present in the full $2\pi$ model. 
%Using restricted domain sizes of $\pi/4$ and $\pi/2$ 
%amplify the saturation level of MRI and affect the turbulent properties.
\end{abstract}


\section{Introduction}
%In sufficiently ionised accretion disk, 
Magneto-rotational instability (MRI) 
can generate MHD turbulence with an outward
directed angular momentum transport driving accretion onto the central
object \citep{bal91,haw91,bal98}.
%For applications on proto-planetary disk evolution and planet formation, 
%the turbulent properties of the gas has to be known. 
%Here, the turbulence could be
%driven by MRI in well-ionised regions of the disk.
A necessary condition is a good coupling 
between the gas and magnetic fields, e.g. a well-ionized gas.
In proto-planetary disks, dust particles and low temperatures will 
reduce the ionisation level and therefor the MRI activity \citep{san00,fle03,inu05,war07,dzy10,tur10}. 
Nevertheless, there are well-ionized regions with possible MRI activity, 
like the coronal region or the inner or outer disk. The inner disk 
will be thermally ionized for temperatures greater then $1000 K$ \citep{ume83}.
The outer disk will be ionized by Cosmic Rays for surface density values 
below $96g/cm^2$ \citep{ume09}.
%A recent work about \citep{bai11} even conclude that an active layer always
%exists. 
In our work we concentrate on well ionized disk regions.
To model the evolution of proto-planetary disks and especially 
to describe the process of planet formation, we need to know 
detailed informations about 
the strength of the turbulence. Several processes, like the MHD dynamo or
the toroidal field MRI, influence the turbulence level. The evolution 
of the magnetic and velocities fields at different scales has to be
investigated.

In the last decades, a large amount of local-box simulations have been
performed to study the small scale MRI turbulence 
\citep{bra95,haw95,haw96,mat95,sto96}.
The MRI works for both, vertical or toroidal seed magnetic fields
\citep{bal91}. The MRI launched with initial toroidal field was 
%successfully 
%studied in several numerical simulations 
analyzed through linear calculations
\citep{haw92,fog95,ter96,pap97} and 
in Taylor-Couette experiments \citep{gel07,gue07b}.
%It plays a key role in determining the saturation level of the turbulence.
%The MRI with initial toroidal field was 
%investigated in Taylor-Couette experiments \citep{gel07,gue07b}.
This experiments showed that most of the energy will be transported 
to the $m=1$ mode.
A similar inverse energy cascade was found in local box simulations
as well \citep{joh09}. Here the turbulent advection term in the induction 
equation drives large-scale radial magnetic field.

The locality and anisotropy of the MRI turbulence is an important aspect
for dust growth and therefor the planet formation.
The eddies are stretched in the azimuthal direction due to the strong shear. 
They have a characteristic 
low tilt angle in the $r-\phi$ plane \citep{gua09}.
%The two-point correlation functions introduced for MRI turbulence 
%in local simulations by \citep{gua09} 
Several works confirmed this tilt angle for the velocity and 
the magnetic fields 
\citep{gua09,fro10,dav10,gua11,sor11}. 
The size of the corresponding correlation wavelengths is dependent on 
resolution \citep{gua09} and converges by using a fixed value of viscous and explicit 
dissipation in unstratified local simulations \citep{fro10}.
%Recently this correlation wavelength were 
%determined in global simulations \citep{bec11} showing wavelengths greater than H.
Unstratified global models interpret the magnetic tilt angle 
as convergence parameter \citep{sor11}. They found convergence 
with tilt angles around $13^\circ$.
\citet{bec11} found tilt angles of $9^\circ$ in global stratified simulations
with spatial structures of the turbulent field in the order of H.

Global disk simulations
\citep{arm98,haw00,arl01,fro06,fro09,dzy10,flo11,bec11,sor11} are used to study 
the MRI evolution on large scales. 
\citet{bec11} found a stronger accretion stress compared to \citet{fro06} and
\citet{flo11} with a stronger initial toroidal field.
Unstratified simulations show a similar correlation between accretion 
stress and the initial plasma beta \citep{haw95}. 
Here a stronger seed field will drive to 
stronger accretion stress.
%He  larger resolution per scale height, a stronger initial magnetic field.
%Differences in stresses or other turbulent properties are mostly 
%due to different disk models, different strength of the seed magnetic fields,
%net or zero-net flux initial magnetic fields or different resolution. 
%In stratified simulations, 
%the accretion stress increases with higher resolution until it converges
%\citep{dav10}.
%Also recent unstratified simulations by \citet{sor11} 
%
The majority of stratified global disk simulations has been done for restricted 
($\phi \le \pi/2$) azimuthal domain sizes.
At first glance, MRI turbulence
behaves similar for both full $2\pi$ and smaller domain sizes \citep{haw00}.

In our previous work we compared stratified simulations of $\pi/4$ and $2\pi$
in azimuth. 
There, we observe stronger azimuthal fields for the $\pi/4$ domain size 
\citep{flo11}.
Recent unstratified global simulations \citep{sor11} do not show large 
differences between domain size of $\pi/4$ and $2\pi$.
This fact indicates a mean field dynamo mechanism.
The stratification is crucial for driving $\alpha\Omega$ dynamo in disks 
and therefor for the creation of large scale magnetic fields \citep{kra80}.
%The fact that stratified simulations show a different mean field 
%behaviour would indicate magnetic dynamo effects.
With this work we perform a detailed study of different azimuthal
domain sizes. We investigate the turbulent and the mean field evolution for
the velocity and magnetic fields.
%Differences in stresses or other turbulent properties are mostly 
%due to different disk models, different seed magnetic fields or numerical
%schemes. %Recent unstratified cylindrical simulations with azimuthal extents 
%of $\pi/4$ and $2\pi$ showed only a small increase of turbulent stress
%\citep{sor11}. 
%The global unstratified simulations
%\citep{sor11} indicate that it could be connected to the dynamo process in
%stratified simulations.\\
%We present a detailed study of the significance of large scale
%azimuthal modes in MHD accretion disks.
%
%
%
%The azimuthal MRI (AMRI) \citep{gue07a,gue07b}, launched from a 
%toroidal field, plays a key role in determining the saturation level of 
%the turbulence .   
%\citet{gue07b} showed that the azimuthal MRI has an inverse cascade, driving 
%the magnetic energy towards the $m=1$ mode. 
%For a domain size of $\pi/4$ 
%these modes are not present and we observed a energy pile up at the 
%domain size \citep{flo11}.\\

%In this work we will investigate 
%the effect of different domain sizes on turbulent evolution.\\
%We want to study the energy pile up and the different large-scale 
%evolution
%
% motivation is also connected to the large-scale development of 
%Bthe azimuthal MRI and .
In stratified disk simulations, there is a periodic change of sign 
for the mean toroidal magnetic field. 
A similar periodicity of toroidal magnetic field, known as 
butterfly diagram, is observed in the sun. It could be explained 
by a MHD dynamo process.
%creating 
%the solar butterfly oscillations. 
The MRI could be self-sustaining by a analogous dynamo process
\citep{haw96,les08,les08b,gre10,sim11}.
Strong shear in accretion disks will wind up 
any radial magnetic field generated by MRI and produce toroidal field. 
This field will act as seed for the MRI again.
%and launch again the
%MRI.
Solutions for $\alpha \Omega$ dynamos in rotating systems
were presented by \citet{rue93,els96}.
Calculations of the dynamo-$\alpha$ for MRI 
have been performed in local box
simulations \citet{bra95,bra97,rek00,zie00,dav10,gre10} showing a negative\footnote{Negative
dynamo-$\alpha$ means a negative correlation
between the turbulent EMF (Electromotive force) and the mean toroidal field in the upper 
(northern) hemisphere.} dynamo-$\alpha$. 
\citet{bra97,rue00} explained the negative sign as an 
effect of vertical buoyancy.    %, different than predicted in \citet{rue93} because of the vertical buoyancy \citep{bra97,rue00}. 
The first indications for a positive dynamo-$\alpha$ 
were found in global disk simulations \citep{arl01,arl01b}. 
Dynamo solutions for positive or negative dynamo-$\alpha$ predict long-term
global mean magnetic fields which become symmetric 
(quadrupole, dynamo-$\alpha^{north} < 0$
 ) or asymmetric (dipole, dynamo-$\alpha^{north} > 0$). E.g. dipole solutions
support the creation of disk wind and jets \citep{rek00}. 
A review of dynamo action in accretion 
disks was presented by \citet{bra05,bra07,bla10}.\\
The connection between the dynamo processes and the large-scale magnetic field 
oscillations was shown by \citet{les08,gre10,sim11}.
These oscillations are universal for stratified MRI simulations
\citep{sto96,mil00} with timescales of ten local orbits, presented
recently in local \citep{gre10,sim11,haw11,gua11} and global
\citep{sor10,dzy10,flo11,bec11} simulations.
%
%The toroidal magnetic fields become buoyant above 1 scale height of the disk 
%\citep{shi10}.\\
%while they are stable to the Parker instability around the midplane 
%region .\\
%
%
%We use different azimuthal domain models to investigate the non-linear evolution and saturation regime 
We use the second order Godunov code PLUTO which was successfully applied 
in recent global simulations \citep{flo10,flo11,uri11,bec11}.
The paper is structured in the following way:\\
First, we describe the disk model and the numerical parameter. For the
results in section 3 we study the turbulent and the mean field evolution for
all azimuthal domain.
Section 4 and 5 present discussion and summary.

\section{Setup}
Our disk model is presented in detail in \citet{flo11}. 
We give here a summary of our physical and numerical 
initial conditions.
\subsection*{Disk model}
The HD initial conditions of density, pressure and azimuthal velocity 
follow a hydrostatic equilibrium. We set
$$\rho = \rho_{0}  R^{-3/2}\exp{}\Bigg(\frac{\sin{(\theta)}-1}{(H/R)^2}\Bigg) $$
with $\rho_{0} = 1.0$, $\rm H/R = c_0 = 0.07$, $R = r \sin{(\theta)}$.
The pressure follows locally an isothermal equation of state: 
$P = c_{s}^2\rho$ with $\rm c_{s} = c_0/\sqrt{R}$.
The azimuthal velocity is set to $$V_{\phi} = \sqrt{\frac{1}{r}}\Bigg(1- \frac{2.5}{\sin(\theta)}c^2_0 \Bigg).$$
The initial velocities $V_{r}$ and $V_{\theta}$ are set to a white noise
perturbation amplitude of $V_{r,\theta}^{Init} = 10^{-4} c_{s}$.
We start the simulation with a pure toroidal magnetic seed field with constant plasma beta
$\beta = 2P / B^{2} = 25$.\\
The radial domain extends from 1 to 10 AU\footnote{We set AU as unit length.
As the simulations are scale invariant, the radial extent could be also from
0.1 to 1 AU or from 10 to 100 AU, more details in \citet{flo11}}.
The $\theta$ domain covers $\pm$ 4.3 disk scale heights, or $\theta = \pi/2 \pm 0.3$.
For the azimuthal domain we use four different models: $\phi^{extent} =
\pi/4$, $\pi/2$, $\pi$
and $2\pi$.
We use a uniform grid in spherical coordinates with an aspect ratio at 5 AU of $1:0.67:1.74$ $(\Delta r:
r\Delta\theta:r\Delta\phi\sin{\theta})$. The resolution is fixed to $N_r:
384$, $N_\theta=192$ , $N_\phi=768\cdot\phi_{extent}/(2\pi)$.
We have around 23 grid cells per pressure scale height.\\
Buffer zones extent from 1 to 2 AU as well as from 9 to 10 AU.
In the buffer zones we use a linearly increasing resistivity to
the boundary. This damps 
the magnetic field fluctuations and suppresses boundary interactions. 
In the buffer zones we use also a relaxation function which reestablishes gently 
the initial value of density over a time period of one local orbit. 
In the buffer zones we set: $\rm \rho^{new} = \rho - (\rho-\rho^{\rm Init})\cdot \Delta
t / T_{Orbits}$.
Our outflow boundary condition projects the radial gradients
in density, pressure and azimuthal velocity into the radial boundary and the
vertical gradients in density and pressure at the $\theta$ boundary. 
We ensure to have no inflow velocities. For an inward pointing velocity
we mirror the values in the ghost cell to ensure no inward mass flux.
The $\theta$ boundary condition for the magnetic field are set 
to zero gradient, which approximates "force-free" - outflow conditions. 
The normal component of the magnetic field in the ghost cells is always 
set to have $\nabla \cdot \vec{B}$ = 0.\\
%The CFL value is 0.7. 

%
%{\bf Using a uniform grid instead of a logarithmic grid, where
%$\Delta r/r$ is constant, has the disadvantage that it will reduce the
%accuracy in the sense that the inner part of the disk is poorly resolved,
%compared to the outer part of the disk: $H(1AU)/\Delta r < H(10AU)/\Delta r  $.\\ 
%However for the uniform grid, the relative broad radial inner 
%buffer zone lies in the poorly resolved disk part and is excluded from
%analysis.
%The outer parts of the disk are, compared
%to a logarithmic grid with the same resolution, better resolved.  
%Logarithmic grid requires a much smaller buffer
%zone, e.g. a logarithmic grid would place one third of the total number of grid
%cells in the first ninth of the domain, between 1 and 2 AU.
%Of course, using a uniform grid will always restrict the range of the
%radial domain and for more radially extended simulation the need of a logarithmic grid is mandatory.}
\subsection*{Numerical setup}
The detailed numerical configuration is presented in \citet{flo10} and was
also successfully used in recent global simulations by \citet{bec11}.
For all runs we employ the second order scheme in
PLUTO with the HLLD Riemann solver \citep{miy05}, piece-wise linear
reconstruction and $2^{nd}$ order Runge Kutta time integration. 
We treat the induction equation with the "Constrained Transport" (CT) 
method in combination with the upwind CT method described in
\citet{gar05}.
All models were performed on a Blue-gene/P cluster for in total over 
3 million CPU hours.

\subsection{Measurement and integration}
For our analysis we use the central domain\footnote{The "central domain" is here 
the domain between 3 and 8 AU to avoid impact of the inner and outer
buffer zones, (see \citet{flo11}).} from 3 to 8 AU. 
Total volume integrations or a variable $F$, as used for the total stress are performed
with $$F^{total} = \int F
dV = \int_3^8
\int_{\theta_{begin}}^{\theta_{end}} \int_0^{\phi^{extent}} F r^2 \sin{\theta} dr
d\theta d\phi.$$ 
In global disk models, the gas dynamics are only self-similar along the azimuth.
Therefor, mean values like $\overline{v_{\phi}}$, are always averaged over azimuth.
This includes the calculation of the turbulent $EMF'$ in Fig. 11. 
For further analysis we always use an 2D dataset of mean values, e.g.
$\overline{v_{\phi}}(r,\theta)$ to construct the 3D turbulent dataset
$v'_{\phi}(r,\theta,\phi) = v_{\phi}(r,\theta,\phi) - \overline{v_{\phi}}(r,\theta)$.
For volume integration over mean values, as $\alpha_{SS}^{mean}$, we use $$\int
dV = \int_3^8 \int_{\theta_{beg}}^{\theta_{begin}} r^2 \sin{\theta} dr
d\theta.$$ 
%All analysis done at 4.5 AU is averaged over azimuth and a small radial extent
%($\pm 0.5$
%H = 0.16 AU). 
Some results are determined in the center of computational domain.
Analysis done at 4.5 AU are the tilt angle calculations, Fig. 6, the mean
field contour plots, Fig. 9, the parity, Fig. 10, the dynamo coefficients in Fig. 11.
This results are averaged over azimuth and a small radial extent ($\pm 0.5$ H = 0.16 AU).
For the time evolution of the tilt angle, Fig. 6, top, we average vertically 
$\pm 0.5 H$ at the midplane.
Radial contour plots are averaged over azimuth and height, between $\rm 0 -
1.5H$.
This applies for the mean toroidal field Fig. 3, the dynamo Fig. 11 and the
mean fields in Fig. 12. 
The parity is averaged over the total disk height at 4.5 AU, Fig. 10.

\section{Results}
In this section we investigate the turbulent and mean field evolution for
the azimuthal MRI for different azimuthal domain sizes.
Table 1 summarises the results of accretion stress, contribution of
mean magnetic field to the total stress, dynamo-$\alpha$ and 
RMS velocities for all models. 
Table 2 summarises results of the two-point correlation function, including
tilt angles, major and minor wavelength.
For all models, the accretion disk becomes unstable to MRI on timescales of 
ten local orbits.
All models develop an oscillating zero-net flux configuration after around
250 inner orbits.
\begin{figure}
\hspace{-0.6cm}
\begin{minipage}{5cm}
\psfig{figure=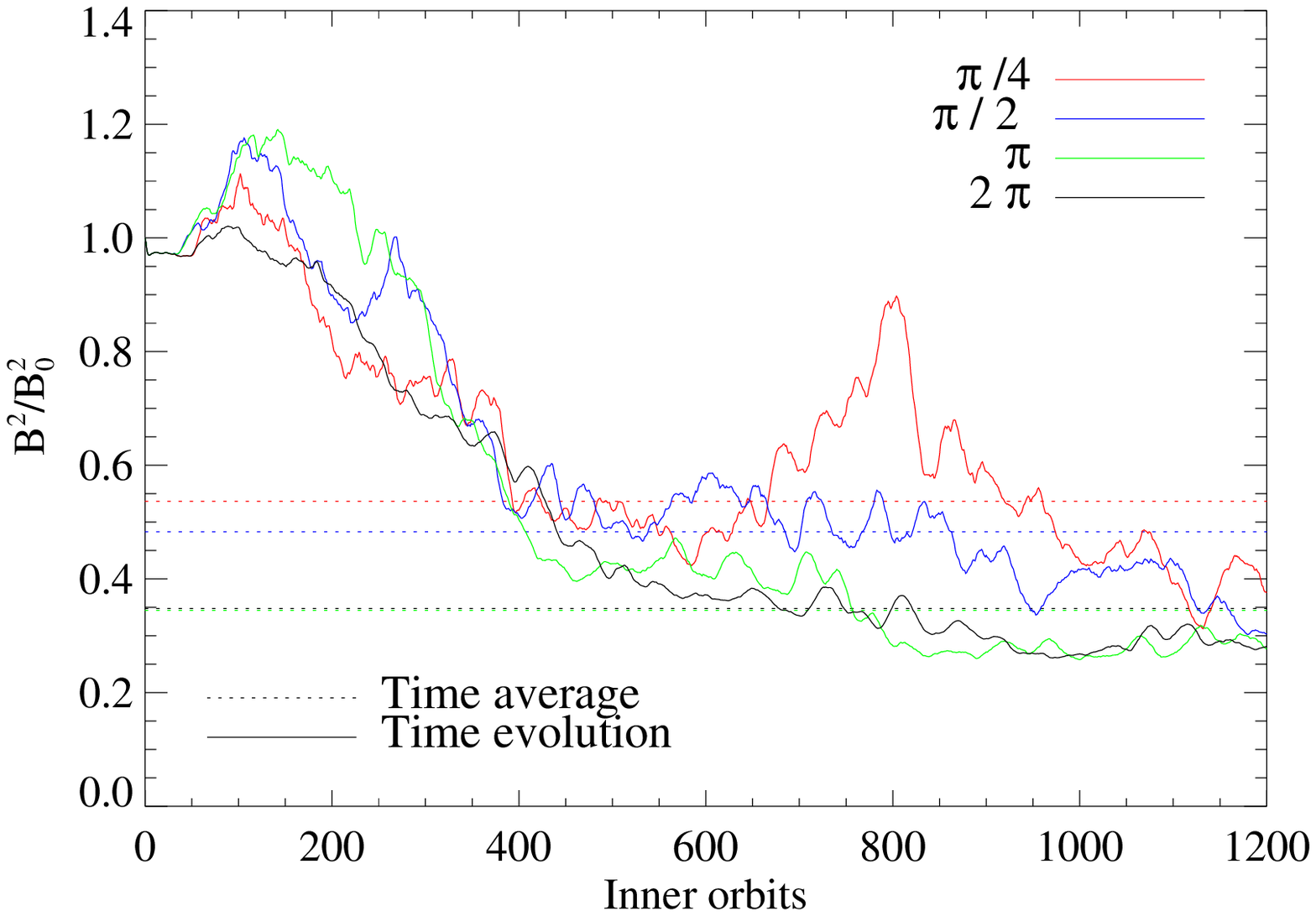,scale=0.52}
\end{minipage}
\hspace{4.0cm}
\begin{minipage}{5cm}
\psfig{figure=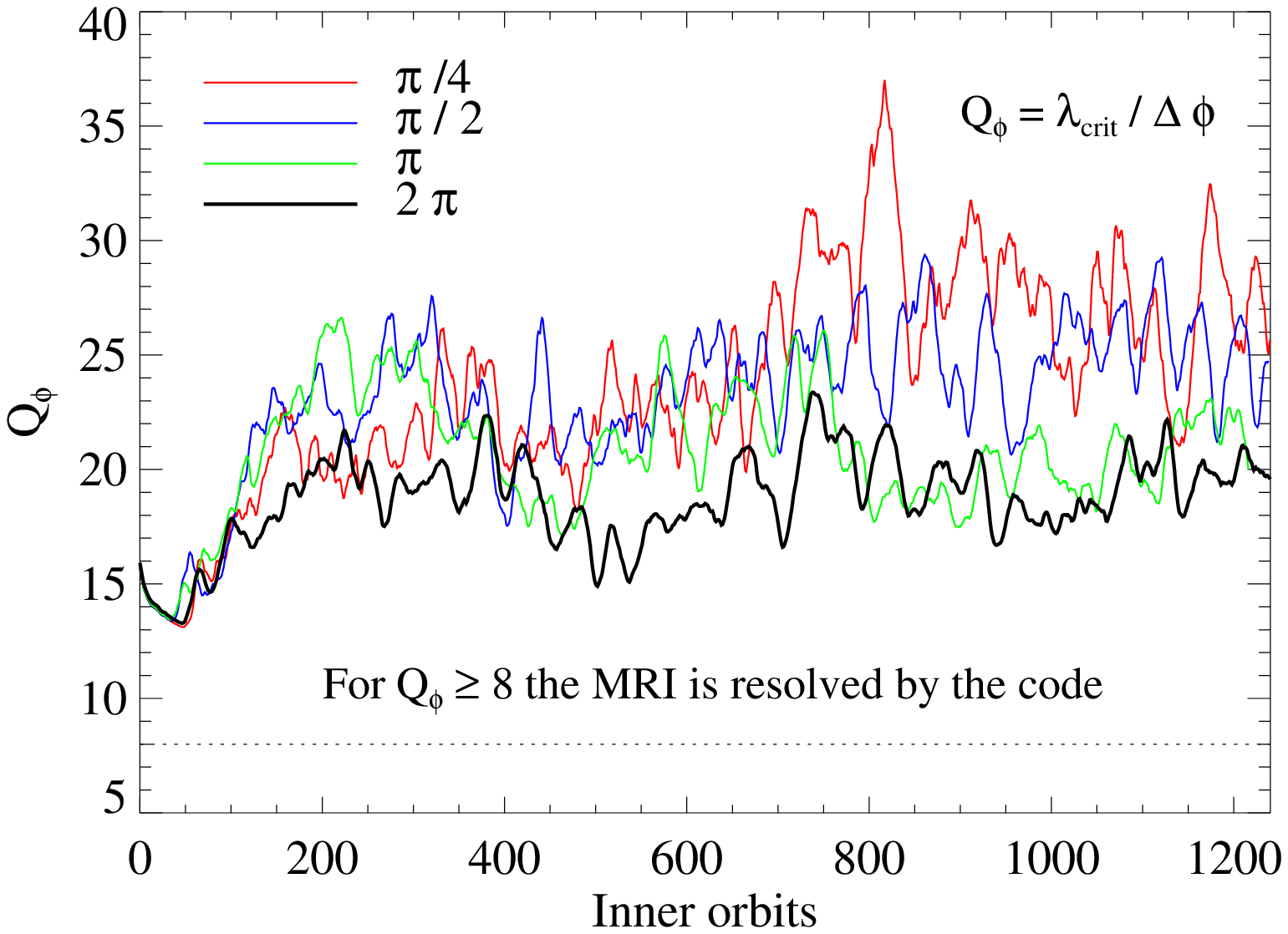,scale=0.52}
\end{minipage}
\label{four}
\caption{Left: Total magnetic energy evolution over time.
Right: Toroidal quality factor $Q_\phi$ over time.
All models show a well resolved MRI.}
%There is more magnetic energy in the $\pi/4$ and $\pi/2$ run.
\end{figure}
The time evolution of total magnetic energy, Fig. 1 left, is normalised over
the total initial magnetic field energy $B_0^2$.
It shows the peak of
magnetic energy shortly after the linear MRI phase around 100 inner orbits.
Between 100 and 400 years, the total magnetic energy decreases due to loss of
the net magnetic flux and mass loss (see also Fig. 13 in \citet{flo11} and Fig.
3 in \citet{bec11}).
After 400 years, $\pi/4$ and $\pi/2$ models show strong fluctuations
while $\pi$ and $2\pi$ models do saturate.
In the saturated state
($\gtrsim$ 800 inner orbits), the total magnetic energy evolution shows a relative
constant level for the $\pi$ and $2\pi$ model.\\
%While $\pi$ and $2\pi$ models show
%a saturated state in the magnetic energy evolution, 
%the $\pi/4$ and $\pi/2$ models strongly fluctuate.\\
%($\gtrsim$ 800 inner orbits) the total magnetic energy evolution shows a relative
%constant level for the $\pi$ and $2\pi$ model.\\
All models have the same resolution per $\phi$ extent ($\phi^{\rm extent}/N_\phi$).
The toroidal quality factor $Q_\phi = \lambda_{crit}/\Delta\phi$ 
shows the quality of resolved MRI ($Q_\phi \ge 8$).
We follow the analysis 
done by \citep{nob10,sor11} and calculate the mean $Q_\phi$ for the central domain
(3 to 8 AU).
The definition is similar to the toroidal quality factor $Q_\phi$ by 
\citet{haw11}.
$$\frac{\lambda_{crit}^{B_\phi}}{\Delta\phi} = 2\pi \sqrt{\frac{16}{15}
\frac{2}{\beta_{\phi}^{B_\phi}}}
\frac{c_0} {\Delta\phi} = 2\pi \sqrt{\frac{16}{15}} \frac{|B_\phi| \cdot r}{\sqrt{\rho}
\Omega \Delta\phi}$$
Fig. 1, right, shows $Q_\phi$ over time. For all models we have $Q_\phi > 8$. 
The $\pi/4$ and $\pi/2$ show a higher $Q_\phi$ due to stronger magnetic fields.
\begin{figure}
\hspace{-0.6cm}
\begin{minipage}{5cm}
\psfig{figure=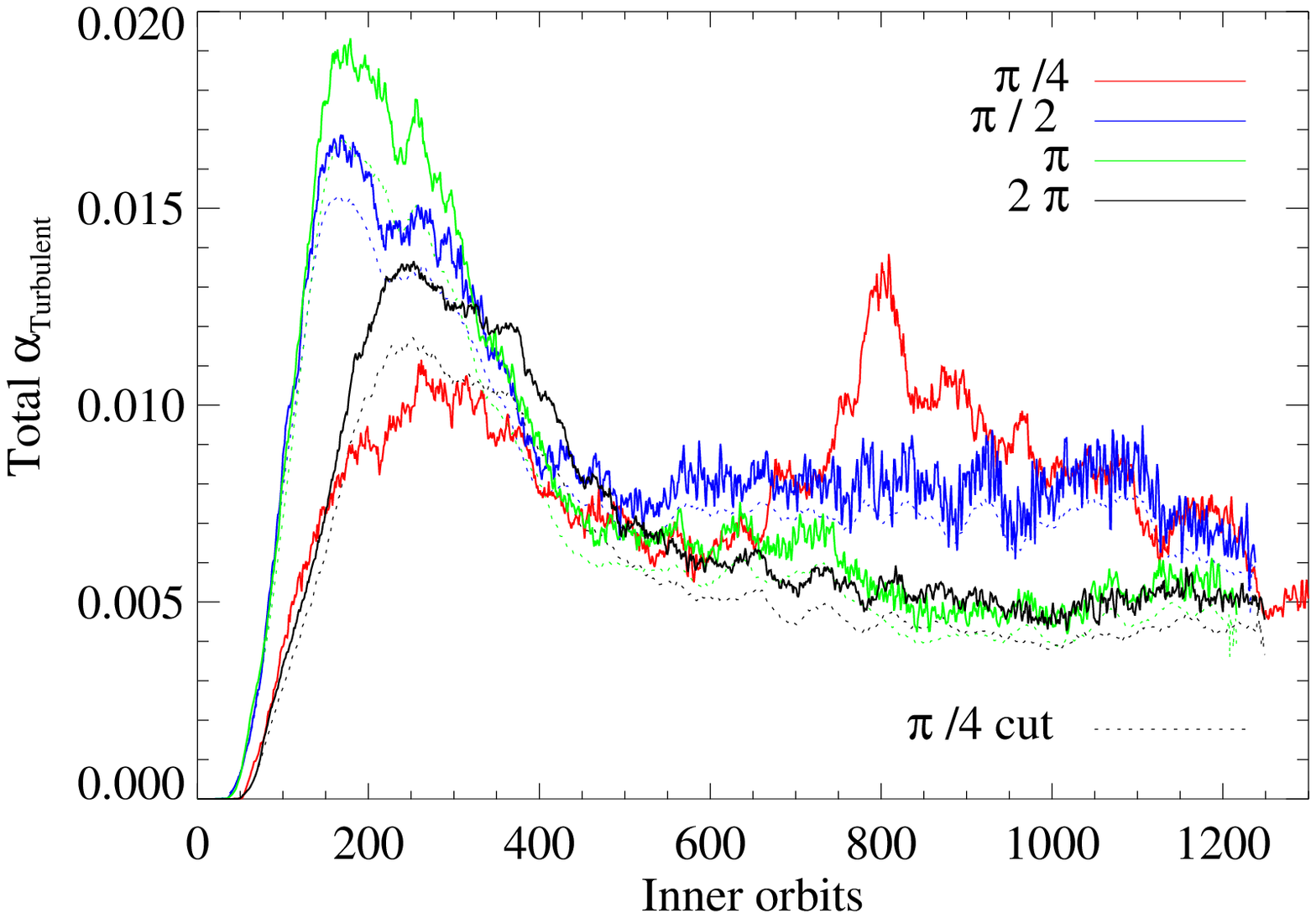,scale=0.52}
\end{minipage}
\hspace{4.0cm}
\begin{minipage}{5cm}
\psfig{figure=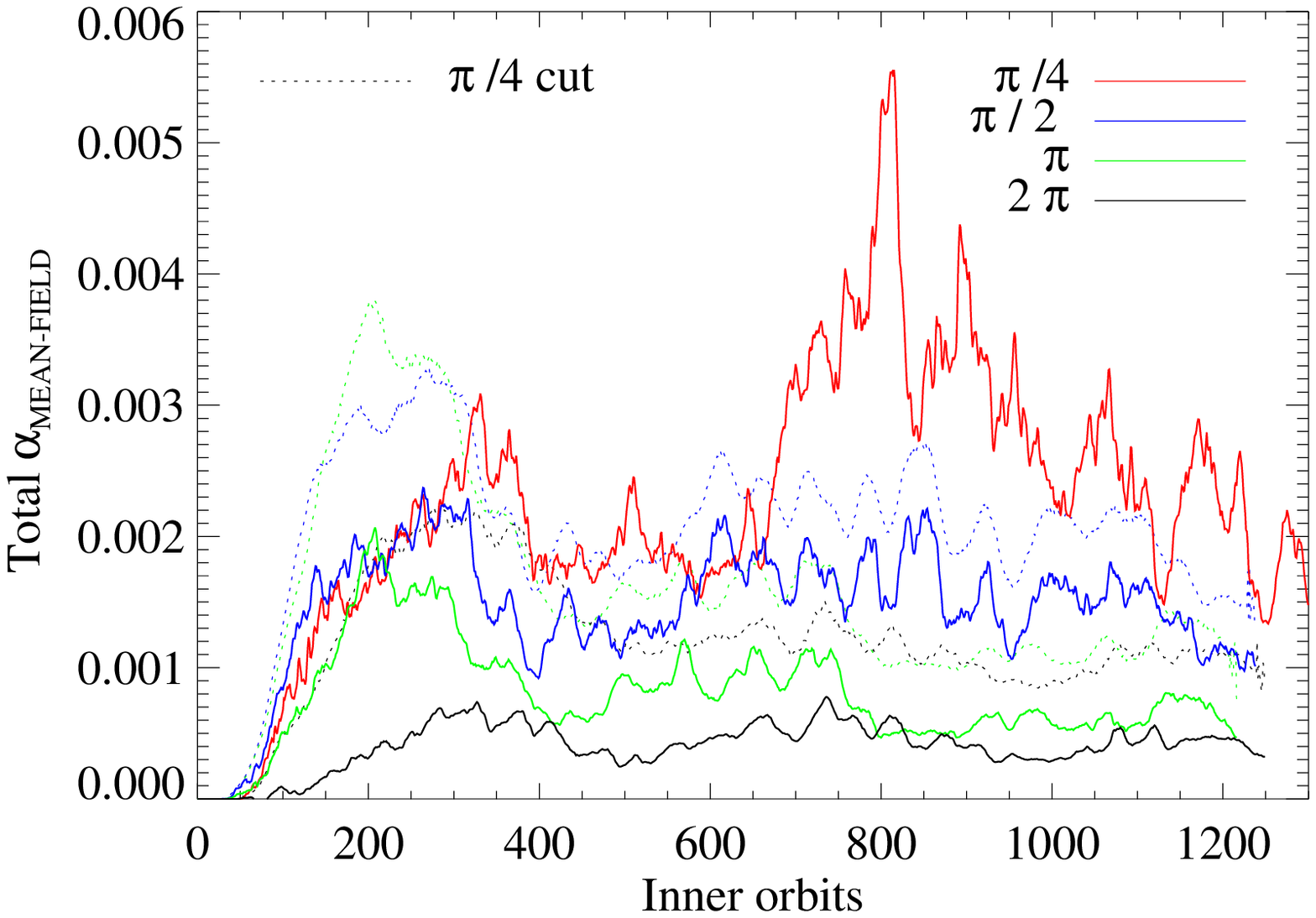,scale=0.52}
\end{minipage}
\label{flux}
\caption{Left: Volume integrated $\alpha_{SS}$ value for all models. 
Right: Volume integrated $\alpha_{SS}$ values using only the
Maxwell component with the mean magnetic fields. 
Dotted lines show same results but for a $\pi/4$ average ($0-\pi/4$) 
instead of whole domain size.}
\end{figure}
\subsection{Turbulent evolution - $\alpha$ value}
%In this section we investigate in detail 
%the turbulent evolution for all four models with different azimuthal domains.
%A closer look is done for the full $2\pi$ model as it presents the most
%realistic case.
We start the comparison with the volume integrated turbulent stress
scaled on the local pressure, e.g. the Shakura-Sunyaev $\alpha_{SS}$.
The $\alpha_{SS}$ value is determined from the turbulent Reynolds stress 
$\rm T_{R}= \overline{\rho v'_{\phi}v'_{R}} $
and Maxwell stress $\rm T_{M}= \overline{B_{\phi}B_{R}}/ 4 \pi $.
%For the Reynolds stress we are only interested in the turbulent velocity, 
% e.g., $\rm v'_{\phi} = v_{\phi} - \overline{v_{\phi}}$.\\
We split the total  
$\alpha_{\rm SS}$ into a mean and turbulent component.
For the Maxwell stress, we split the magnetic field components into the
turbulent and mean component, e.g. $\rm B_{\phi} = B'_{\phi} + \overline{B_{\phi}}$.
This leads to a second Maxwell stress component, e.g. 
the mean Maxwell stress 
$\rm T^{mean}_{M}= \overline{B_{\phi}} \cdot \overline{ B_{R}}/ 4 \pi $.
%For all different azimuthal runs, the Maxwell stress is about three times
%higher then Reynolds stress.
For the volume integrated turbulent $\rm \alpha_{SS}^{turb}$ value we integrate the mass weighted stresses over
the central domain
$$\rm \alpha_{SS}^{turb} = \frac{ \int \rho \Bigg( \frac{v'_{\phi}v'_{R}}{c^2_s} -
\frac{B'_{\phi}B'_{R}}{4 \pi \rho c^2_s}\Bigg)dV} {\int \rho dV}.$$
The same is done for the mean Maxwell stress
%We normalise the mean Maxwell stress also over the pressure
$$\rm \alpha_{SS}^{mean} = \frac{ \int \overline{\rho} \Bigg( -\frac{ \overline{B_{\phi}}
\cdot \overline{B_{R}} }{4 \pi \overline{\rho} \overline{c^2_s}}\Bigg)dV} {\int
\overline{\rho} dV}.$$\\
The volume integrated $\rm \alpha_{SS}^{turb}$ (Fig. 2 left - solid line) 
and the volume integrated $\rm \alpha_{SS}^{mean}$ (Fig. 2 right - solid line) 
are plotted versus time.
We are interested in the steady state and we use the time period between
800 and 1200 inner orbits for averaging. 
%In the first 400 inner orbits during the linear MRI stage, 
%the stresses develop differently because the growth rate of the MRI depends 
%on the size of the azimuthal domain. The azimuthal MRI 
%has different growth rates for different azimuthal modes \citep{gue07a} 
%with a maximum for the m=1 mode.
Fig. 2 (left) shows that the $\pi/4$ and $\pi/2$ models present higher
$\alpha_{SS}$ value than the $\pi$ and $2\pi$ models.
The mean magnetic fields provide a significant
contribution to the total stress for the restricted azimuthal domains, see Fig. 2,
right.
The time averaged ratio between the turbulent Maxwell stresses and 
the mean Maxwell stresses is up to 33 $\%$ for the $\pi/4$ model while it decreases in the full $2\pi$ model
down to 8 $\%$, see Table 1.
In Table 1 we summarise the results of $\rm \alpha_{SS}^{mean}$, $\rm \alpha_{SS}^{turb}$ and
$\rm \alpha_{SS}^{total}$. The standard deviation is determined by the
temporal fluctuations.
For model $\pi/4$ we determine $\rm \alpha_{SS}^{total} = (11.8\pm2.3)\cdot 10^{-3}$.
For model $\pi/2$, $\rm \alpha_{SS}^{total}$ reduces to $(9.3\pm0.9)\cdot 10^{-3}$.
%The results show that the $\pi/4$ and $\pi/2$ models present a higher 
%$\alpha$ value than the $\pi$ and $2\pi$ models.
The stress of the two largest azimuthal
domain sizes, $\pi$ and $2\pi$, matches within the standard deviation.
For model $\pi$, the time averaged $\rm \alpha_{SS}^{total}$ is $(5.6\pm0.5)\cdot10^{-3}$
and $(5.4\pm0.4)\cdot10^{-3}$ for model $2\pi$. 

To verify the results we made the same analysis in the same azimuthal
extent for every model.
Instead using the full azimuthal dataset for the analysis, we use here the azimuthal extent between $0 - \pi/4$ 
in every model.
%Instead using the full azimuthal domain we use now the azimuthal extent between $0 - \pi/4$ 
%in every model.
The results are shown in Fig. 2, dotted lines.
%We note that the stress calculated from $0 - \pi/4$ of models $\ge \pi/2 $ is
%still different from the $\pi/4$ model. 
In Fig. 2, left, these $\alpha_{SS}$ values 
are only slightly lower than the total 
domain integration. This indicate that most of the turbulent stress 
is generated by the small scale turbulence ($m \le 8$). 
%Later, we will investigate the
%spatial turbulent scale with the two-point correlation function. 
In Fig. 2, right, these $\alpha_{SS}$ values represent the stress for one specific mode $(m=8)$. 
We see again that the smaller scales contribute more to the $\rm \alpha_{SS}^{total}$
than the larger scales. We summarise that the turbulence is amplified in case for the $\pi/2$ and
$\pi/4$ model. These models present higher $\rm \alpha_{SS}^{turb}$ and
$\rm \alpha_{SS}^{mean}$ values than the $\pi$ and $2\pi$ runs.
%The results show that there is indeed a different turbulent evolution
%for the $\pi/2$ and $\pi/4$ model compared to the $\pi$ and $2\pi$ run. 

\subsection*{Accretion burst due to mean fields}
The $\pi/4$ run presents another exceptional behaviour. Around 800 inner
orbits, the $\alpha$ value increases quickly up to $\alpha = 0.013$. The reason
for this increase is connected to 
strong mean toroidal field oscillations. In Fig. 3 we plot contour lines of
the resolved $\lambda_{crit}^{\overline{B_\phi}}$ from the mean toroidal field
$\overline{B_\phi}$ with
$\lambda_{crit}^{\overline{B_\phi}}/\Delta\phi \ge 8$. 
$$\frac{\lambda_{crit}^{\overline{B_\phi}}}{\Delta\phi} = 2\pi \sqrt{\frac{16}{15}} \frac{|\overline{B_\phi}| \cdot r}{\sqrt{\rho}
\Omega \Delta\phi}$$
The definition is equivalent to the definiton of the toroidal quality factor $Q_\phi$
but calculated from the mean toroidal field instead from the total field (see Fig. 1, right).
There is clear correlation between the rise of the $\alpha_{SS}$ value and
resolved mean toroidal field. At the same time there is a superposition 
of strong mean field along radius, see Fig. 3 red solid line.
The amplifications are present in the $\pi/4$ model, Fig. 3 top, and the 
$\pi/2$ model, Fig. 3 bottom. For the larger domains, $\pi$ and $2\pi$ 
(Fig. 4), the mean field stays at lower values and
$\lambda_{crit}^{\overline{B_\phi}}$ is not resolved.

\subsection*{Turbulent magnetic and velocity fields}
We investigate the spatial distribution of magnetic energy with Fourier 
analysis.
%The spatial distribution of magnetic energy shows also
%the level of turbulence.
The magnetic field amplitudes, $\sqrt{B(m)^2}$ are plotted 
in Fourier space along azimuth at the midplane and for all
models, Fig. 5, left.
%The plots show that most of the magnetic energy is placed at the largest
%domain size.
The plots show that the highest amplitudes of the magnetic fields 
are at the largest scales.
The $\pi/4$ and $\pi/2$ model show systematically increased amplitudes compared to the
$\pi$ and $2\pi$ model. 
This is true for all modes and for all three magnetic field components.
It is also visible in the time averaged total magnetic energy, Fig. 1 left
dotted lines.
Time averaged values, in units of the initial total magnetic energy, 
are $B^2/B_0^2 = 0.54\pm0.12$ for model $\pi/4$, 
$0.48\pm0.09$ for model $\pi/2$, $0.34\pm0.07$ for
model $\pi$ and $0.35\pm0.07$ for model $2\pi$. Here, time average is done
between 400 and 1200 inner orbits.
We present the velocity field in Fourier space $\sqrt{V(m)^2}$ 
in Fig. 5, right. We observe increased turbulent velocities for the restricted
domain models. %All models show the same profile for smaller scales. 
%Similar to the magnetic fields, we present the velocity field in Fourier space $\sqrt{V(m)^2}$ 
%in Fig. 5, right. We see the increasing of turbulent velocities for the restricted 
%azimuthal domains. 
The radial velocity (dashed line) dominates in the range between $ 2 \lesssim m \lesssim
40$. The peak turbulent velocity is $V_r$ at $m=4$ for the $\pi/2$, $\pi$
and $2\pi$ run. Coincidentally, this
mode matches the domain size of $\pi/2$. 
The $\pi/4$ does not include this mode. This lack of large scale turbulent
radial fields becomes again visible in the velocity tilt angle.
The peak at $m=4$ ($\rm 22H$) is connected to spiral density waves. After 
\citet{hei09} we should observe the peak at $m=14$ ($\rm 6H$). 
This could be a resolution issue as the domain size of 
$\pi/4$ ($\rm 11H$) should be large 
enough to include spiral density waves.

\begin{figure}
\psfig{figure=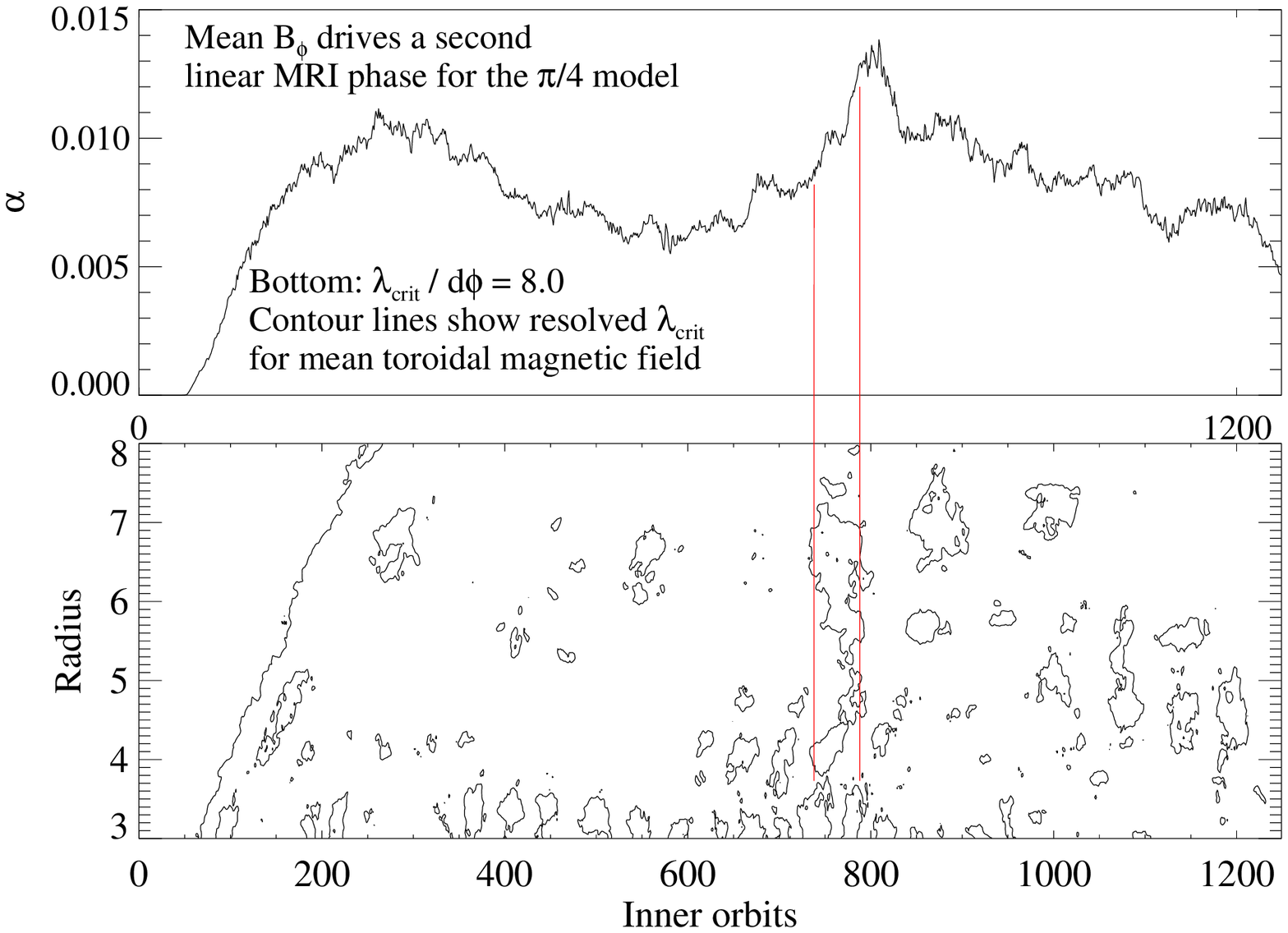,scale=0.76}
\psfig{figure=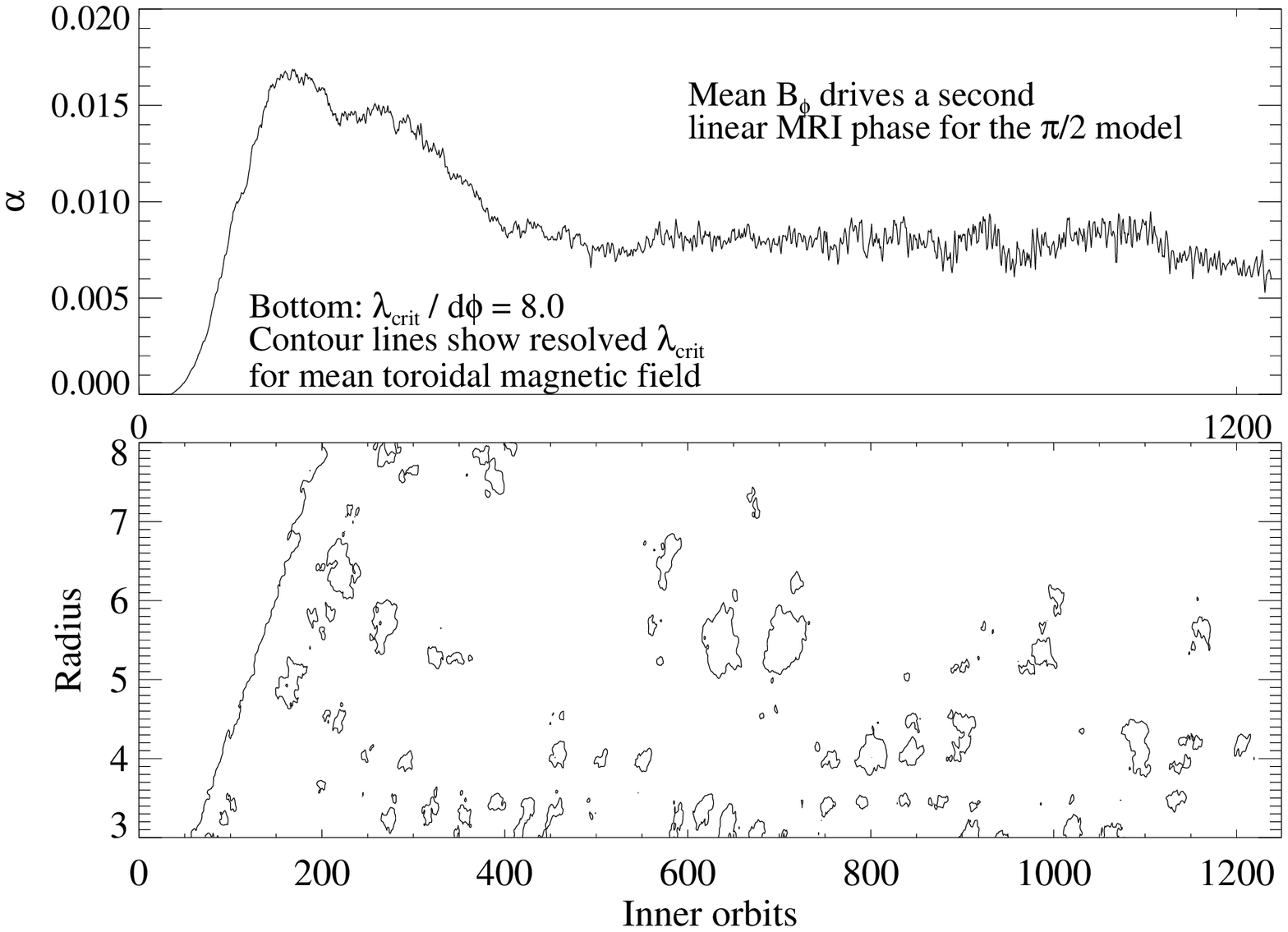,scale=0.76}
\label{dyn-cor2}
\caption{Contour lines of 
the resolved MRI from the mean toroidal field $\lambda_{crit}^{\overline{B_\phi}}$ with
the evolution of the $\alpha$ value for the models $\pi/4$ (top) and $\pi/2$
(bottom). The contour lines show $\lambda_{crit}^{\overline{B_\phi}}=8$. 
The strong mean toroidal field amplifies the turbulence.}
\end{figure}

\begin{figure}
\psfig{figure=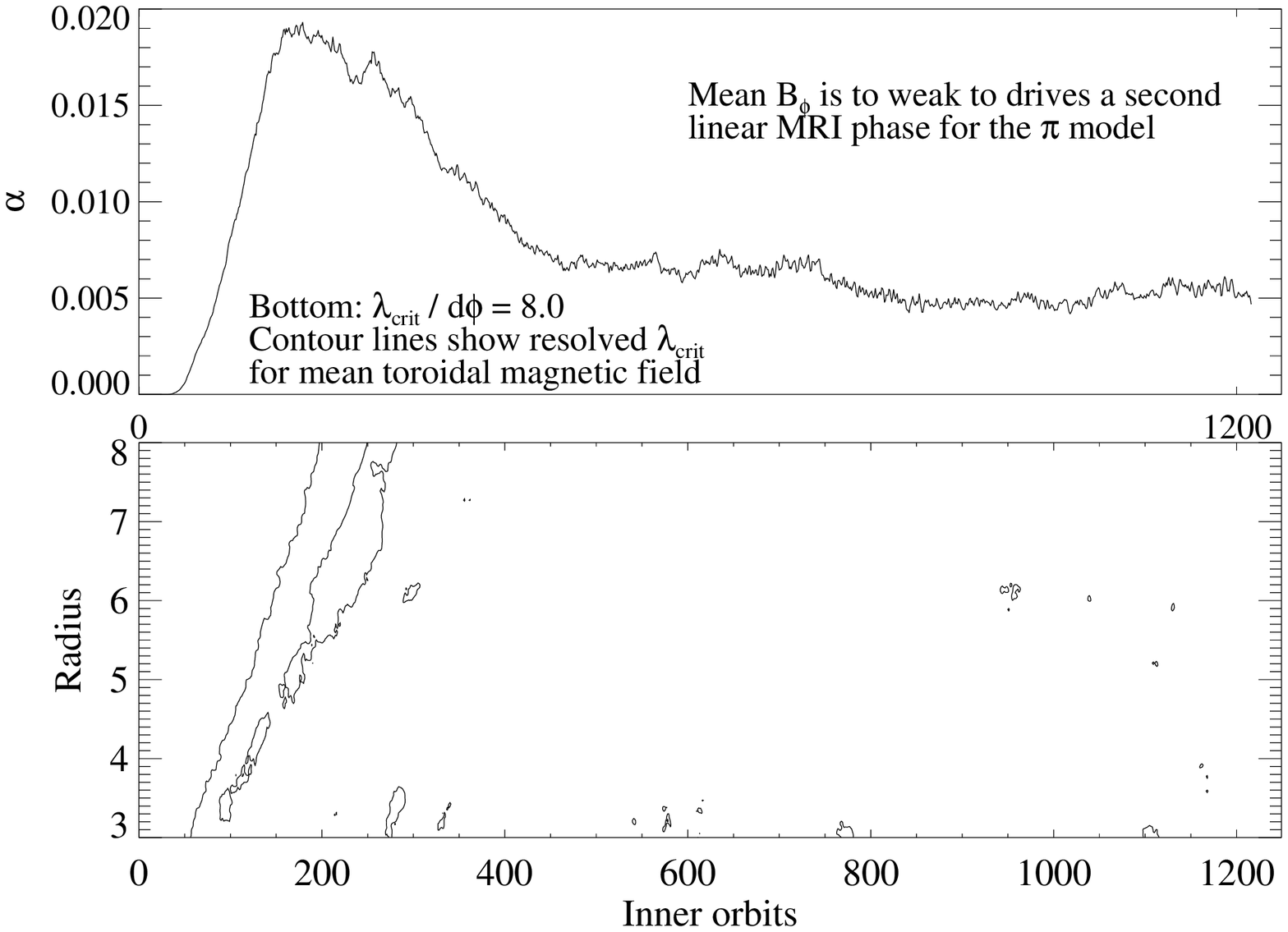,scale=0.76}
\psfig{figure=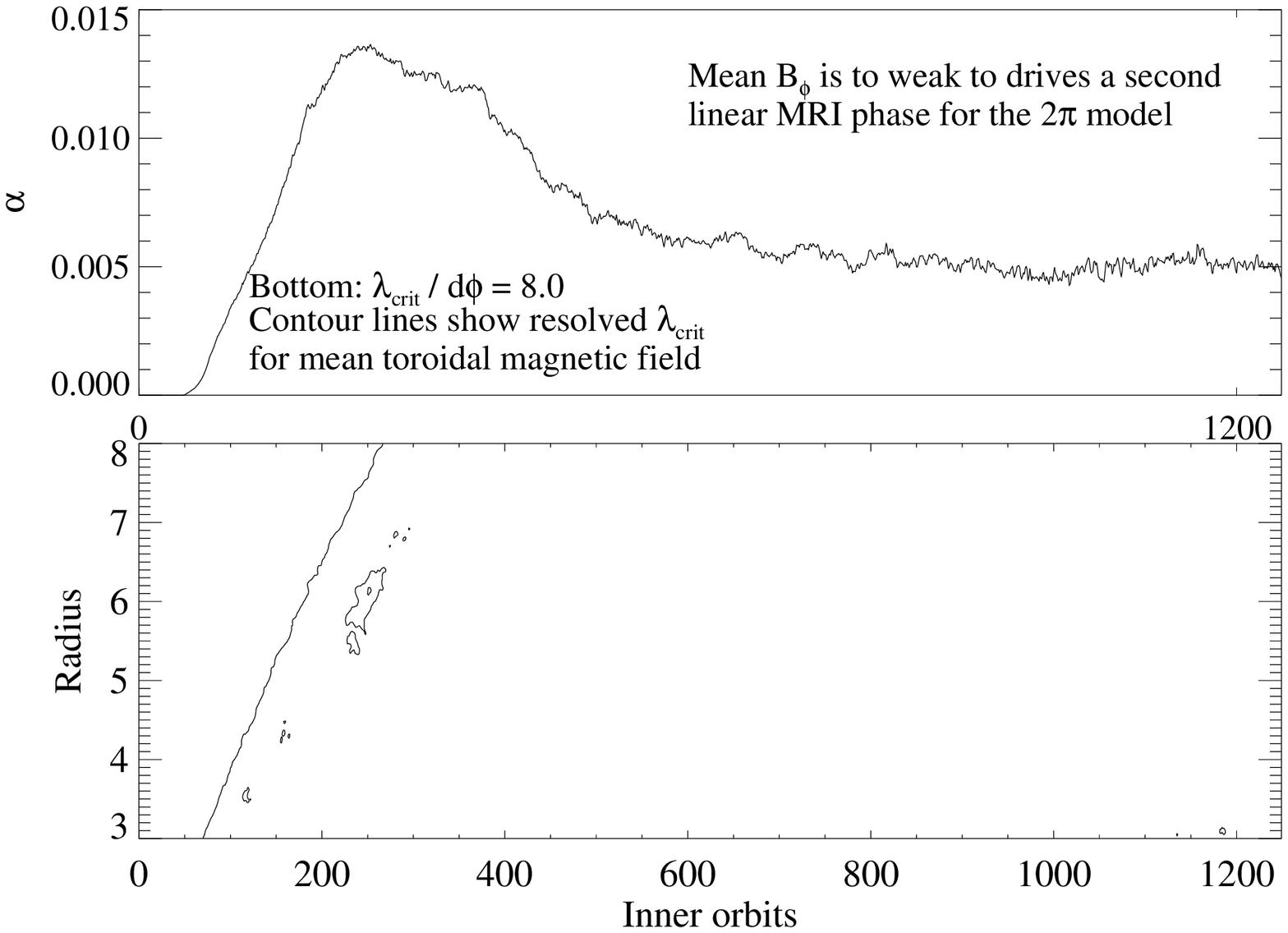,scale=0.76}
\label{dyn-cor3}
\caption{Contour lines of 
the resolved MRI from the mean toroidal field $\lambda_{crit}^{\overline{B_\phi}}$ with
the evolution of the $\alpha$ value for the models $\pi$ (top) and $2\pi$
(bottom). The contour lines show $\lambda_{crit}^{\overline{B_\phi}}=8$.
Here, the mean toroidal field is weaker and not resolved by the code.}
\end{figure}

\begin{figure}
\hspace{-0.6cm}
\begin{minipage}{5cm}
\psfig{figure=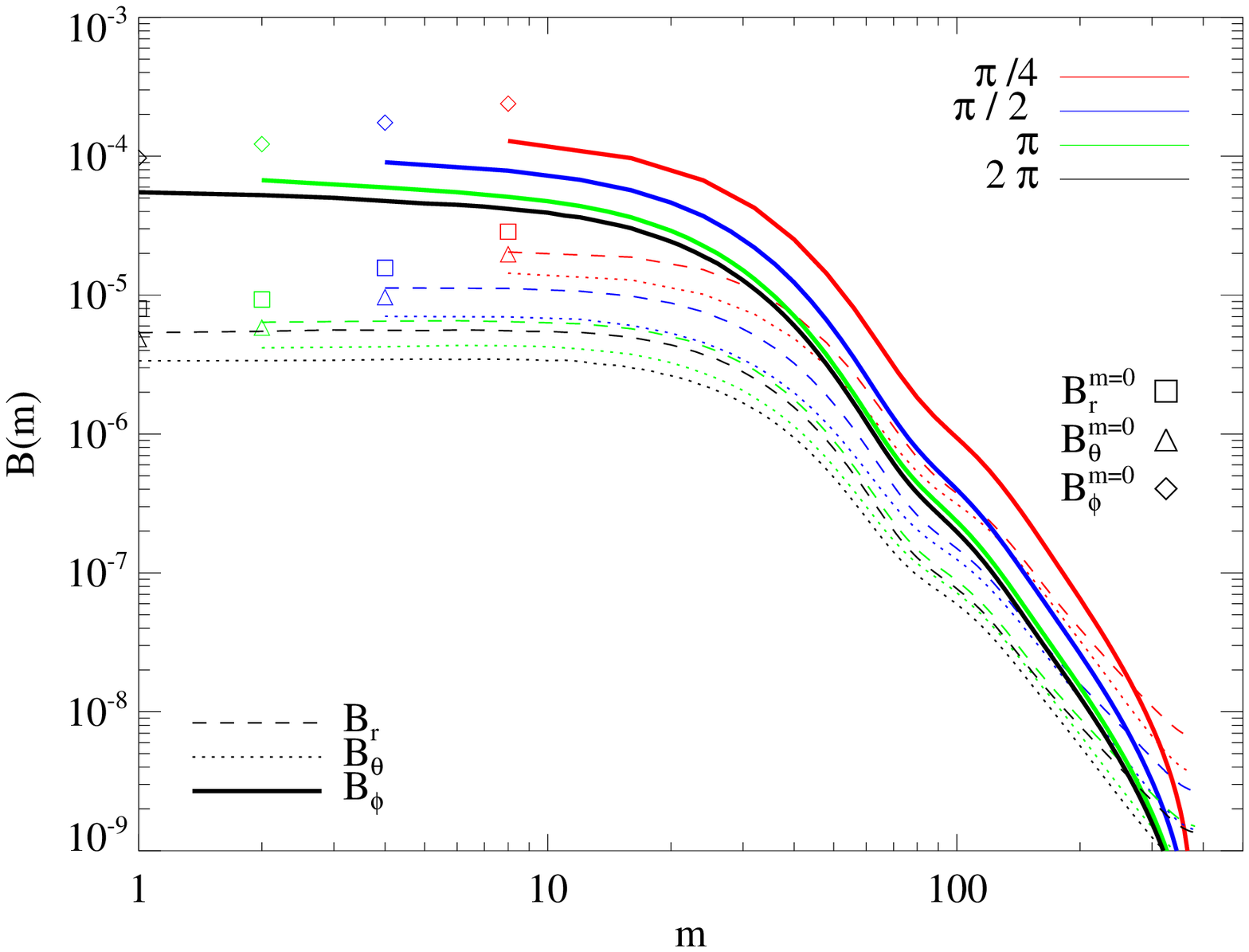,scale=0.46}
\end{minipage}
\hspace{4.0cm}
\begin{minipage}{5cm}
\psfig{figure=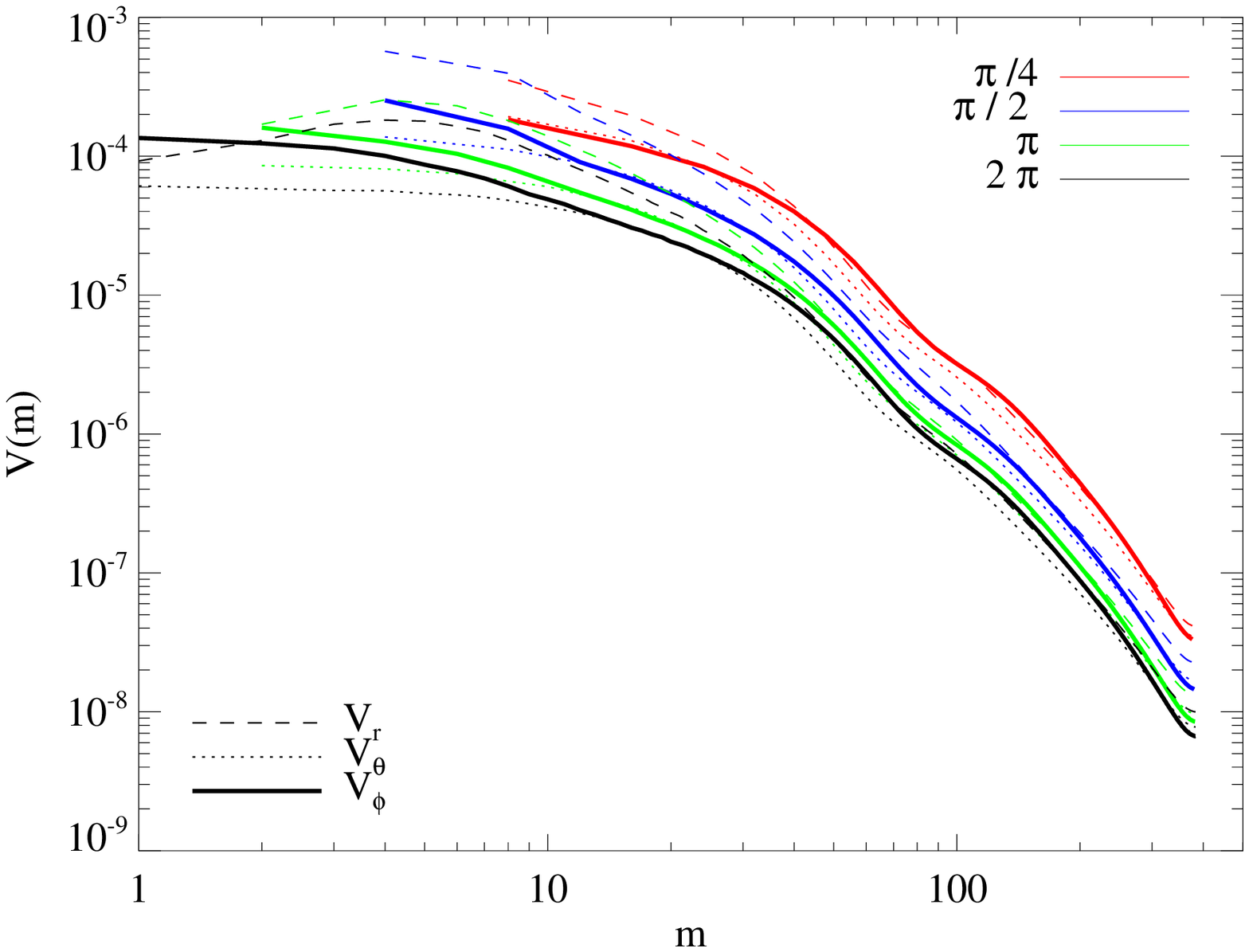,scale=0.46}
\end{minipage}
\label{dyn-cor5}
\caption{Left: Magnetic field distribution in Fourier space over 
azimuthal wave number for all models and magnetic field components. 
Right: Same for the velocity field.
Values are from the midplane and 
time averaged between 800 and 1200 inner orbits.}
\end{figure}

%Therefor, e.g. the mean field of the $\pi/4$ run represent the
%$\rm m=8$ mode in the full $2\pi$ run.
%In Fig. 1, right, we plotted the Maxwell stress only for the the mean field
%values for $B_r$ and $B_\phi$. Again, the comparison is done also with the
%split-ted azimuthal domain values, over-plotted with the dotted line.
%The contribution of the mean component of the Maxwell Stress to the total
%value is presented in Table 2.
%In contrast are again the split-ted domain values (Fig.1, right, dotted
%line). Even the values are bigger what is expected because most of the turbulence
%acts on the smallest resolved scale, they do not show the same strength of the
%$\pi/4$ run.\\
%To recall again, the only difference for the restricted domains are the
%exclusion of larger modes. For model $\pi/4$ only modes $m \le 8$ can
%occur, for model $\pi/2$ it holds for modes $m \le 4$ and so on.
%The amplitudes for the magnetic fields increases with
%decreasing the azimuthal domain. 
%This is also seen in the total magnetic energy, Fig. 2, right, compared for
%all models in a time interval between 800 and 1200 inner orbits.\\
\begin{table}
\begin{center}   
\begin{tabular}{cccccccc}
$\Delta \phi$ & $\rm \alpha_{SS}^{total}10^{-3}$ & $\rm \frac{
\alpha_{SS}^{mean}}{\alpha_{SS}^{total} }$ & $\rm \alpha_{SS}^{turb}10^{-3}$ &
$\rm \alpha^{SH}_{\phi\phi}10^{-3}$ 
& $\rm \alpha^{NH}_{\phi\phi}10^{-3}$ & Parity & $\rm V_{RMS}$
$[c_s]$ \\
\hline
\hline
$\pi/4$ & $ 11.8 \pm 2.3  $  & 0.33  &  $8.9 \pm 1.6  $  &  $-3.4 \pm 0.9 $   & $ 3.3 \pm 0.8 $ & 
$-0.2 \pm 0.4$ &
$0.125 \pm 0.009$ \\
$\pi/2$ & $ 9.3 \pm 0.9 $ & 0.19 &  $7.8 \pm 0.7  $  &$-2.8 \pm 0.6  $   & $ 3.1 \pm 0.7 $ &
$-0.2 \pm 0.5$&
$0.148 \pm 0.006$ \\  
$\pi$   & $5.6 \pm 0.5$ & 0.12 &   $5.0 \pm 0.4  $   &$-2.4 \pm 0.3  $  & $ 2.1 \pm 0.3 $ &
$-0.1 \pm 0.5$&
$0.112 \pm 0.005$ \\
$2\pi$  & $5.4 \pm 0.4 $ & 0.08 &   $5.0 \pm 0.3  $  &$-2.3 \pm 0.2  $ & $ 2.1 \pm 0.2 $ &
$0.2 \pm 0.4$&
$0.113 \pm 0.005$ \\ 
\hline
%$\pi/4^{Pm=5500} $ & $5.1 \pm 1.2$ & 0.36 &  $3.4 \pm 1.1 $ & $1.7 \pm 0.4$  & 
%$-2.2 \pm 0.4 $ & $0.064 \pm 0.008 $  \\
%$2\pi^{Pm=5500} $ & $2.0 \pm 0.7 $ & 0.03 & $2.0 \pm 0.7$ & $0.8 \pm 0.1$  & 
%$-1.0 \pm 0.1 $ & $0.064 \pm 0.007 $  \\
%\hline
\\
\end{tabular}
\caption{Model overview. From left to right: Azimuthal
domain; Volume integrated total stress; Relation between 
$\rm \alpha_{SS}^{mean}$ to $\rm \alpha_{SS}^{turb}$; 
$\rm \alpha_{SS}^{turb}$ stress;
Value of dynamo $\rm \alpha^{SH}_{\phi\phi}$ for southern hemisphere (lower disk); 
Value of dynamo $\rm \alpha^{NH}_{\phi\phi}$ for northern hemisphere (upper disk).}
\end{center}
\end{table}

\begin{figure} 
\vspace{-0.6cm}\hspace{-0.6cm}
\begin{minipage}{5cm}
\psfig{figure=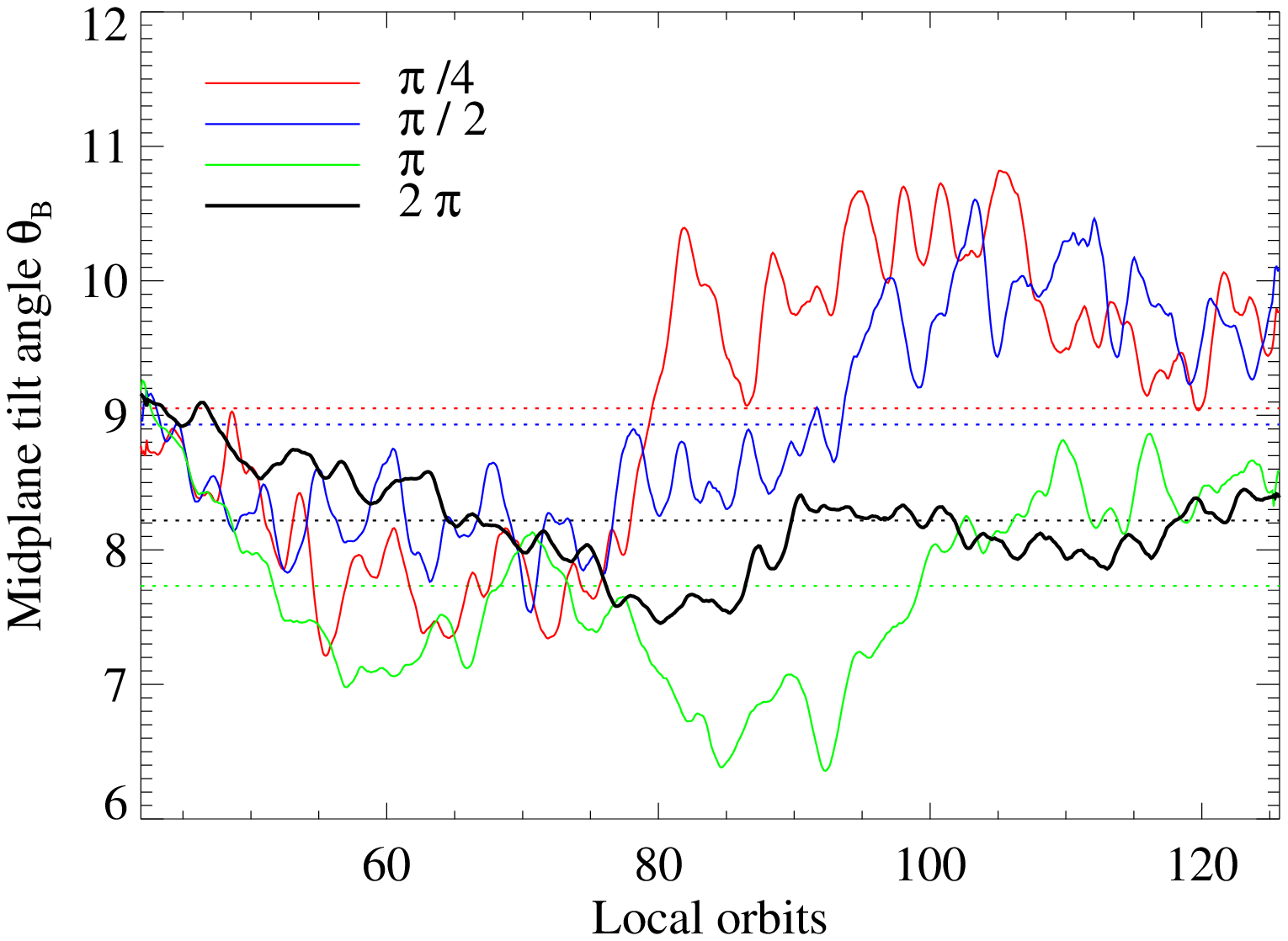,scale=0.46}
\psfig{figure=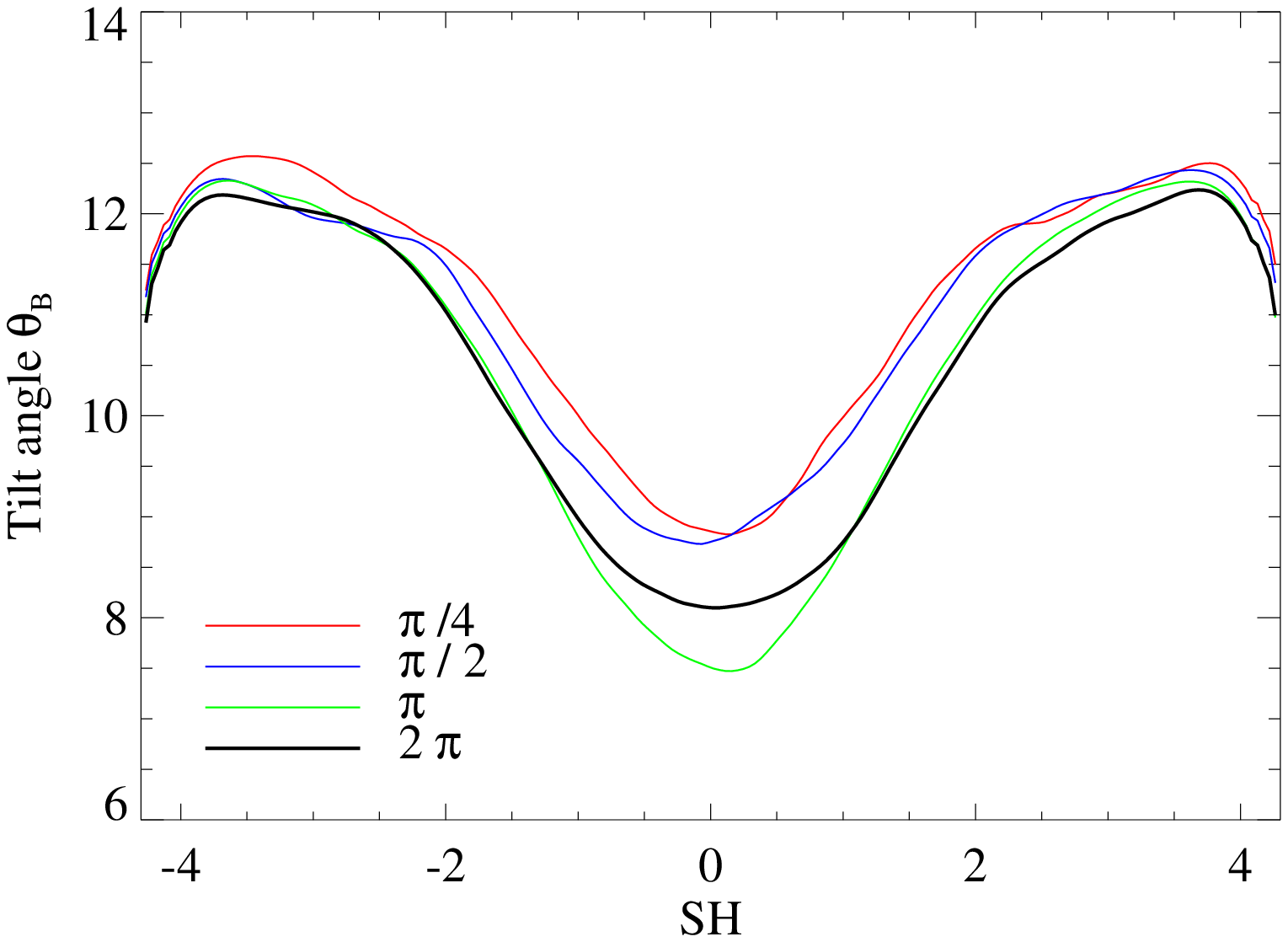,scale=0.46}
\end{minipage}
\hspace{4.0cm}
\begin{minipage}{5cm}
\psfig{figure=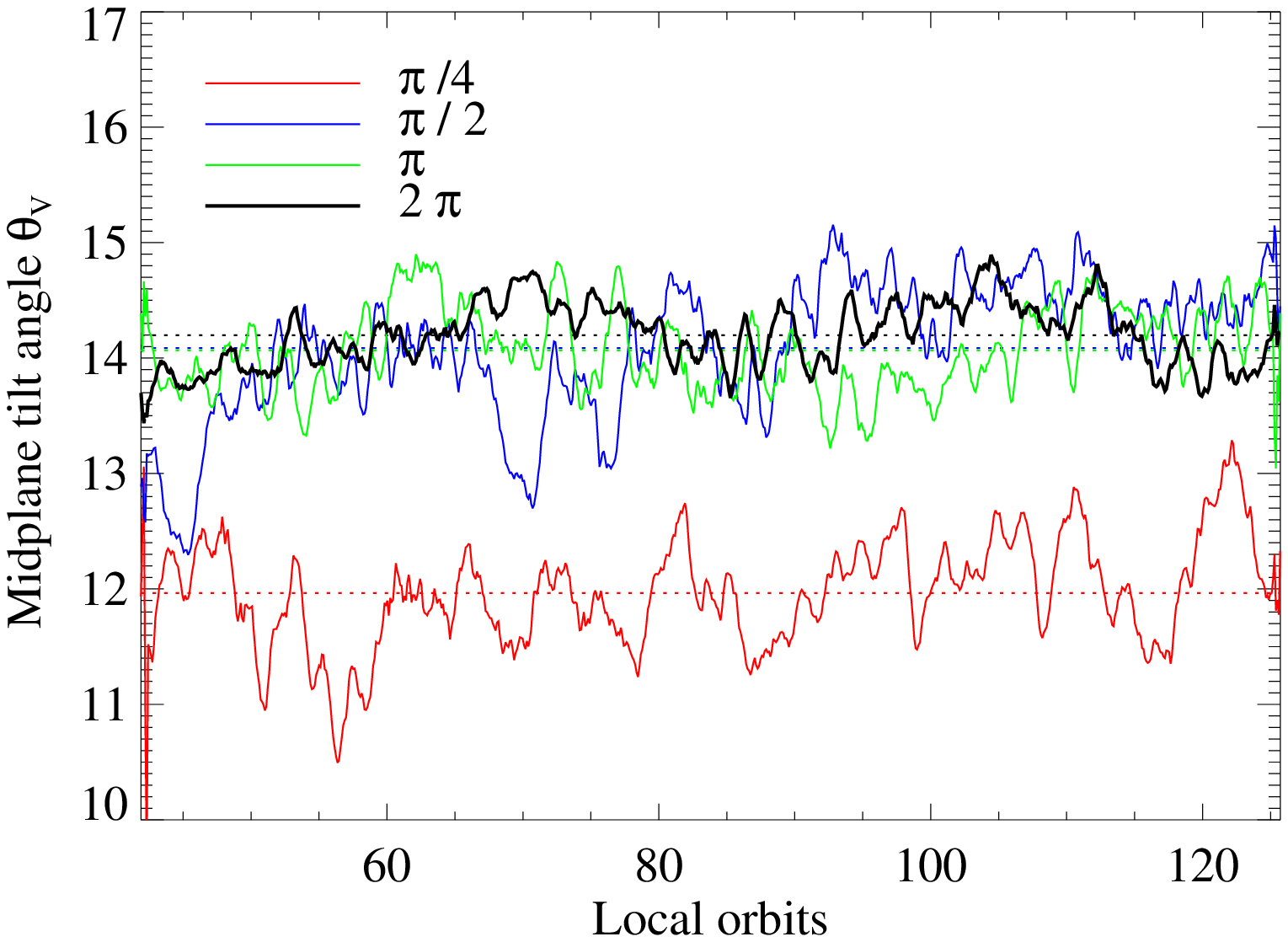,scale=0.46}
\psfig{figure=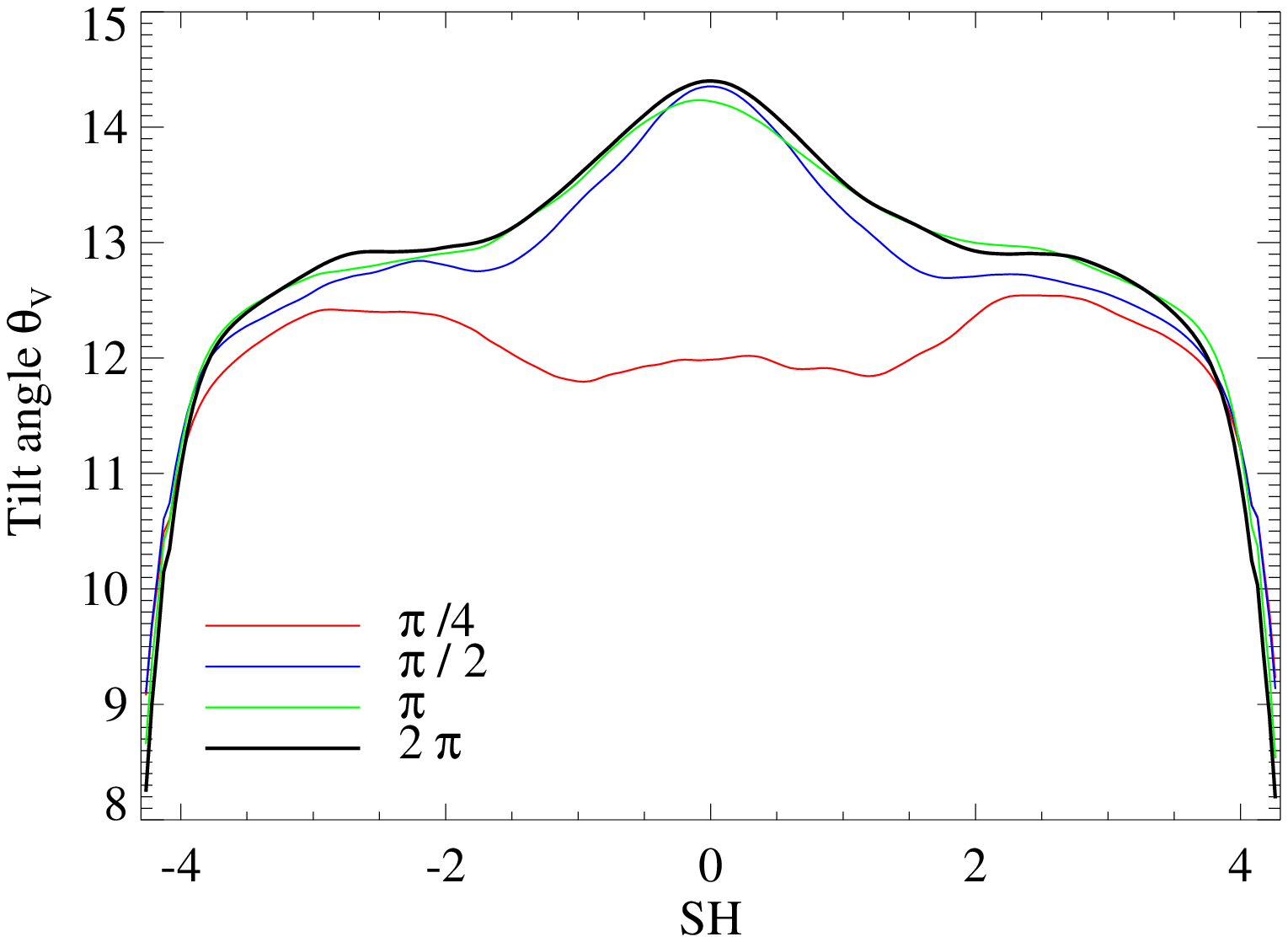,scale=0.46}
\end{minipage}
\label{cor-vel}
\caption{Top left: Midplane magnetic tilt angle over time for all models. 
Bottom left: Time averaged magnetic tilt angle for all models.
Top right: Midplane velocity tilt angle over time for all models. 
Bottom right: Time averaged velocity tilt angle for all models.} 
\end{figure}

\subsection*{Two-point correlation function}
The two-point correlation function, specified for MRI by \citet{gua09}, 
allows to study the locality and anisotropy of the turbulence.
%specific for MRI 
%was presented by \citet{gua09}
%using two-point correlation functions of the velocity and magnetic
%field. 
We measure the tilt angle for the 
magnetic $\sin{2\theta_B} = |B_rB_\phi|/B^2$ and 
the turbulent velocity field $\sin{2\theta_V} = |V'_rV'_\phi|/V'^2$ at 4.5 AU.
In Fig. 6, we plot the time evolution, top, and the vertical distribution,
bottom, of the magnetic tilt angle $\theta_B$, left, and the velocity
$\theta_V$, right. 
The time evolution of the magnetic tilt angle $\theta_B$ is plotted in 
Fig. 6 top left. The $\pi/4$ and $\pi/2$ model show higher tilt angles 
($\theta_B \sim 9^\circ$) with much higher time deviations as the
$\pi$ and $2\pi$ model ($\theta_B \sim 8^\circ$).
%The $\pi/4$ and $\pi/2$ model show at the same time higher fluctuations.
The $\pi/4$ model shows sudden increase of the tilt angle at 80 local orbits.
At this time, the turbulence gets amplified due to strong axisymmetric
fields, see Fig. 3.
%For $\pi/4$ we note also the increase of the tilt angle at 80 local orbits 
%as the turbulence increases again.
%At the midplane, the time averaged magnetic tilt angle $\theta_B$ of the
%$\pi/4$ and $\pi/2$ model is with $\sim 9^\circ$ slightly higher compared 
%to the $\pi$ and $2\pi$ model $\sim 8^\circ$. The time evolution shows also the   
The time averaged vertical profile of $\theta_B$ is plotted in Fig. 6,
bottom left. The tilt angle present the highest values in the coronal
region. 
Here, we see again higher $\theta_B$ values for the $\pi/4$ and
$\pi/2$. 
The $\pi$ model shows smaller $\theta_B$ at the midplane compared
to $2\pi$ which is an artefact of the selected time average. 
Both models present equal values after 100 local orbits, see 
Fig. 6, top left.

We do the same analysis for the velocity tilt angle $\theta_V$.
The time evolution for $\theta_V$ does not show strong fluctuations. 
At the midplane, 
we measure a time averaged velocity tilt angle of $\theta_V \sim 14^\circ$ for
all models except of $\pi/4$. The $\pi/4$ model shows a systematic 
lower tilt angle $\theta_V^{\pi/4} \sim 12^\circ$. This becomes also visible
in the vertical profile. Here all models, except $\pi/4$, show a peak of $\theta_V$ at the
midplane. The reason is unresolved density waves. 
%The $\pi/4$ model does not resolve the peak of density waves 
The $\pi/4$ model does not resolve the density waves with $\rm m=4$. 
At $\rm m=4$, all models show the highest turbulent amplitude in the radial
velocity.
For model $\pi/2$ it matches the size of the domain and it is not captured
by model $\pi/4$.
The fast drop of magnetic and velocity tilt angles above 4 scale height could be
due to boundary effects.

We calculate the two-point correlation functions in the $r-\phi$ plane:
$\epsilon_V = < \delta V_i(\vec{x}) \delta V_i(\vec{x}+\Delta \vec{x} )>$
and $\epsilon_B = < \delta B_i(\vec{x}) \delta B_i(\vec{x}+\Delta \vec{x}
)>$ with $\vec{x}={r,\phi}$.
%All results are calculated around 5 AU.
In Fig. 7 and Fig. 8 we present the two-point correlation function at 5 AU at 1 scale height 
with $\Delta r = 2H = 0.7 AU$ and the total $\phi$ domain 
$r\Delta\phi = \phi^{Domain}/0.07 H$. For the $2\pi$ model we have around
$\rm 90H$ ($2\pi/0.07$).
The corresponding major and minor wavelength are calculated using the half
width at half maximum (HWHM) in units of $\rm H$ $(\rm H|_{\rm 5AU}=0.35AU)$.
It measures the distance between the center $\epsilon = 1.0$ and $\epsilon =
0.5$ along the major $\lambda_{maj}$ and minor $\lambda_{min}$ axis, see 
footnote 7 in \citet{gua09}.
We measure the two-point correlation function at different heights.
The results between $\rm \pm 2 H$ are similar and we present the values at 1 scale
height.
For the velocity, the $\rm \lambda_{maj}$ of the $\pi/4$ run is 
$1.1 H$. The $\pi$ and $2\pi$ run present both a value of $1.9 H$.
We find a similar increase for the $\rm \lambda_{min}$, from $0.19 H$ for
$\pi/4$ to $0.24 H$ and $0.23 H$ for model $\pi$ and $2\pi$.
The values of the $\pi/2$ model present the highest values, $\rm \lambda_{maj}
= 2.0H$ and $\rm \lambda_{min}= 0.29H$. This is again due to the peak 
of turbulent radial velocity at domain size, see Fig. 5, right. It is 
visible in the magnetic fields too.
%The relative low tilt angle show that the turbulent eddies are stretched
%along azimuth. The $\lambda_{maj}$ which is pointing mostly in the azimuthal
%direction is around one order of magnitude larger
%then the $\lambda_{min}$, which points mostly in the radial direction.
The $\rm \lambda_{min}$ value for the magnetic fields are 0.14 H, except 
the $\pi/2$ model with 0.16 H.
The $\rm \lambda_{maj}$ 
increases with increasing the azimuthal domain, the $\pi/4$ model with $1.1 H$ to $1.4
H$, $1.6 H$ and $1.7 H$ for the full $2\pi$.
All results of the tilt angels, major and minor wavelengths 
are summarised in Table 2.
%We present the contour plot of the two-point correlation function for the
%velocity in Fig. 3 and for the magnetic field in Fig. 4. 
%All models show the ellipsoidal shape which is very similar to unstratified 
%and stratified local box simulations \citep{gua09,fro10,dav10} and recent 
%global simulations \citep{bec11}.
%The angle $\theta_t$ for the velocity field is between
%$5^\circ$ and $8^\circ$ degree dependent on the model. The $\theta_t$ for the 
%magnetic field lies between $4^\circ$ and $7^\circ$. 
%The $\pi/4$ and $\pi/2$ model show an artificially increased tilt angle for
%the velocity and magnetic field. The major and minor wavelengths are thereby reduced.
%
%
%The tilt angels, major and minor
%wavelengths are summarized in Table 1 in units of $\rm H$ $(\rm H(5AU)=0.35AU)$.
%With increasing the azimuthal domain the angle decreases for the velocity
%and the magnetic fields.
%The corresponding major and minor wavelength are calculated using the half
%width at half maximum (HWHM) \citet{gua09}. 
%These method produces only slightly higher
%values for the wavelengths, as shown by \citet{gua09}.
%The tilt angels, major and minor wavelengths 
%are summarized in Table 1 in units of $\rm H$ $(\rm H(5AU)=0.35AU)$.

%  0.13 H could be due to the grid resolution
%  represent 2 grid cells at 5 AU
\begin{table}
\begin{center}   
\begin{tabular}{ccccccc}
$\Delta \phi$ & $\rm \theta_V$ & $\rm \lambda_{maj}^{Vel.}$ & $\rm \lambda_{min}^{Vel.}$
&               $\rm \theta_B$ & $\rm \lambda_{maj}^{Mag.}$ & $\rm \lambda_{min}^{Mag.}$ \\
\hline
\hline
$\pi/4$ & 12.0 & 1.1 H  & 0.19 H & 9.1 & 1.1 H & 0.14 H \\
$\pi/2$ & 14.1 & 2.0 H  & 0.29 H & 8.9 & 1.4 H & 0.16 H \\  
$\pi$   & 14.1 & 1.9 H  & 0.24 H & 7.7 & 1.6 H & 0.14 H \\
$2\pi$  & 14.2 & 1.9 H  & 0.23 H & 8.2 & 1.7 H & 0.14 H \\ 
\hline
\\
\end{tabular}
\caption{Two-point correlation values for all runs. From left to right:
Azimuthal domain, correlation angle for the velocity, wavelength
of the major axis, wavelength of the minor axis, correlation angle for the 
magnetic field, wavelength of the major axis, wavelength of the minor axis.}
\end{center}
\end{table}
\begin{figure} 
\vspace{-0.6cm}\hspace{-0.6cm}
\begin{minipage}{5cm}
\psfig{figure=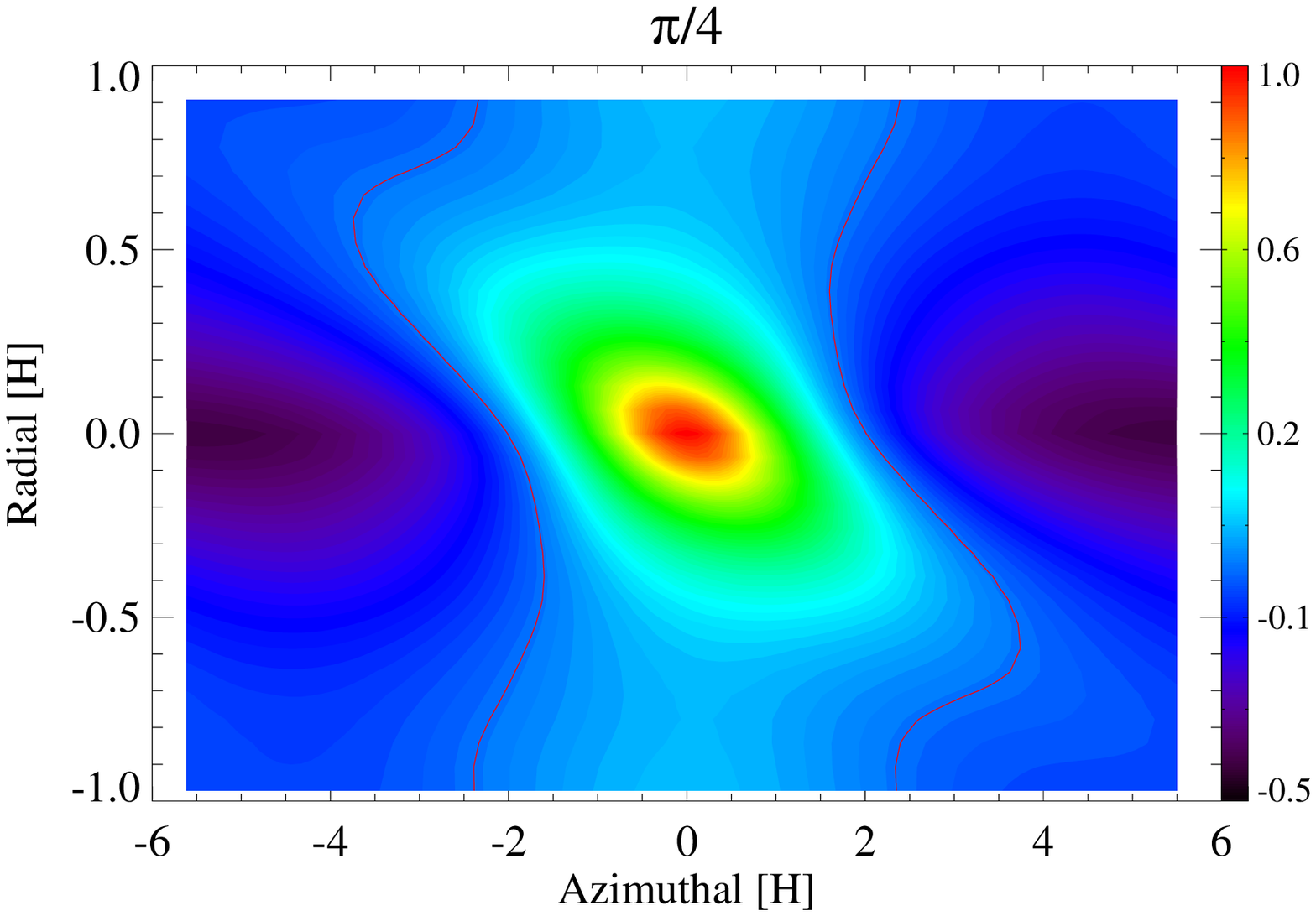,scale=0.46}
\psfig{figure=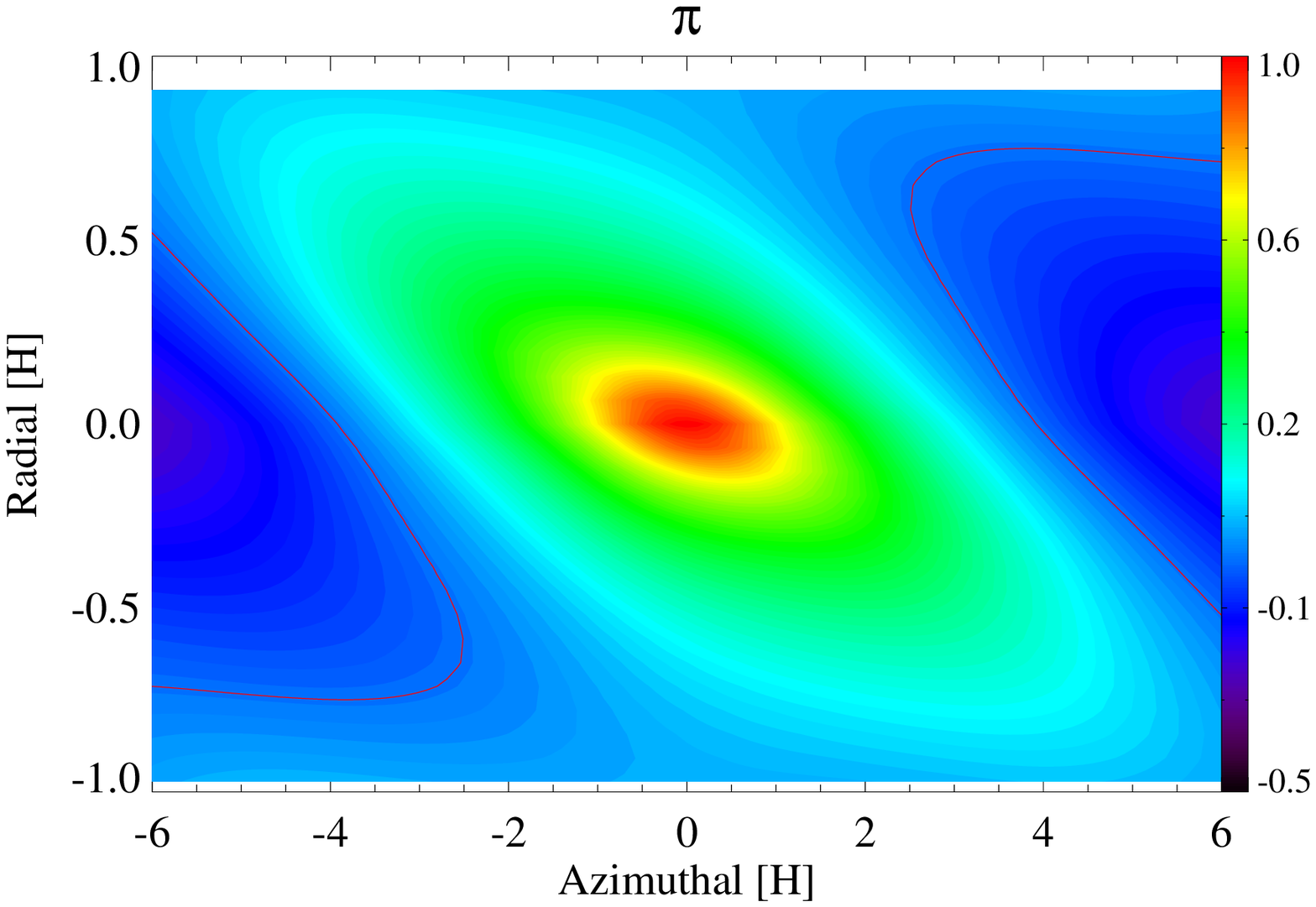,scale=0.46}
\end{minipage}
\hspace{4.0cm}
\begin{minipage}{5cm}
\psfig{figure=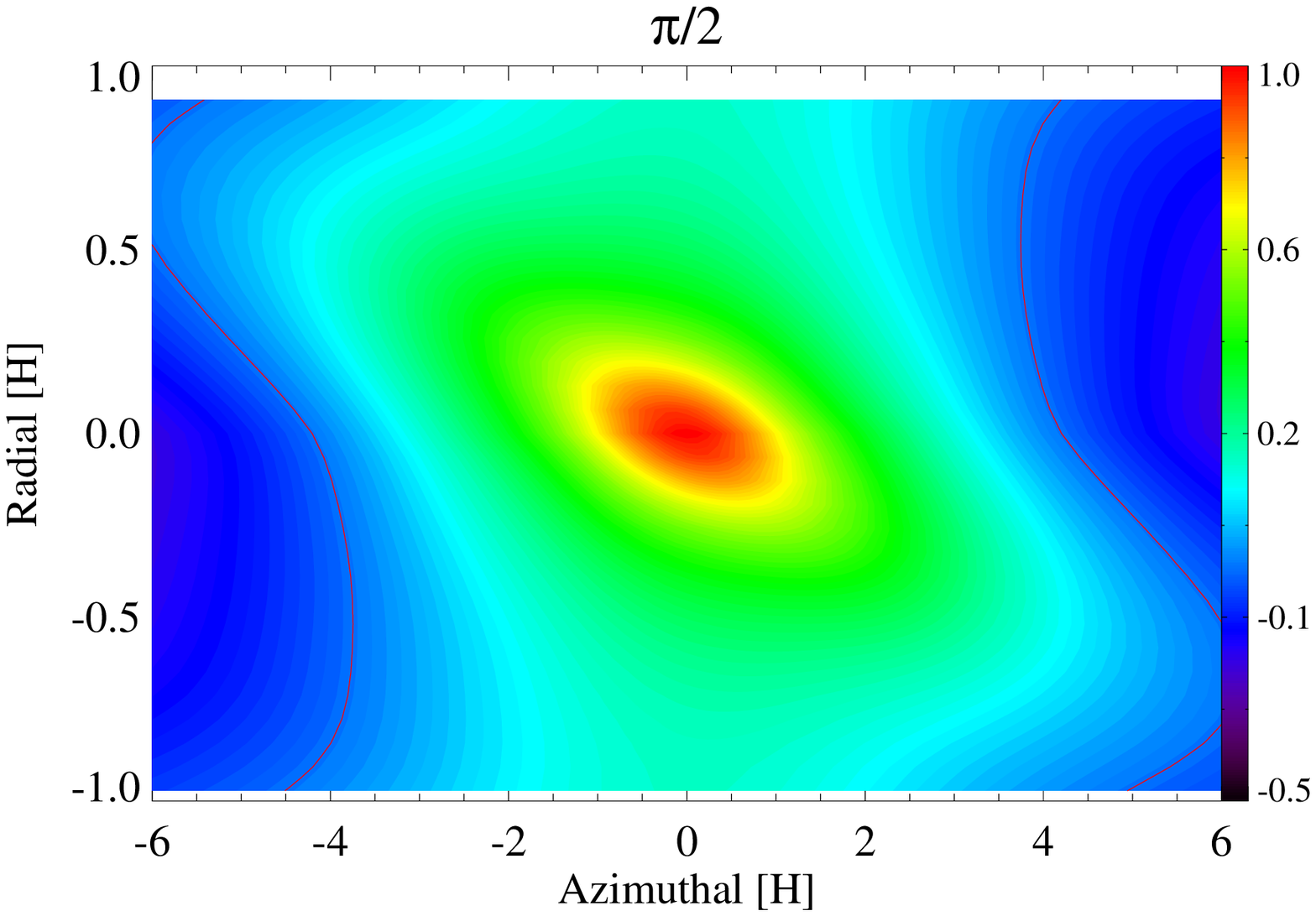,scale=0.46}
\psfig{figure=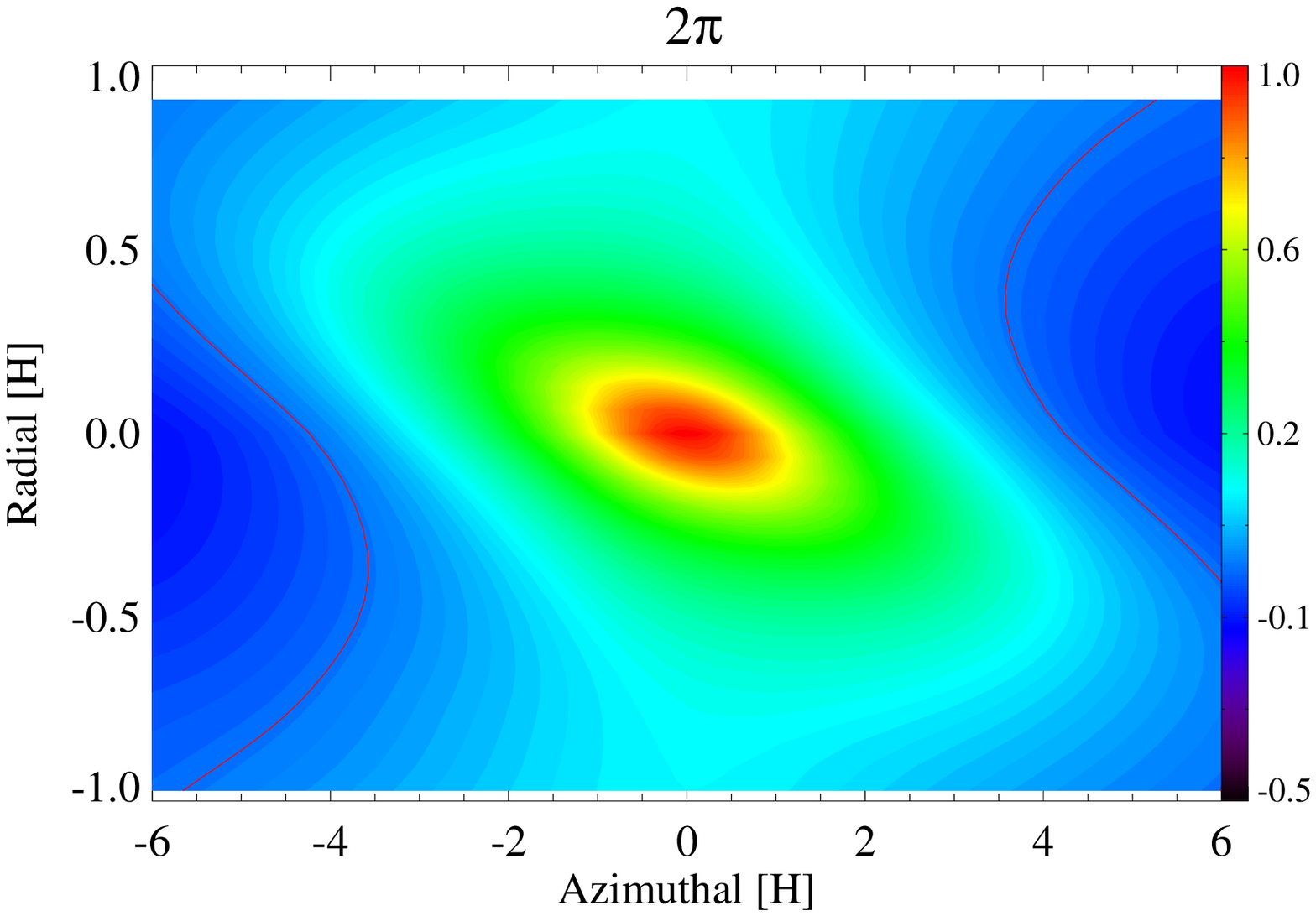,scale=0.46}
\end{minipage}
\label{cor-vel}
\caption{Contour plot of the two-point velocity correlation function 
at 1 scale height at 5 AU. The red line shows zero contour.} 
%Bottom right: $2\pi$. Bottom left: $\pi$. Top right:
%$\pi/2$. Top left $\pi/4$.

\end{figure}
\begin{figure} 
\vspace{-0.6cm}\hspace{-0.6cm}
\begin{minipage}{5cm}
\psfig{figure=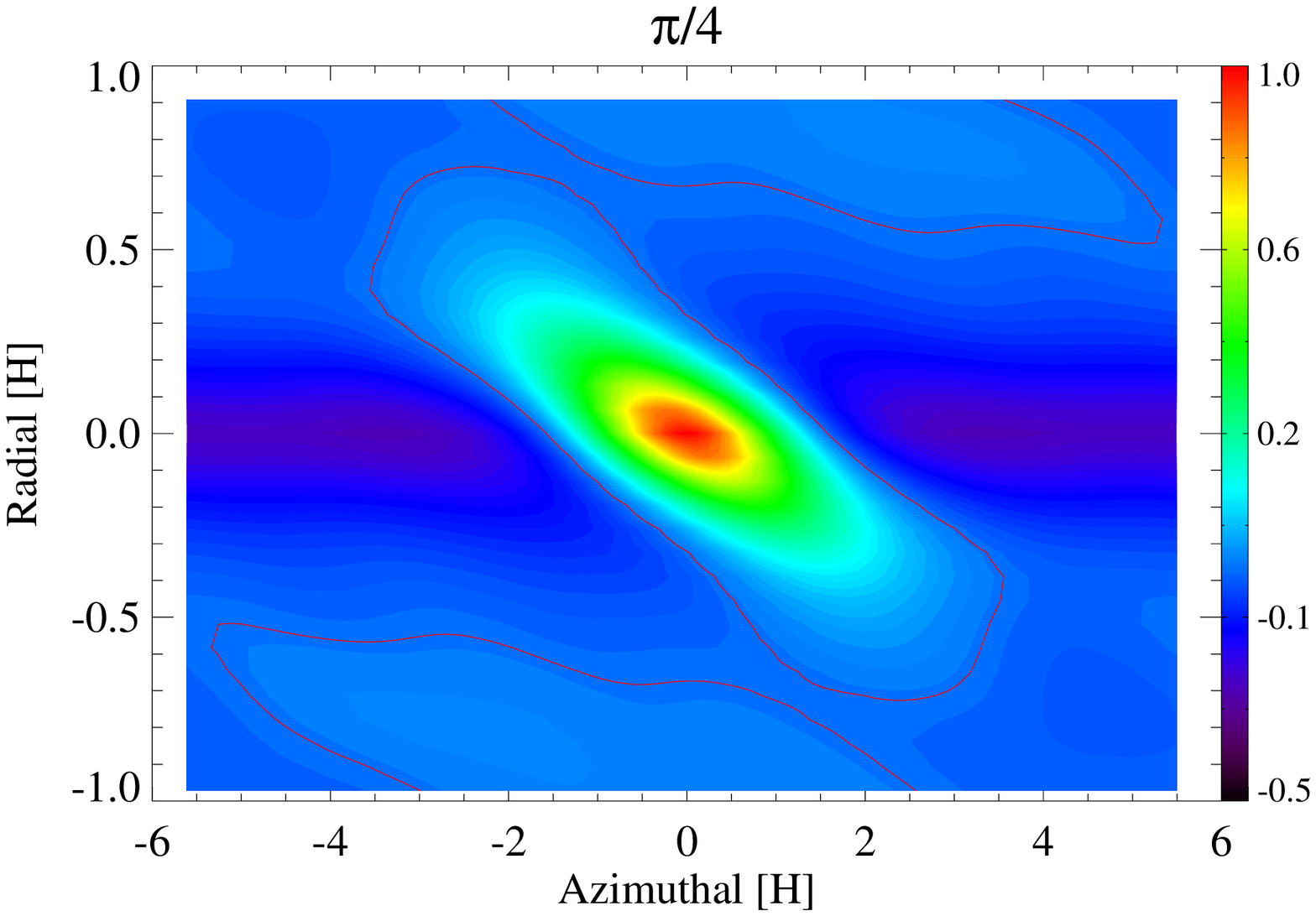,scale=0.46}
\psfig{figure=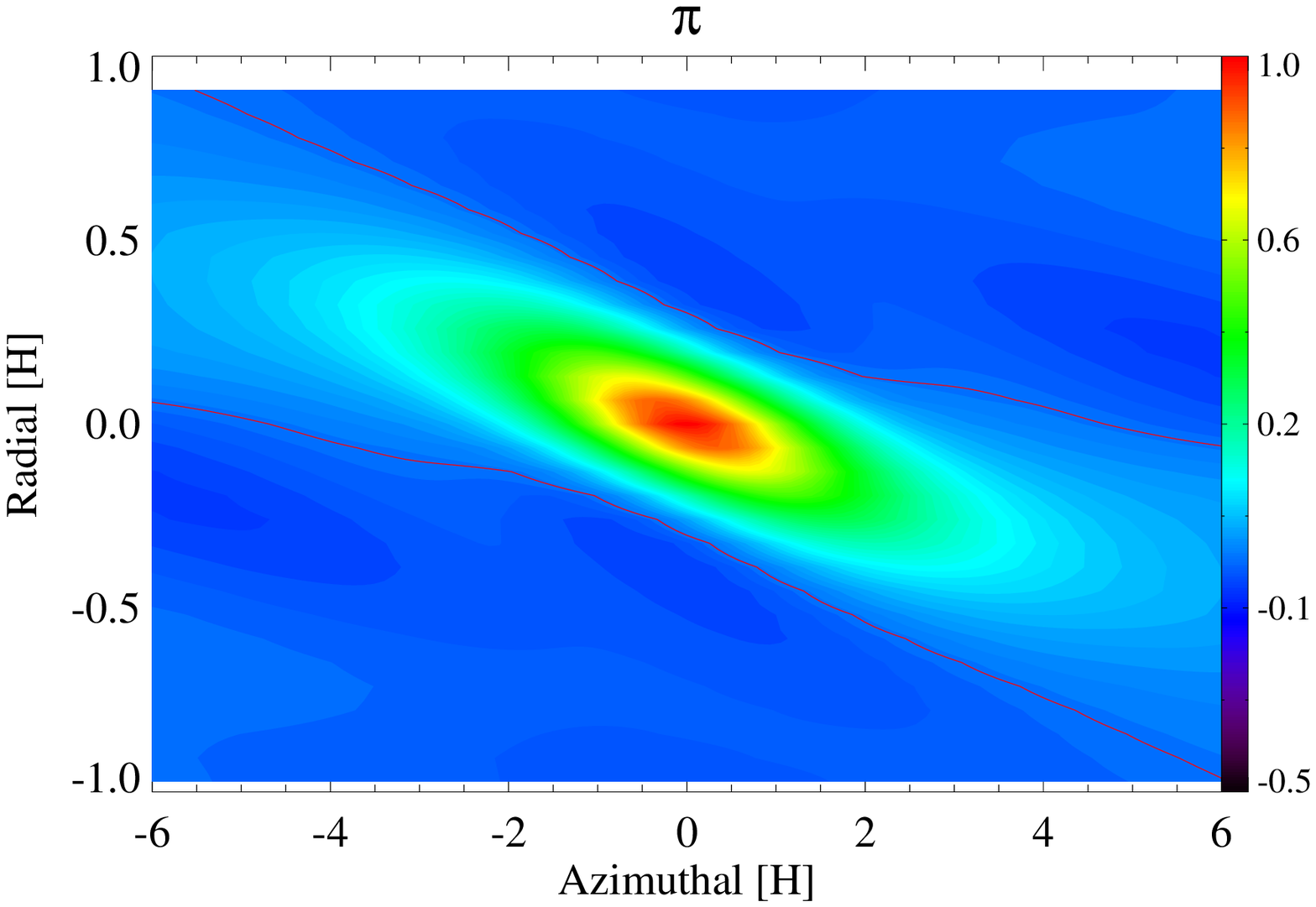,scale=0.46}
\end{minipage}
\hspace{4.0cm}
\begin{minipage}{5cm}
\psfig{figure=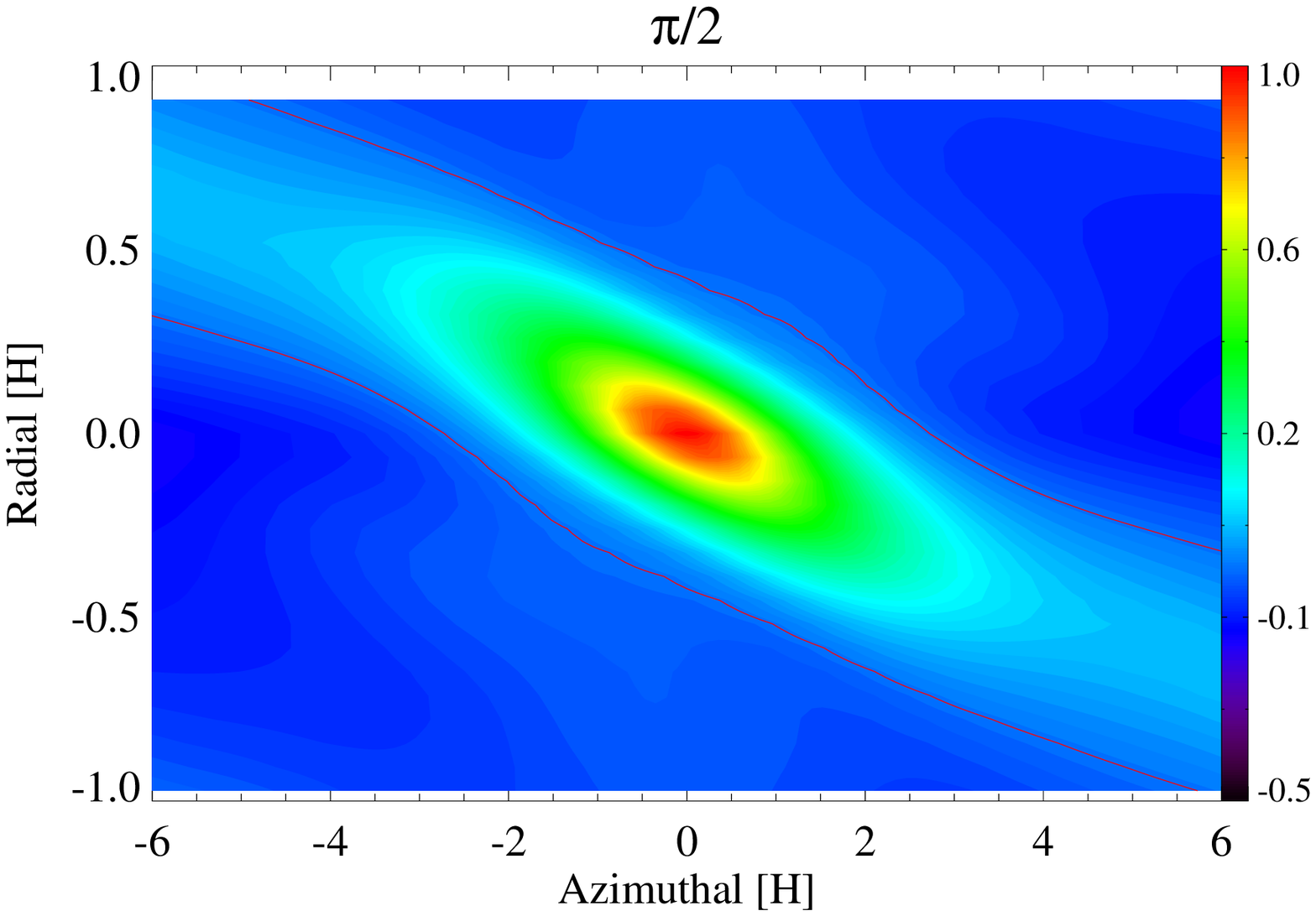,scale=0.46}
\psfig{figure=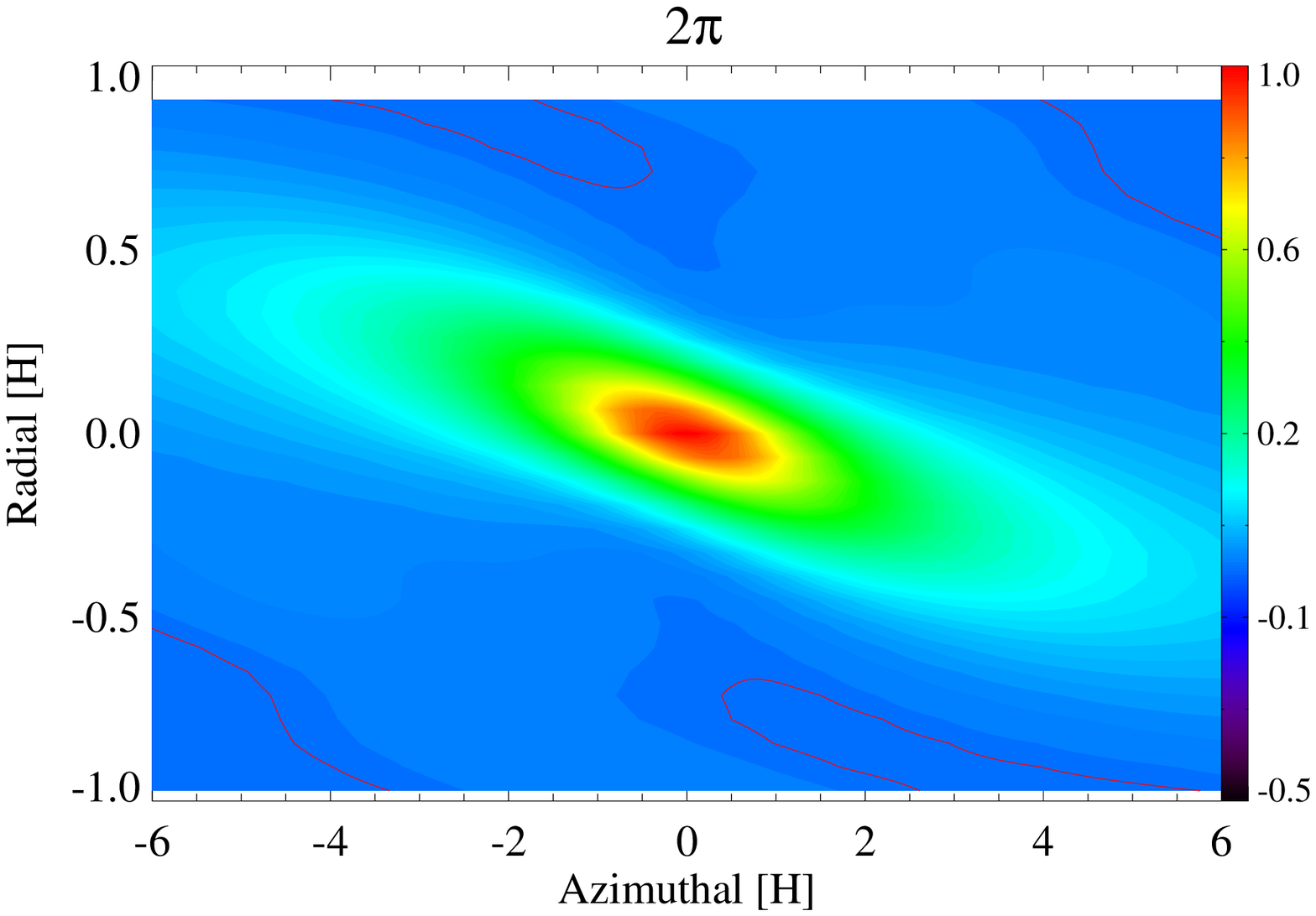,scale=0.46}
\end{minipage}
\label{cor-vel}
\caption{Contour plot of the two-point magnetic field correlation function 
at 1 scale height at 5 AU. The red line shows zero contour.} 
%Bottom right: $2\pi$. Bottom left: $\pi$. Top right:
%$\pi/2$. Top left $\pi/4$.}
\end{figure}
%
%The reduced azimuthal domain runs $\pi/4$ and $\pi/2$ do also
%affect the small scale evolution.
%effects the
%turbulent evolution. 
The models with $\pi/4$ and $\pi/2$ show an amplified turbulence. The $\phi^{extent}$
affects the large scale and small scale turbulent properties.
%A combination of self-coupling of MRI \citep{les08}
%and magnetic energy pile-up at the domain size \citep{flo11} could be
%responsible for this. 
%Restricting the azimuthal domain means restricting the largest possible
%mode which can growth, e.g. in the $\pi/4$ run the modes larger then $\rm m=8$ mode
%cannot occur. 
%A different magnetic energy level at these largest scales (domain
%size) effect the small scale turbulent evolution showed by the
%two-point correlation function of the velocity and magnetic field.
%that by reducing the azimuthal extent the tilt angle is increased.
Only an azimuthal domain of $\pi$ does
reproduce similar large scale and small scale turbulent properties as in the full $2\pi$ run.
The strong mean field generated by the $\alpha \Omega$ dynamo are
responsible for the MRI amplification. 
%This includes a similar evolution of the magnetic energy, the total $\alpha$
%parameter and a similar correlation length in the velocity and magnetic
%field structure.
%%
\subsection{Mean field evolution}
A typical feature of MRI in stratified disks is an oscillating toroidal
magnetic field, generated by oscillating radial magnetic field.
%This feature is well known as 'butterfly' pattern due to the buoyant
%movement of the toroidal field from 1 scale height to upper layers.
This feature is well known as 'butterfly' pattern, which wings appear due to the buoyant
movement of the toroidal field from the midplane to upper layers.
The timescale of these oscillation is around ten local orbits.
%In the previous section we investigated the overall turbulent evolution. 
%We saw that for the restricted domains $\pi/2$
%and mostly $\pi/4$, the
%magnetic fields at domain size can influence the small scale turbulent
%evolution.
Recent work in local box simulations showed the context between this oscillating magnetic field
and a dynamo process \citep{gre10,sim11,haw11,gua11}. 
In this section we investigate the evolution of this axisymmetric
magnetic fields and the connection to the dynamo process.
%
%One typical feature of MRI in stratified disks is the 
% oscillating radial mean magnetic field which again
%generates a strong oscillating toroidal magnetic field.
%The buoyant movement of the toroidal field 
%from 1 scale height to upper layers in the disk is well known 
%as 'butterfly' pattern.
\subsection*{The parity and butterfly pattern}
In Fig. 9, top, we present the time evolution of axisymmetric 
radial and toroidal magnetic field over height. The values are normalised over the initial
toroidal field.
The generated toroidal magnetic field, Fig. 9 (second from top) 
is around one order of magnitude higher than the radial magnetic field.
We observe a change of sign every 5 local orbits.
%This change of sign is also present at the midplane but better visible in
%the upper layers of the disk. 
%This radial magnetic field generates 
%the toroidal magnetic field due to up-winding, Fig. 6, second from top.
%In this section we will investigate the mean field evolution. 
%
%We start the investigation with the mean field evolution in the southern and
%northern hemisphere. 
The butterfly wings are mostly antisymmetric with respect to the midplane.
To quantify the symmetry we determine the parity of the mean magnetic field.
We calculate the symmetric (S) and asymmetric
(AS) magnetic field component: $B^{\rm S}_{r,\theta,\phi} = 0.5 (
B_{{r,\theta,\phi}}^{\rm NH} +
B_{{r,\theta,\phi}}^{\rm SH})$ and $B^{\rm AS}_{r,\theta,\phi} = 0.5 (
B_{{r,\theta,\phi}}^{\rm NH} -
B_{{r,\theta,\phi}}^{\rm SH})$ 
with the values of the northern (NH) and southern
(SH) hemisphere (SH)\footnote{The northern hemisphere is placed on the upper disk
if the azimuthal velocity is positive. Then if one looks at the north pole
, the disk is rotation counter-clockwise in the northern hemisphere, e.g.
mathematically positive.}.
The parity $$Parity = \frac{E^{D} - E^{Q}}{E^{D} + E^{Q}}$$
is determined with total dipole and quadrupole energy components
$\rm E^{D} = (B^{AS}_r)^2 + (B^{S}_\theta)^2 + (B^{AS}_\phi)^2$ and
$\rm E^{Q} = (B^{S}_r)^2 + (B^{AS}_\theta)^2 + (B^{S}_\phi)^2$.
%determine the parity as $Parity = \frac{E^{D.} - E^{Q.}}{E^{D} + E^{Q}}.$
%- AS: asymmetric. S: symmetric\\
The toroidal field is much larger then the radial and theta magnetic field.
It is possible to define a symmetric (Quadrupole) or antisymmetric
(Dipole) configuration as the total parity is set by the toroidal field.
%dominates mostly the magnetic energy and controls the
%total parity.
%Therefor, one can alternatively define a symmetric (Quadrupole) or antisymmetric
%(Dipole) configuration for the toroidal field.
Then, a parity of -1 defines a pure symmetric configuration (Quadrupole)
while a parity of +1 defines a pure antisymmetric configuration
(Dipole).
The time evolution of the total parity is plotted in Fig. 10, top, for all
models.
The total parity starts with -1 as the initial field $B_\phi$ is symmetric.
The parity of only $B_r$ and $B_\theta$ is plotted in Fig. 10, bottom, and
present a similar time evolution.
Both parities change sign several times during the simulation for all
models. The time averaged values (400 - 1200 inner orbits) show strong 
deviations around zero parity, see Table 1.
%The parity is averaged over the total disk height at 4.5 AU.
%The time evolution of the total parity is plotted in Fig. 5, left, for all models.
%The time evolution of the parity of $B_r$ and $B_\theta$ 
%is plotted in Fig. 5, right.
%Azimuthal and vertical average is done at 4.5 AU.
%The parity is changing several times during the simulation time for all
%models. 
%For a time average of the total parity (between 400 and 1200 inner orbits) 
%we get $Parity^{\pi/4} = -0.2\pm0.4$, $Parity^{\pi/2} = -0.2\pm0.5$,
%$Parity^{\pi} = -0.1\pm0.5$ and $Parity^{2\pi} = 0.2\pm0.4$.
Only the $2\pi$ model is mostly antisymmetric for the simulation time.
The contour plot of total parity over height, Fig. 9 third plot from top,
shows the correlation between the parity and the 'butterfly' pattern.
The symmetry of the mean toroidal field in respect to the midplane sets the total parity.
Even the total parity is mostly antisymmetric (yellow, +1)
there is a change of the parity to symmetric for two butterfly cycles
between 80 and 100 local orbits (also visible in Fig. 10, solid
line).
%A closer look at the distribution of the parity over height, Fig. 6 third
%plot from top, shows that the parity correlates with the 'butterfly' pattern.
%Here the parity is plotted as contour colour at 4.5 AU over height and time. 
%We see a mostly 
%evolution in the butterfly wings.
%
%In general, only in the sign changing phases 
%the configuration is symmetric (blue). 
%Even the total parity is mostly antisymmetric (yellow, +1)
%$there is a change of the parity to symmetric for two butterfly cycles 
%between 80 and 100 local orbits (visible in Fig. 5, solid
%line).
%
%The parity of the turbulent field remains close to zero.
%
%To summarise: there is no clear defined parity of the mean fields. Even the
%antisymmetric parity seems to dominate in the full $2\pi$ run, it can change
%to a more symmetric configuration. Still we can conclude that the mean field
%oscillation, know as "butterfly" pattern, sets the parity.

\begin{figure} 
\psfig{figure=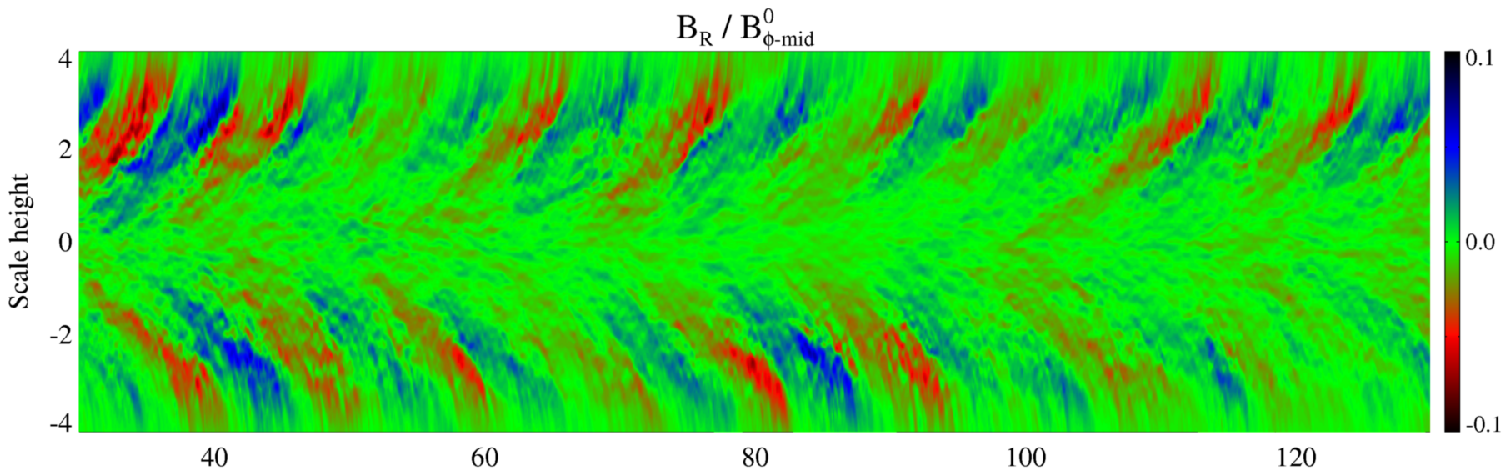,scale=0.91}
\psfig{figure=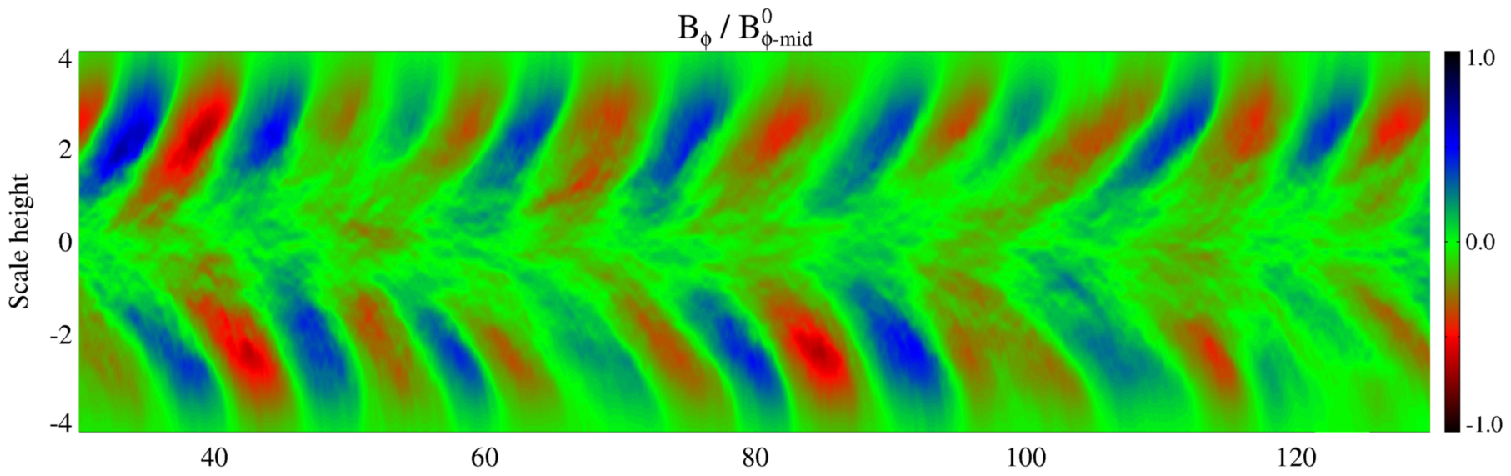,scale=0.91}
\psfig{figure=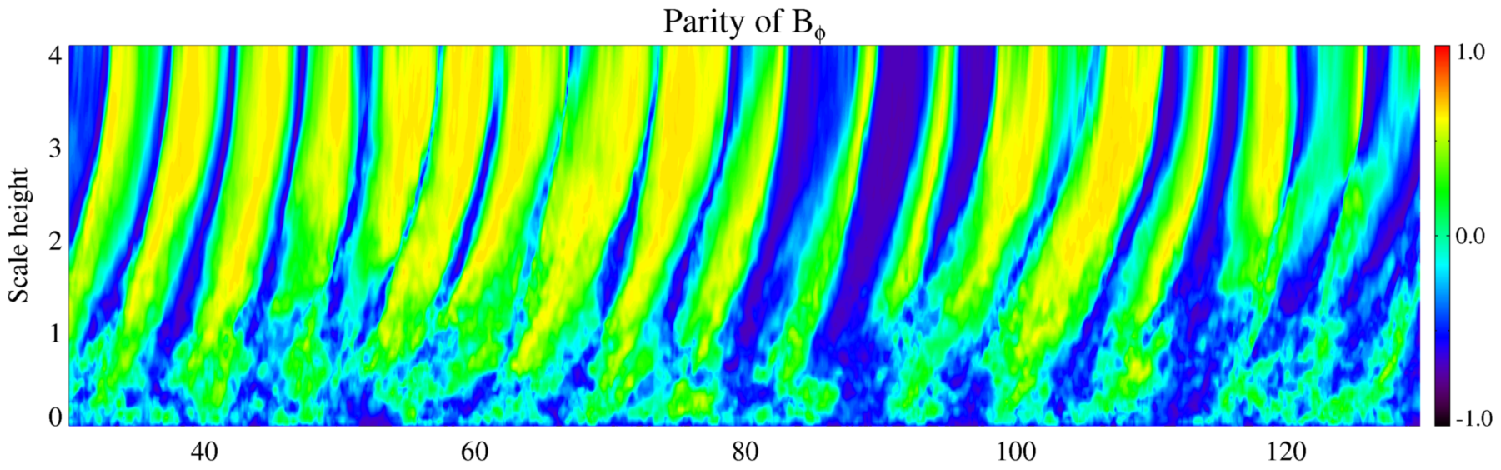,scale=0.91}
\psfig{figure=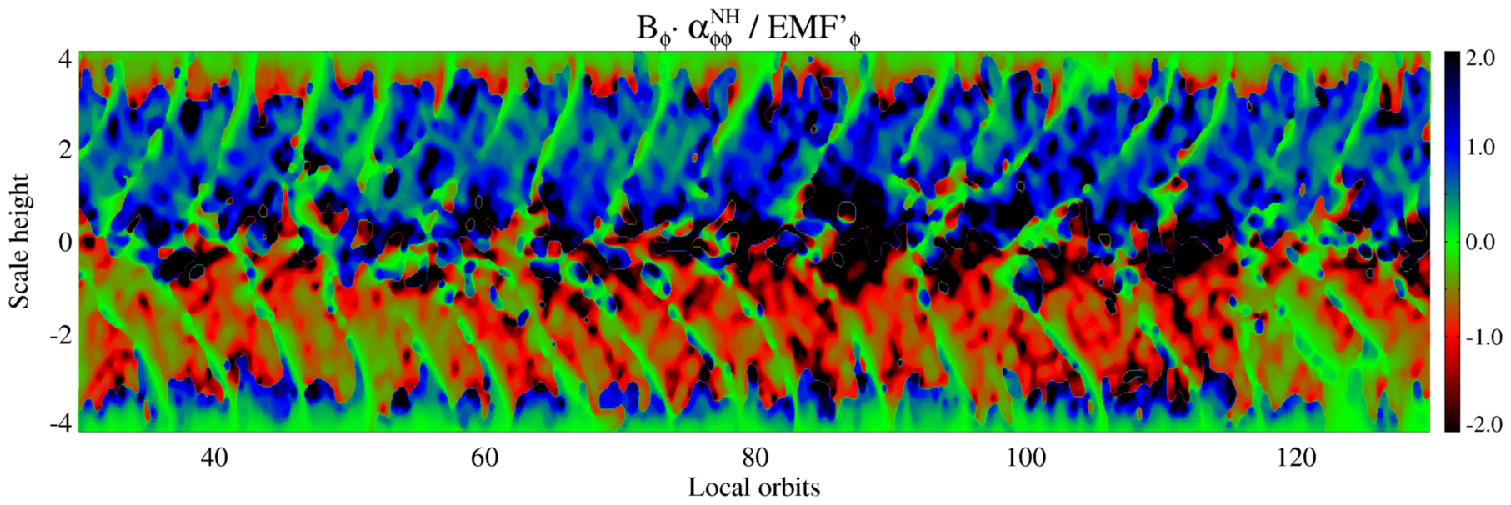,scale=0.91}
\label{dyn-cor}
\caption{Top to bottom: 
1. Mean radial magnetic field over height and time. 
2. Mean toroidal magnetic field over height and time. 
3. Contour plot of the parity over height and time. 
%There is no clear preference of specific parity. 
%The butterfly wings
%are mostly antisymmetric (yellow) but can occur also symmetric (blue).
4. Contour plot of $\overline{B_\phi} \alpha_{\phi\phi}^{NH}/EMF'_\phi$  
over height and time.
All plots are made for model $2\pi$ at 4.5 AU.}
%There is a clear change of sign
%between the two hemisphere.}
\end{figure}

\begin{figure}
\hspace{-1.6cm}
\begin{minipage}{5cm}
\psfig{figure=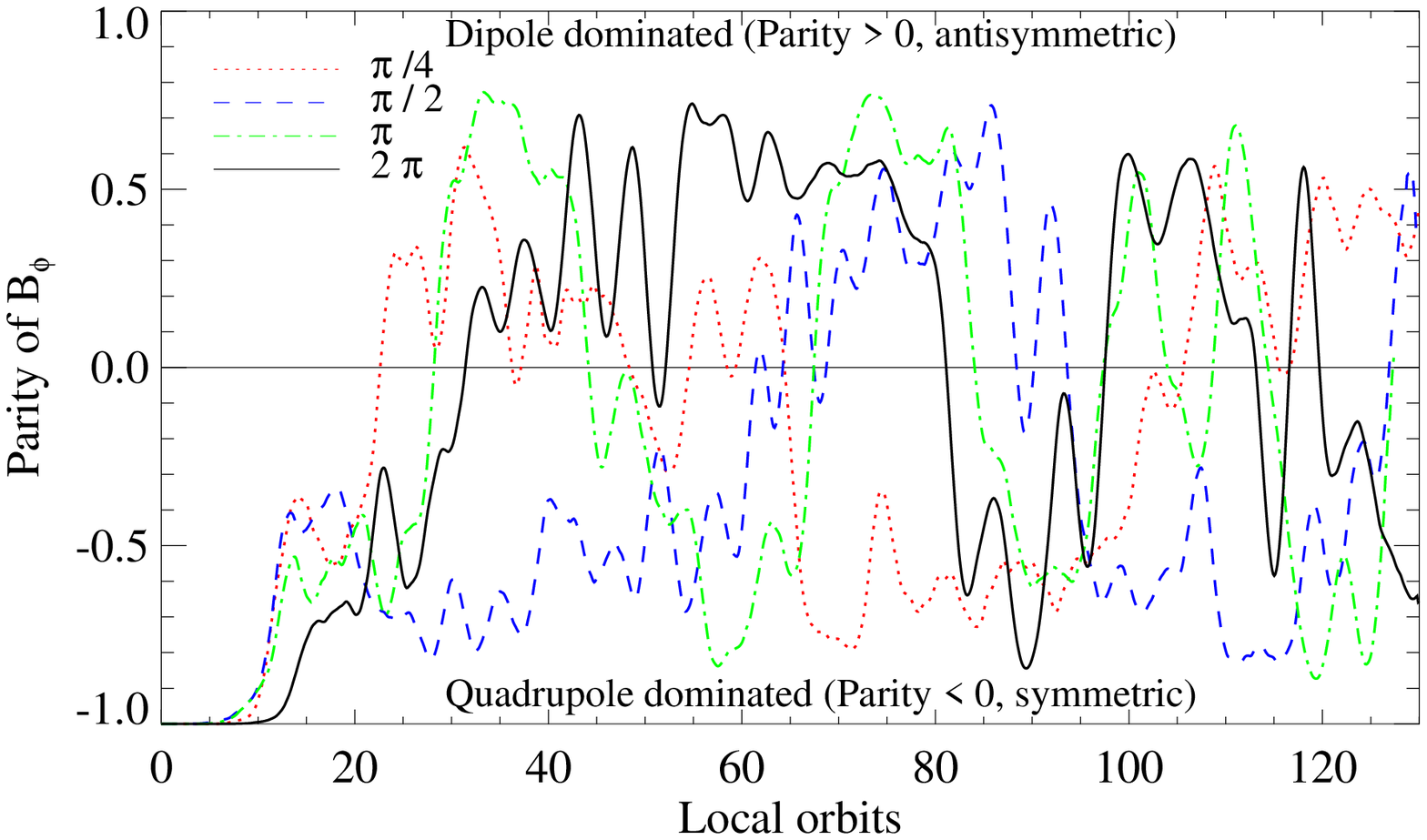,scale=0.56}
\psfig{figure=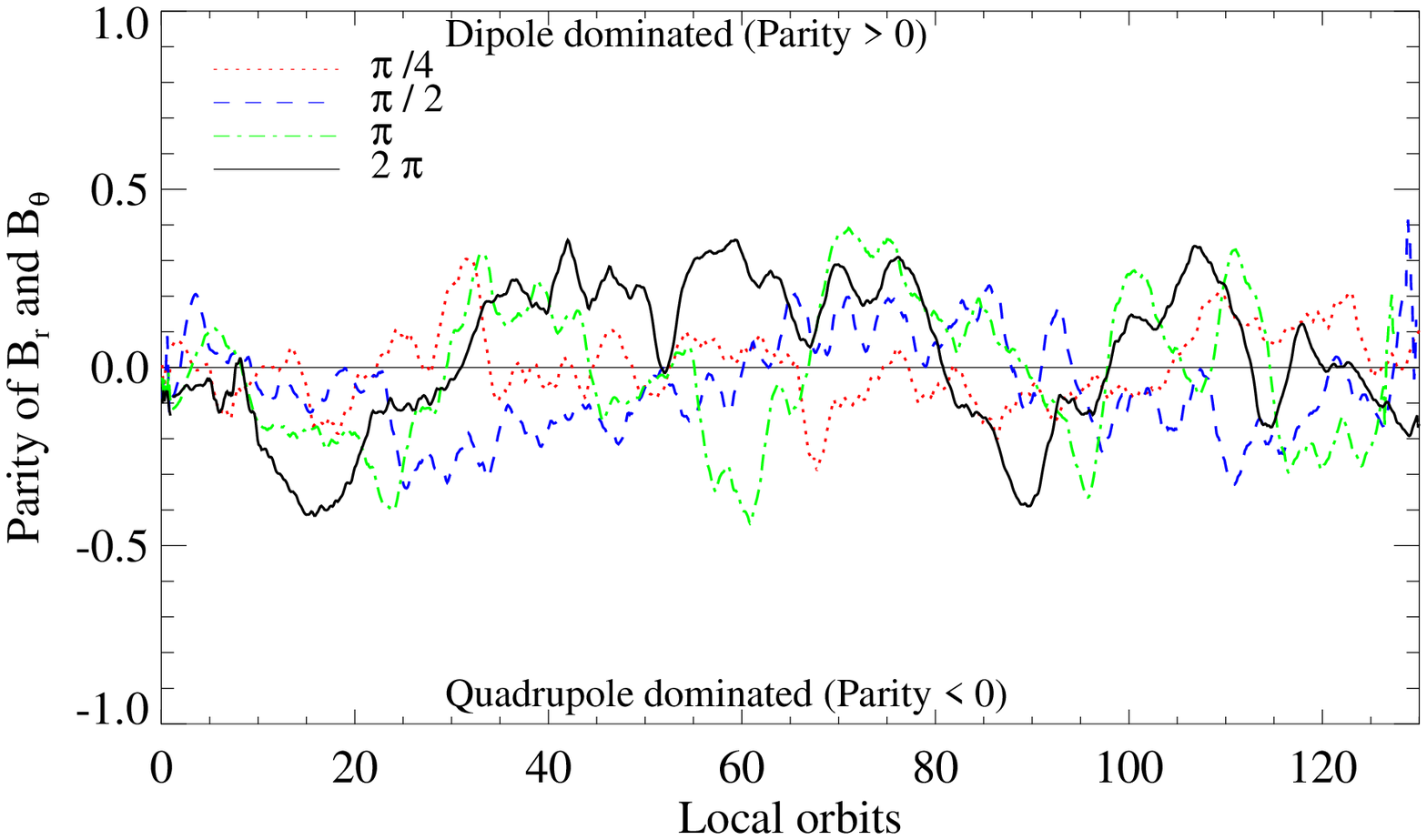,scale=0.56}
\end{minipage}
\caption{Parity of mean toroidal magnetic field (top) and of mean
poloidal field (bottom).}
%There is no clear dipole or quadrupole symmetry visible.}
\end{figure}

\begin{figure}
\hspace{-0.6cm}
\begin{minipage}{5cm}
\psfig{figure=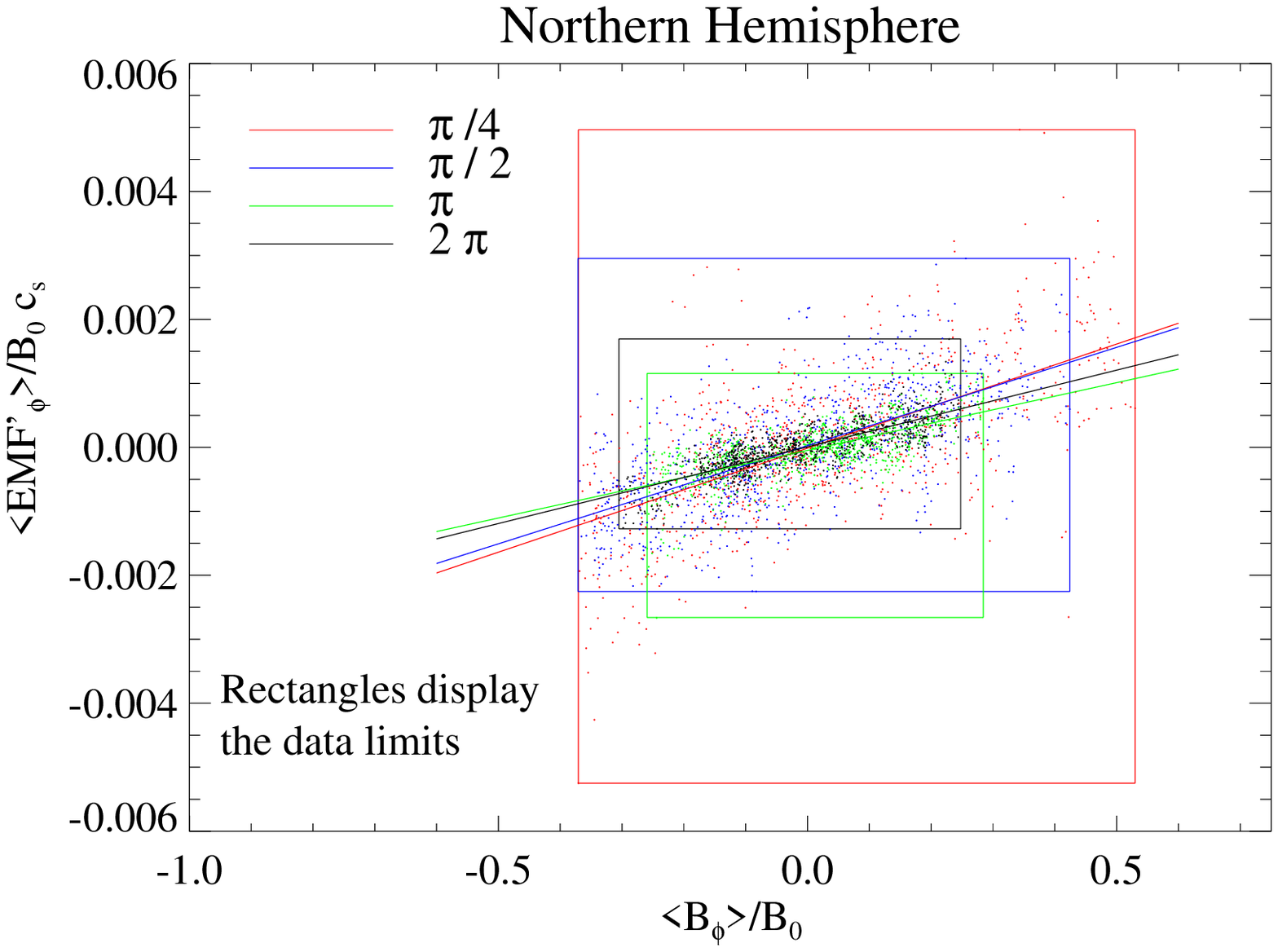,scale=0.46}
\psfig{figure=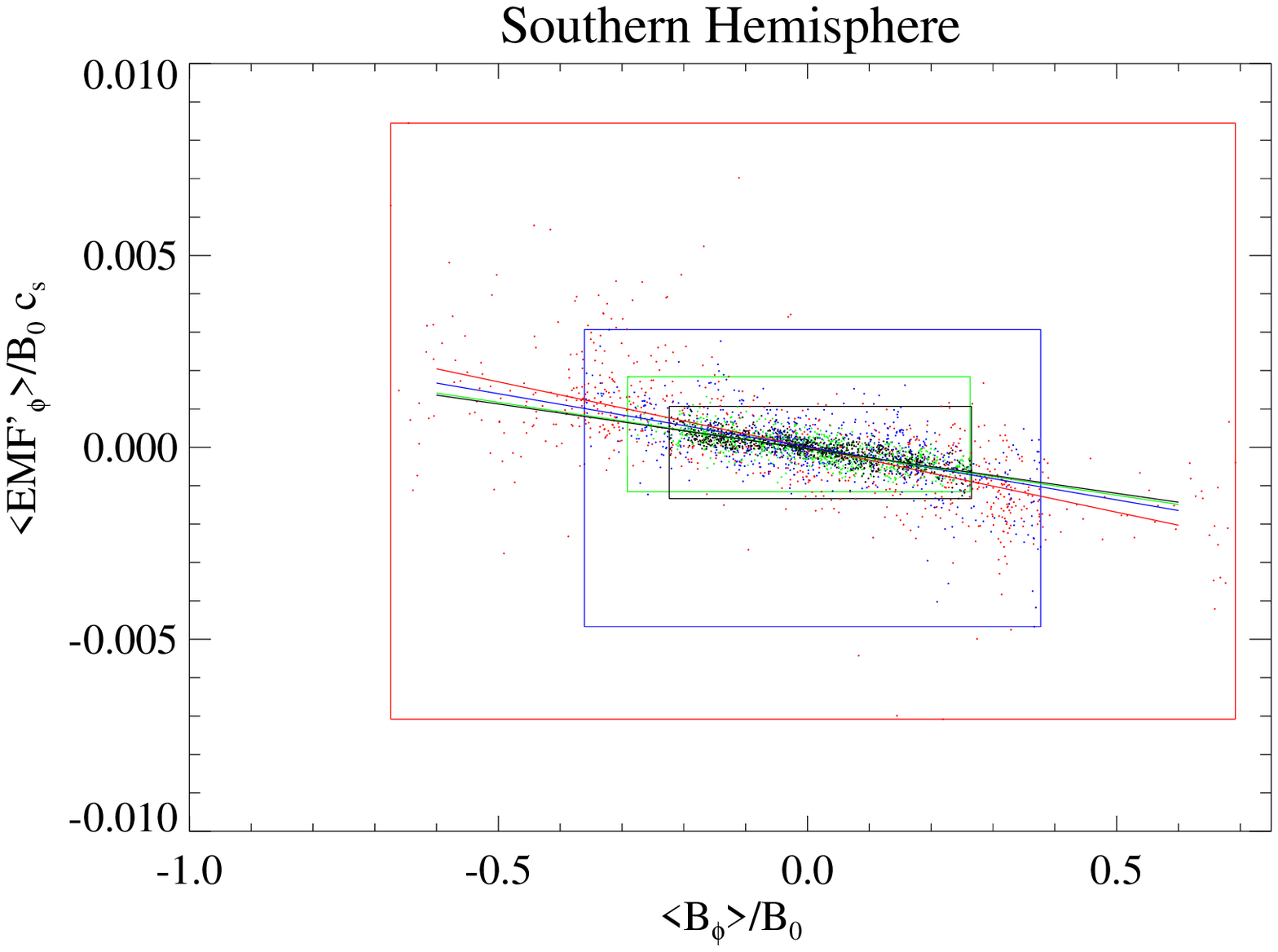,scale=0.46}
\end{minipage}
\hspace{4.0cm}
\begin{minipage}{5cm}
\psfig{figure=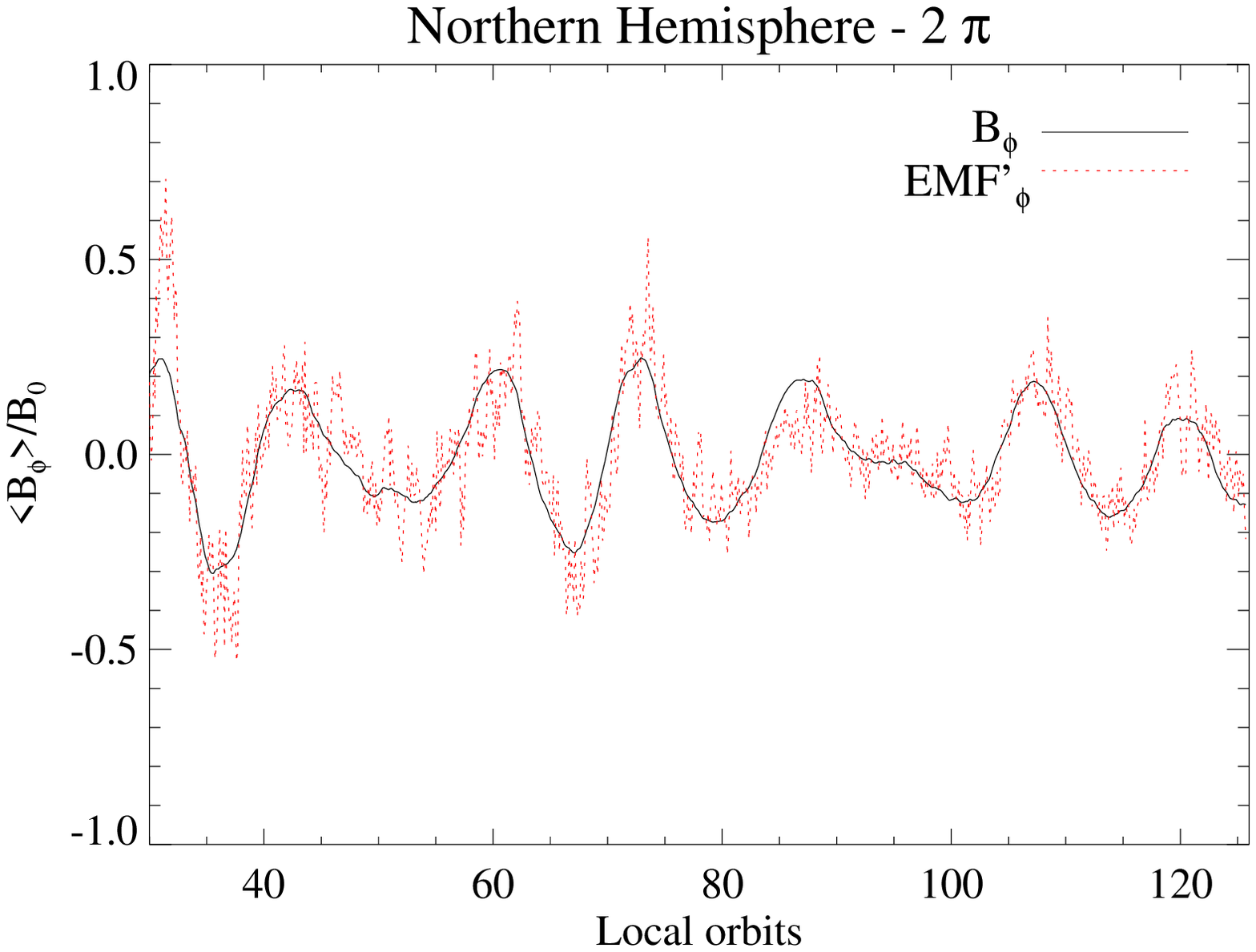,scale=0.46}
\psfig{figure=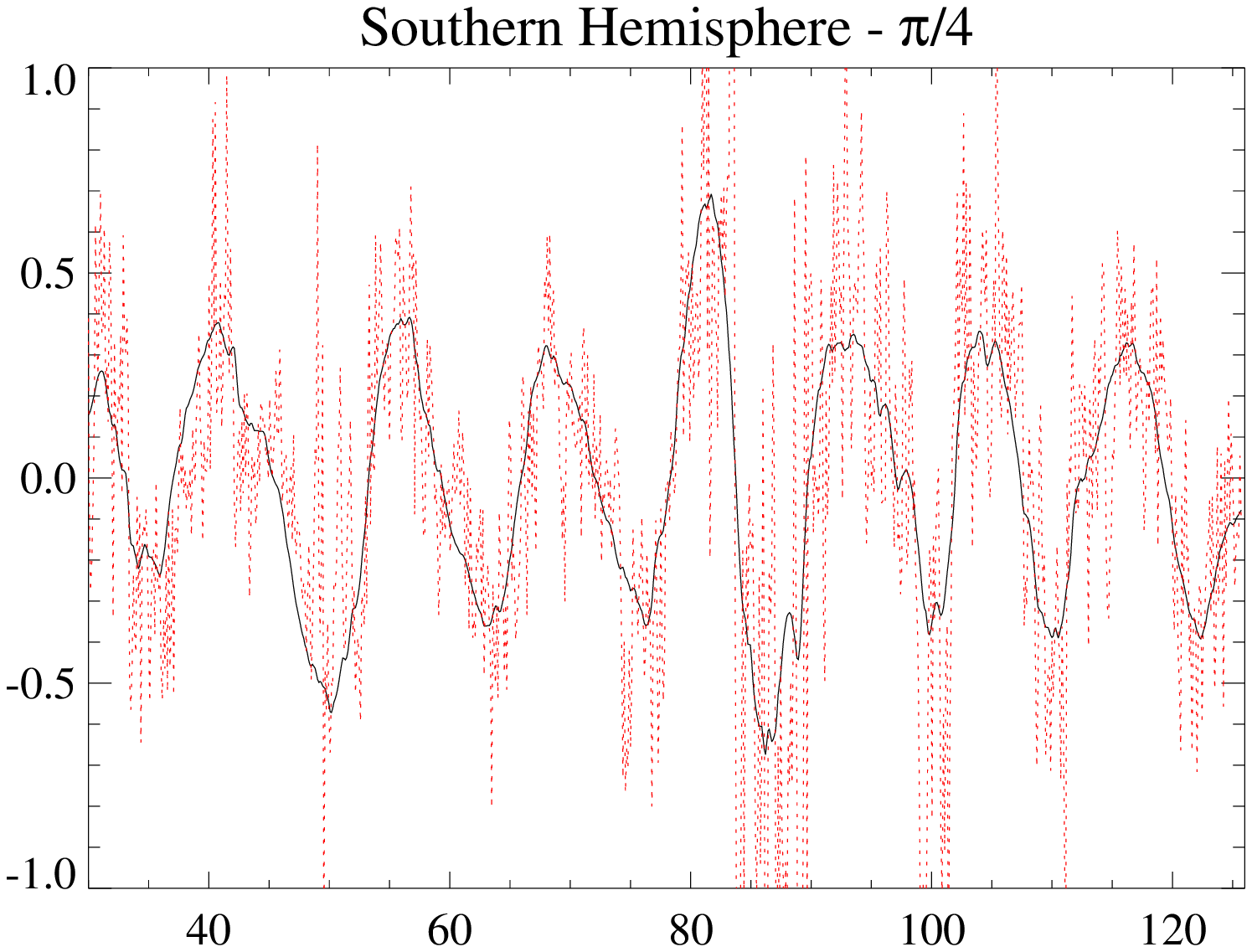,scale=0.46}
\end{minipage}
\label{corel}
\caption{Top left: Correlation between the mean toroidal magnetic field 
and the turbulent EMF component $EMF'_\phi$ for the northern (upper) hemisphere  
of the disk and for all models. 
Rectangles show the limits of the data values.
Bottom left: Correlation between the mean toroidal magnetic field
and the turbulent EMF component $EMF'_\phi$ for the southern hemisphere of the
disk and for all models.
Top right: Time evolution of mean toroidal field (solid line), 
over-plotted with the turbulent EMF (red dotted line) divided by
$\alpha_{\phi\phi}^{NH}$ for model $2\pi$.
Bottom right: Time evolution of mean toroidal field (solid line),
over-plotted with the turbulent $EMF'_\phi$ (red dotted line) divided by
$\alpha_{\phi\phi}^{SH}$ for model $\pi/4$.}
\end{figure}
\subsection*{$\alpha \Omega$ Dynamo}
%The MRI turbulent disk shows an oscillating mean
%field evolution. 
%In the upper layers of the disk the mean oscillating fields
%become buoyant and rise up, leaving continuously the domain.
%From mean field theory it is known that there could exist a magnetic dynamo
%in such 3D MHD stratified disk simulations \citep{rue93,bra97,rue00}.
In mean field theory, there is a mechanism 
to generate large-scale magnetic fields 
by a turbulent field.
In case of an $\alpha \Omega$ dynamo \citep{kra80}
there should be a correlation between the 
turbulent toroidal electromotive force ($EMF'_{\phi}$) 
component and the mean toroidal magnetic field,
$$\rm \overline{EMF'_{\phi}} = \alpha_{\phi \phi} \overline{B_{\phi}}\rm +\ higher\ derivatives\ of\ \overline{B}$$ 
with $EMF'_{\phi} = v'_r B'_\theta - v'_\theta B'_r $.
%For this correlation between $E'_{\phi}$ and $\overline{B_{\phi}}$, 
The sign of $\rm \alpha_{\phi \phi}$ has to change for the southern and
northern hemisphere. % this could indicate an existing dynamo process. 
%The calculation of $\alpha_{\phi\phi}^{dyn}$ was done in local 
%box simulations \citet{bra95,zie00}.
%
%We calculate the turbulent $EMF'_\phi$ and the mean field
%$\overline{B_{\phi}}$ at 4.5 AU, $\pm$ 1.5 SH around the midplane. 
The correlation is plotted in Fig.
11, left,  
for the northern hemisphere (top) and the southern hemisphere (bottom).
We get a positive sign for the $\rm \alpha_{\phi \phi}$ in the northern
hemisphere ($\rm \alpha_{\phi\phi}^{NH}$) of the disk (Fig. 11 top) and a negative sign in the southern
hemisphere ($\rm \alpha_{\phi\phi}^{SH}$). This result was predicted for stratified accretion disks
\citep{rue93} and also indicated in global simulations \citep{arl01}. 
%Local box simulations show the opposite sign
%\citep{bra95,bra97,rue00,zie00,gre10,dav10}.
%All our models show a positive $\rm \alpha_{\phi\phi}$, see Fig. 11 left.
%There exists a positive $\rm \alpha_{\phi\phi}$ in the
%northern hemisphere and negative in the southern hemisphere.
Each dot in Fig. 11 left, represent a result from a single time snapshot. The boxes show the
limits of the values for each model.
The $\pi/4$ and $\pi/2$ model show 
higher amplitudes in the mean field $\overline{B_{\phi}}$ as well as in the
$EMF'_\phi$ fluctuations.
All values of $\rm \alpha_{\phi \phi}$ are determined using a robust regression method
and summarized in Table 1.
%The fit, solid lines in Fig. 7 left, 
%shows that each azimuthal
%domain run has a good correlation.
%
A time evolution of the mean field and the turbulent $EMF'_\phi$
is presented in Fig. 11, right, for model $2\pi$, top, and model $\pi/4$,
bottom.
In Fig. 11 right, we divide the turbulent $\rm EMF'_\phi$ with the measured
$\rm \alpha_{\phi \phi}$ (see also Table 1).
%For the northern hemisphere we choose the $2\pi$ run (Fig. 7, top) and for the southern hemisphere the $\pi/4$ run.
The $\pi/4$ run shows higher fluctuations compared to the $2\pi$ run.
%As for the parity, the $\alpha^{dyn}$ also correlate with the 'butterfly' 
%structures. 
%
A time evolution of $\rm B_\phi\cdot \alpha_{\phi\phi}^{NH} / EMF'_\phi$ over height is presented in Fig.
9, bottom. We see that the sign of $\rm \alpha_{\phi\phi}$ is well defined for the
two hemispheres, reaching up to 3 scale heights of the disk.
%In Fig. 6, bottom, we plot the $\alpha_{\phi\phi}$ over height and
%time. One can clearly see that the two hemispheres have different signs.
%We see a not vanishing $\alpha_{\phi\phi}$ at a location 
%where mean toroidal magnetic field is present.
%
%
\begin{figure}
\begin{minipage}{5cm}
\psfig{figure=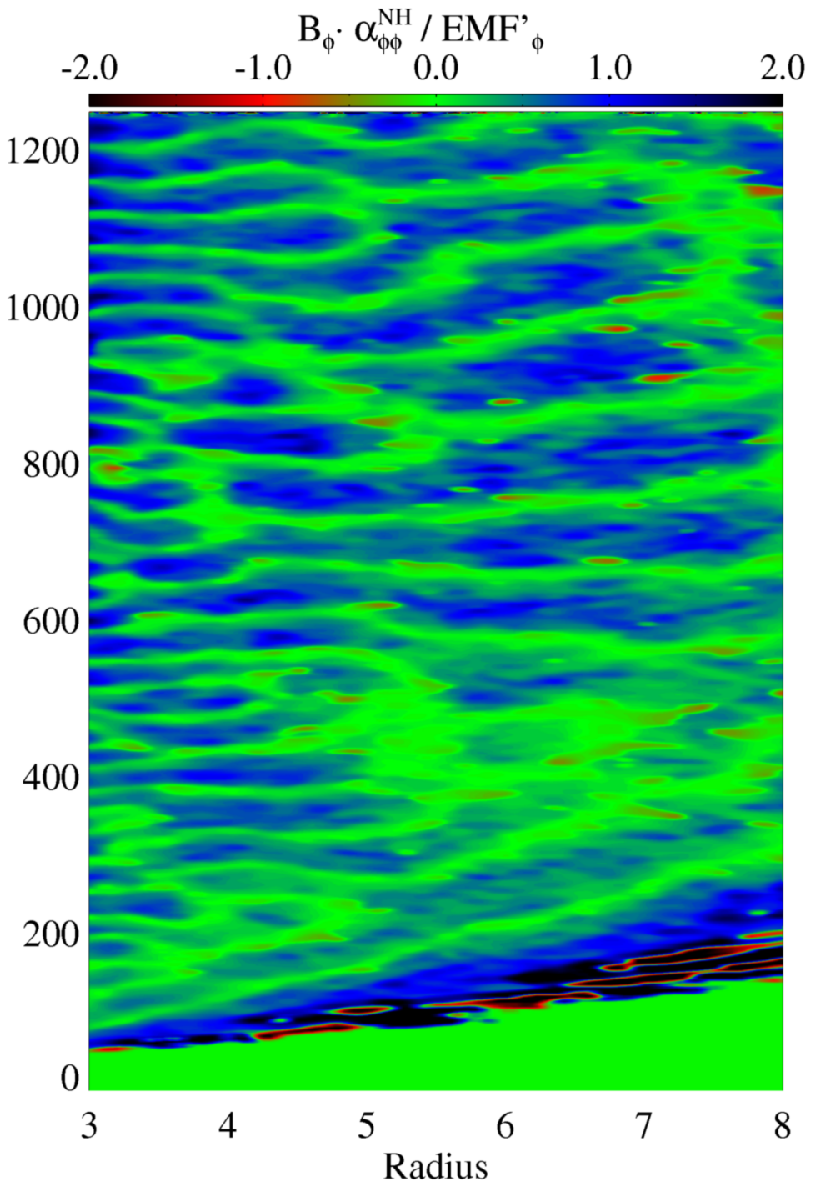,scale=0.86}
\psfig{figure=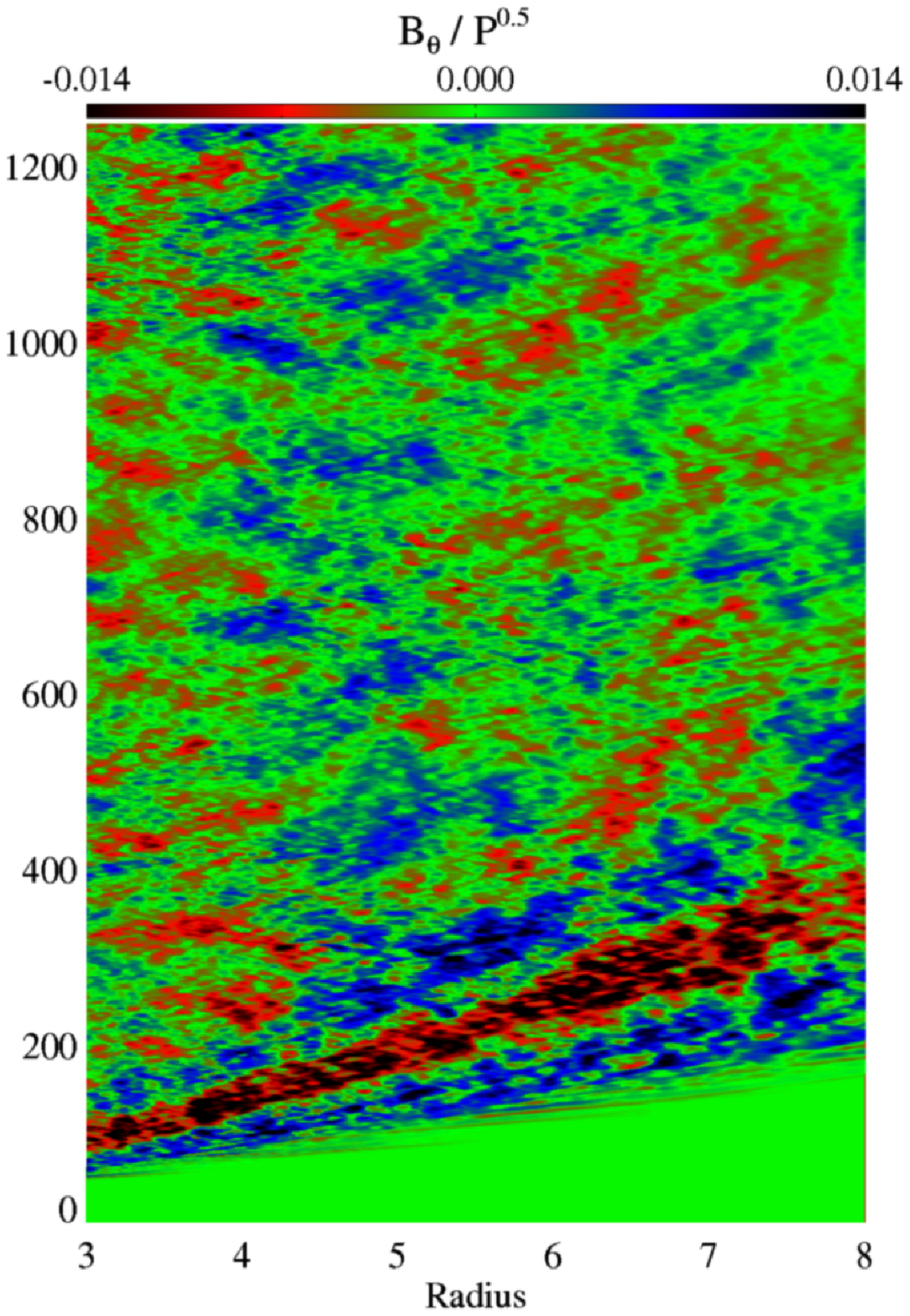,scale=0.62}
\end{minipage}
\hspace{3.0cm}
\begin{minipage}{5cm}
\psfig{figure=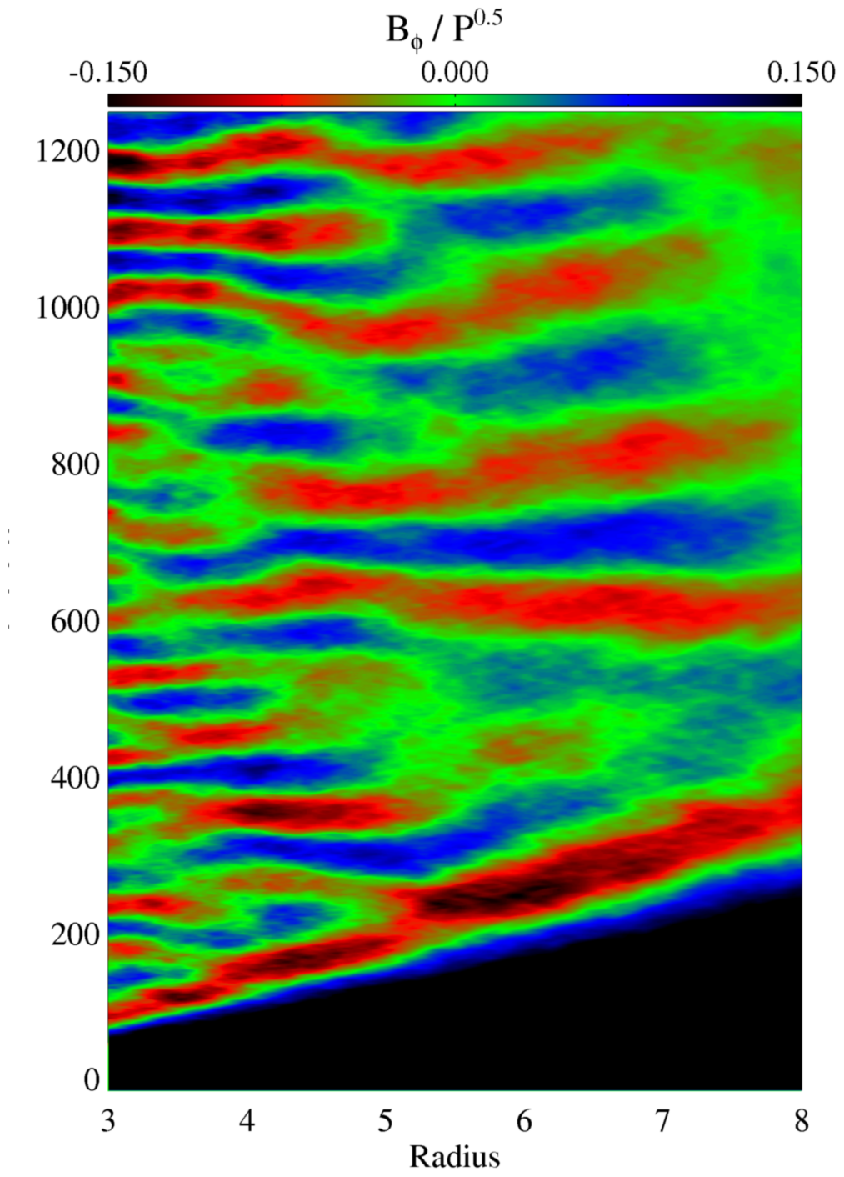,scale=0.86}
\psfig{figure=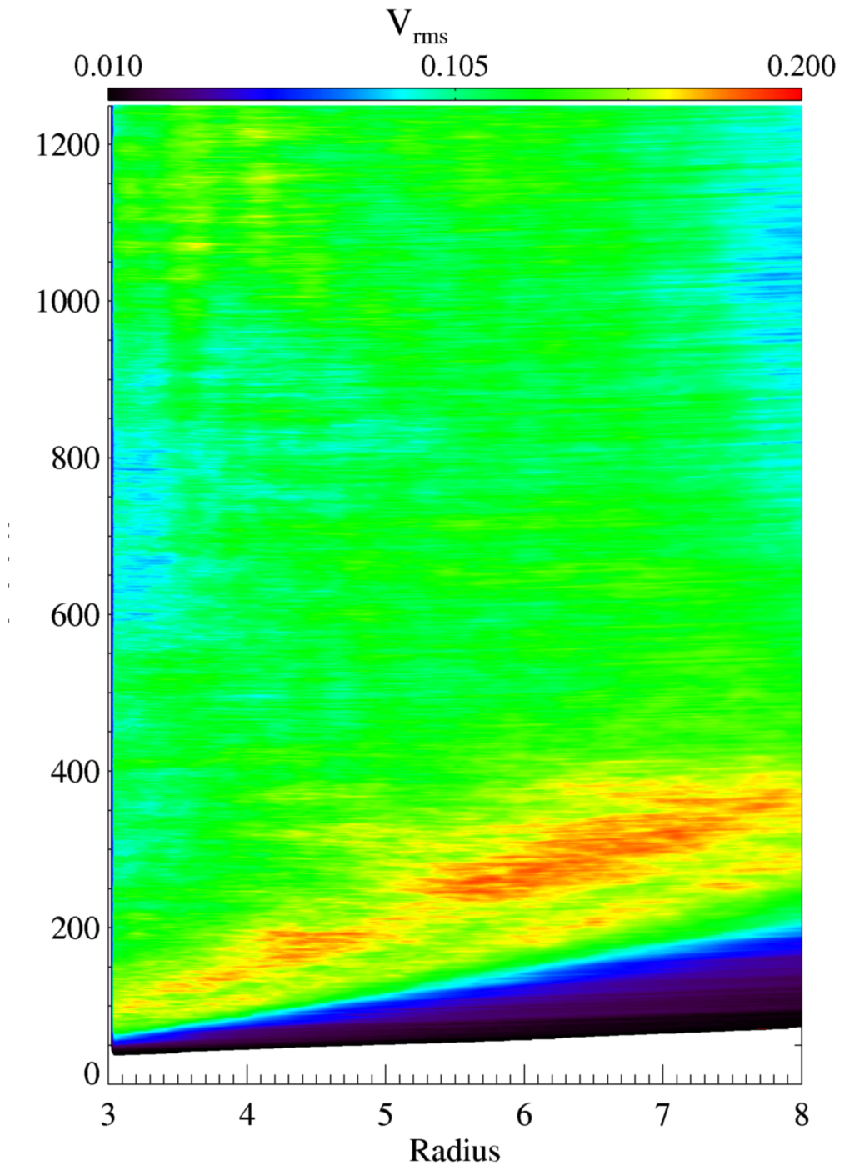,scale=0.86}
\end{minipage}
\label{flux}
\caption{Top left: Contour plot of
$\overline{B_\phi}\cdot\alpha_{\phi\phi}^{NH}/EMF'_\phi$
over radius and time (see also Fig 9, bottom). 
Top right: Mean toroidal magnetic field 
over radius and time.
Bottom left: Mean $\theta$ magnetic field over radius and time. 
Bottom right: Turbulent RMS velocity over radius and
time. All plots are made for model $2\pi$ in the northern hemisphere. }
\end{figure}
\begin{figure}
\hspace{-0.6cm}
\psfig{figure=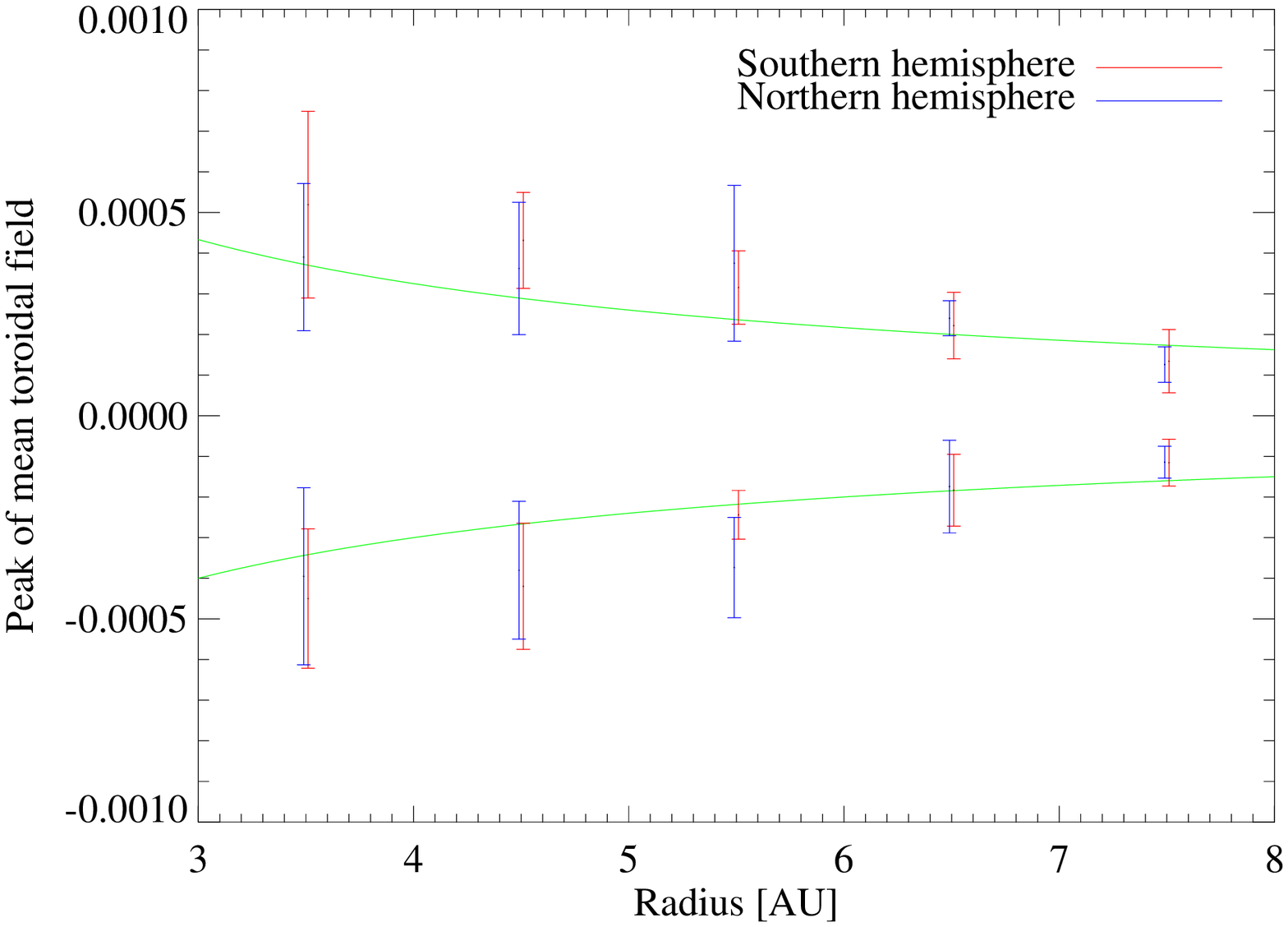,scale=0.56}
\caption{Radial distribution of the peaks of mean toroidal
 magnetic field. Values from the northern hemisphere are in red and from 
the southern hemisphere in blue, (see also Fig. 12, top right). }
\end{figure}
\subsection{Mean fields over radius}
%So far we used a local disk patch for most of our analysis.
%For most of our analysis we took a local disk patch to study the different
%physical effects. 
In this section we study the development of the mean magnetic fields 
along radius.
We show results from our full $2\pi$ model as it represents the most
realistic physical domain size.
% the turbulent properties along radius,
%like the $1/r$ profile for the turbulent magnetic field or the 
%The radial extension in our global models allows also to study the properties along radius. 
%
A contour plot of mean toroidal field, normalised over the square root of
the pressure, is presented in Fig. 12, top right, over radius and time.
All results in Fig. 12 are averaged along azimuth and along $\theta$ between
the midplane and two disk scale heights in the northern hemisphere.
Fig. 12, top right, shows the irregular change of sign for the mean toroidal magnetic field
along radius. 
The timescale of the "butterfly" oscillations at a given radius
can change because of radial interactions. 
The timescale of reversals of the toroidal magnetic field 
does vary from the ten local orbital line 
(see Fig. 12, top right, horizontal homogeneous $B_\phi$).
The mean field configuration along radius can strongly affect the 
accretion stress, see Fig. 3.
%The timescale of reversals of the toroidal magnetic field 
%is not strictly placed along the ten local orbital line 
%(see Fig. 12 top right horizontal homogeneous $B_\phi$).
%Due to radial interactions, there is no clear timescale  the
%local oscillations.
The distribution of mean $B_\theta$ over radius is more irregular compared
to the toroidal field, see Fig. 12 bottom left, although we observe a preferred sign of mean $B_\theta$
for a specific radial location, e.g. positive over time between 4 and 5 AU.
%In Fig. 8, bottom left, % we plot the mean $\theta$ magnetic field.
%we plot for the northern hemisphere the mean $\theta$
%field over radius and time. 
%One could see a preferred sign of $B_\theta$
%for a specific radial location, e.g. positive between 4 and 5 AU.
A time evolution over radius of $\rm B_\phi\cdot \alpha_{\phi\phi}^{NH} / EMF'_\phi$
, Fig. 12 top left, shows again the positive sign of $\rm \alpha_{\phi\phi}$ in the
northern hemisphere (see also Fig. 9, bottom).
By definition, the $\rm \alpha_{\phi\phi}$ presents the same distribution along radius
as the mean toroidal magnetic field.
In contrast we do not find a correlation between the turbulent velocity of
the gas and the distribution of mean magnetic fields.
Fig. 12, bottom right, presents $\rm v_{rms}$ over radius and time
for the northern hemisphere.
The RMS velocity is about $0.1 c_s$, nearly constant over radius and time.
A time average of $\rm v_{rms}$ is given in Table 1 for all models.
We again emphasize the lower turbulent velocity in the $\pi/4$, 
compared to $\pi/2$, due to the lack of the radial velocity peak (see Fig.
5, right).
%The values of $v_{rms}$ are obtained by averaging over azimuthal and between 
%0 and 2 scale heights in the northern hemisphere.
%$v_{rms}$ is plotted over radius and time
%for the northern hemisphere, Fig. 8, bottom right.
%There is no correlation visible between
%the RMS velocity and the mean toroidal magnetic field.    
%The RMS velocity is about $0.1 c_s$, constant over radius.
%A time average of $v_{rms}$ is given in Table 2 for all models.\\
%In Fig. 8, top, we plot the mean toroidal field (top right) 
%and $\alpha_{\phi\phi}$ (top left) over radius and time. 
%The sign of the mean toroidal magnetic field does change irregular along
%radius. 
%The timescale of these "butterfly" oscillations 
%is not well defined at a given radius because of radial interactions and
%large-scale modes.
%The reversals of magnetic field can overlap (see Fig. 8 top right 
%horizontal stripes).\\
% 
%

In our previous work we have shown the $1/r$ profile for the turbulent
magnetic fields \citep{flo11}.
Because of the time oscillations, it is difficult to estimate a radial profile
for the mean magnetic field.  
%The mean magnetic field is oscillating along time and it is not possible to
%average 
To determine a time averaged radial profile of the mean toroidal field we measure the 
amplitude values of the oscillations. 
We use five different radial locations to measure the peak values of the
mean toroidal field.
The results are plotted in Fig. 13 for the southern (blue) and northern
hemisphere (red).
%
%In Fig. 9, we measure at five
%different radial locations the peaks of the mean toroidal field for the 
%southern (blue) and northern hemisphere (red). 
The amplitudes of mean toroidal field decreases with radius.
The relative low number of values and their high standard
deviation makes it difficult to fit. A 
$1/r$ profile would apply (Fig. 13, green solid line).
The values in both hemispheres look quite symmetric (Fig. 9, blue and red) 
and we do not see a preferred hemisphere for the mean field generation.
%
%In Fig. 6, bottom, we saw that the $\alpha^{dyn}$ correlate with the
%mean toroidal field over disk height. 
%This is also true in radius. 
%In Fig. 8, top left, 
%Similar to the mean toroidal field evolution is the contour 
%plot over radius and time
%we plot $\alpha_{\phi\phi}$ at the northern hemisphere over radius and time. 
%As expected, the distribution follows the mean toroidal field, see
%Fig. 8 top right.
%which remains roughly constant over radius. The profile follows mostly the
%mean toroidal field evolution.
%
%In Fig. 8, bottom left, % we plot the mean $\theta$ magnetic field.
%we plot for the northern hemisphere the mean $\theta$
%field over radius and time. One could see a preferred sign of $B_\theta$ 
%for a specific radial location, e.g. positive between 4 and 5 AU.
%but the time evolution is  clear as seen for the
%toroidal field.
%
%The $v_{rms}$ presents only small time fluctuations
%and the $\pi/4$ and $\pi/2$ model present only slightly higher turbulent RMS
%velocities as the $\pi$ and $2\pi$ run.
%The turbulent RMS velocity is plotted over radius and time 
%for the northern hemisphere, Fig. 8, bottom right. 
%There is no correlation visible between
%the RMS velocity and the mean toroidal magnetic field. 
%The RMS velocity is about $0.1 c_s$, constant over radius. 
%The values do not correlate with changes of sign in
%the mean field.
%
\section{Discussion}
After the saturation of MRI, the initial magnetic field configuration 
is lost. Each model develops oscillating mean magnetic fields which appears
to be strongest in the $\pi/4$ and $\pi/2$ run.
The strength of turbulence follows this trend.
The mean fields are generated by a dynamo process which relies on the
symmetry and on the strength of the turbulent field. We measure higher dynamo
coefficient $\alpha_{\phi\phi}$ for the $\pi/2$ and $\pi/4$ model as well as
higher Maxwell stresses. This agrees with the correlation between
Maxwell stress and dynamo coefficient, found by \citet{rek00}.
The effect of increased magnetic energy at domain size seems to be
independent of resolution in stratified simulations (compare Fig. 12, bottom
left, model FO and PO in \citet{flo11}) but not present in unstratified
simulations (compare Fig. 9b in \citet{sor11}) as they do not develop a
dynamo.

\subsection*{Energy pile up and magnetic dynamo}
%In our simulations, the magnetic field peaks at the domain size. For reduced
%azimuthal domain models, we see
%higher amplitudes for the same mode,   
%independent of resolution (compare Fig. 12, bottom left, model PO and FO
%\citep{flo11}).
%Recent unstratified simulations did not find an energy pile up 
%(compare Fig. 9,b, in \citet{sor11}).
%The energy pile
%up at domain size is connected to a dynamo process.
Which physical process is sensitive to the domain size and lead to the 
increased mean toroidal fields in $\pi/4$ and $\pi/2$ models ?
The first mechanism leads to the dynamo process as it generates axisymmetric
magnetic fields out of the turbulence.
Another way to transport magnetic energy at domain size could be due to
an inverse energy cascade.
\citet{joh09} showed in local box simulations that the Keplerian advection
term in the induction equations drives an inverse energy cascade. This will
lead to a transport of energy to larger scales. Also \citet{gue07b} found 
in Taylor-Couette experiments that MRI, launched from a toroidal field, 
will have most of magnetic energy at the $\rm m=1$ and $\rm m=0$ mode.
% Mean magnetic fields ($\rm m=0$) support or hinder the formation of 
%disk winds and jets \citep{rek00,fer06}, e.g.
%
%Self generating mean magnetic fields can be both, 
%the seed for MRI and a support for disk wind and jets \citep{rek00,fer06}. 
%a dipolar magnetic field supports the generation of jets \citep{rek00,zie00}. 
%In $\alpha \Omega$ dynamo, a negative sign\footnote{In
%the northern hemisphere} of
%$\rm \alpha_{\phi\phi}$ would support a dipolar symmetry \citep{rek00}. 

Another open question is the sign of $\alpha_{\phi\phi}$ in
global simulations.
We find a positive $\rm \alpha_{\phi\phi}$, independent of the azimuthal
domain size.
This positive $\rm \alpha_{\phi\phi}$ has been indicated for 
global simulations by \citet{arl01}.
Local simulations show a negative $\rm \alpha_{\phi\phi}$ \citep{bra95,bra97,rue00,zie00,dav10,gre10}.
The reason of stronger mean fields in reduced azimuthal models as well as
the positive sign of $\rm \alpha_{\phi\phi}$ in global simulations 
has to be investigated in future work. 
One possibility would be to implement the 'Test field' method and to measure other
components of the dynamo and diffusivity tensor, as it was done in
\citet{gre10}.\\

\subsection{Time variability of accretion stress.}
Oscillating mean field are organized in elongated radial patches, normally
following the time-line of ten local orbits. It can occur that for a given
time, mean toroidal
field of one sign covers the whole radial extent (3 - 8 AU).
In such a case, temporal linear MRI will lead to a peak in
accretion stress, Fig. 3.
%The restricted azimuthal domain simulations, $\pi/2$ and $\pi/4$, develop stronger
%axisymmetric magnetic fields compared to the full $2\pi$ model.
%turbulence for global 3D MHD stratified simulations of accretion
%disks. 
%The turbulence level is amplified by strong axisymmetric oscillating toroidal
%magnetic fields. 
%To observe this effect of amplification there is a need of 
%high resolution. 
%Linear MRI of toroidal magnetic field will 
%The $\lambda_{crit}$ of the mean toroidal field has to be resolved. 
The effect of mean toroidal field, stretching over the whole radius, is independent
on the
azimuthal domain size, compare Fig. 12 top right.
The amplification of accretion stress due to linear MRI, is visible only in the $\pi/4$
model, as it present strongest amplitudes in the mean toroidal
magnetic field.
%Here, the $\lambda_{crit}$ is resolved 
%and the turbulence is amplified. 
%These fields are upwinded from axisymmetric radial fields.
%Axisymmetric radial fields are generated due to the $\alpha \Omega$ dynamo.
%Such dynamo models in stratified disk models successfully reproduce these
%oscillations \citep{les08,les08b,gre10,sim11}.
%The stronger toroidal fields in the $\pi/2$ and $\pi/4$ run, 
%The $\pi/2$ and $\pi/4$ present larger dynamo coefficients.
%A larger $\alpha_{\phi\phi}$ could generate stronger radial mean fields 
%generation leading to stronger toroidal fields.
%The question remains: 
%why there is a stronger $\alpha \Omega$ dynamo in the
%restricted domains. 
% makes a correlation between
%$\alpha_{\phi\phi}$ and $\alpha_{SS}$. A stronger turbulence leads to a
%stronger dynamo action. 
%Further investigations are needed to solve this chicken and
%egg problem. 
%So can we expect further amplification also in the $2\pi$ models with higher
%resolution ?
%
%We think there should be temporal amplifications.
%Mean toroidal fields of one specific sign can occur along radius.
%The superposition of linear MRI will drive a stronger but temporal 
%accretion stress (compare Fig. 12,
%top right, between 600 and 800 inner orbits).
%A similar effect of resolution have been found by \cite{haw11} in 
%stratified torus simulations of $\pi/2$. 
%They find higher poloidal magnetic
%fields for higher azimuthal resolution.\\ 
\subsection*{Correlation functions}
We confirm the results of recent stratified global simulations by 
\citet{bec11}. We find similar correlation angles (around $9^\circ$) 
and wavelengths (around H) for the magnetic field.
A larger correlation length is expected because of the 
relative low resolution per scale height compared to local simulations 
\citep{gua09,haw11,sor11}.
Recent unstratified global simulations \citet{sor11} suggest a magnetic 
tilt angle of around $13^\circ$ for converged MRI turbulence. It remains still unclear how
this could be applied for stratified disks with a minimum of $\theta_B$ at
the midplane. We found a magnetic tilt angle of around $13^\circ$ above 2 scale
heights. As discussed in \citet{flo11} we believe to find convergence
with resolutions around 32/64 grid cells per pressure scale height. 
Here, a Fargo
MHD approach as used in \citet{sor11} would be helpful.
%\subsection*{Sign of dynamo coefficient}

%k
\section{Summary}
We have studied the impact of different azimuthal extents in 3D global
stratified MHD simulations of accretion disks onto the saturation level
of MRI with an initial toroidal magnetic field. % having a zero-net flux toroidal magnetic field. 
%
%
%For the short azimuthal domains $\pi/2$, but especially $\pi/4$,
%the influence of modes having the size of
%domain, effects the overall turbulent evolution of AMRI.
\begin{itemize}
\item Turbulence in restricted domain sizes like $\pi/2$ and $\pi/4$ 
is amplified due to strong toroidal mean field oscillations.
For these runs, the $\lambda_{crit}$ of the mean field is resolved 
leading to a temporal magnification of the $\alpha_{SS}$ value and 
increased total magnetic energy.
In addition, radial superpositions of such strong mean fields 
can drive to a strong episodic increase of accretion.
%The effect is strongest in the $\pi/4$ model but 
%it is also present for the $\pi/2$ model. 
The time averaged total $\alpha_{SS}$ is $ 1.2 \pm 0.2 \cdot
10^{-2}$ for model $\pi/4$, $ 9.3 \pm 0.9 \cdot 10^{-3}$ for model $\pi/2$ 
and converge to $5.5 \pm 0.5 \cdot 10^{-3}$ for both models $\pi$ and $2\pi$.   
\item We find a positive dynamo $\alpha_{\phi\phi}$ for all models, 
a positive correlation between the turbulent $EMF'_{\phi}$ and the mean
toroidal magnetic field in the upper (northern) hemisphere. 
%For the $\pi$ and $2\pi$ models we observe a much
%less noise-induced error compared to the restricted domain runs
%$\pi/2$ and $\pi/4$. 
For the $2\pi$ model we found 
$\rm \alpha_{\phi\phi}^{North} = 2.1 \pm 0.2 \cdot 10^{-3}$.
The $\pi/2$ and $\pi/4$ present higher $\rm \alpha_{\phi\phi}$ values but
with stronger fluctuations in $EMF'_{\phi}$ and mean $B_\phi$.
\item The $\pi/4$ and $\pi/2$ models show higher tilt angles and smaller
correlation wavelengths in the two-point correlation of velocity and magnetic field
compared to the $\pi$ and $2\pi$ models.
We find $\rm \theta_t^{vel} = 14^\circ$ for models $\ge \pi/2$ and 
$\rm \theta_t^{vel} = 12^\circ$ for model $\pi/4$. The $\pi/4$ model 
does not resolve the
peak radial velocity at $\rm m=4$.
The tilt angles for the magnetic fields are smaller.
At the midplane we observe time averaged magnetic tilt angles between 
$\rm \theta_B = 8-9^\circ$ increasing up to $\rm \theta_B = 12-13^\circ$ in the
corona.
For the full $2\pi$ model we found
$\rm \lambda_{maj}^{vel} = 1.9 H$ and $\rm \lambda_{maj}^{mag} = 1.7 H$.
%This results confirm recent stratified global simulations by \citet{bec11} with 
%similar correlation angles and wavelengths for the magnetic field. 
%
\item The parity of the mean magnetic fields is a mixture 
of dipole and quadrupole for all models. 
The total parity is set by the oscillating toroidal field.
%We have shown that there is no preferred symmetry 
%of mean magnetic fields in all models. 
The timescale of symmetry change
between dipole and quadrupole is around 40 local orbits.
The time evolution of the parity is distinct in each model.
The $2\pi$ model remains longer in a dipole (antisymmetric) dominated 
configuration for the simulation time.
%
%The restricted domains show also higher amplitudes in the mean field 
%evolution and stronger fluctuations in the turbulent EMF.
%
%It also effects the major and minor wavelength of both fields. 
%The restricted domains show an increases the tilt angle for
%the magnetic field and for the velocity.
%
%
\end{itemize}
We conclude: In global MRI simulations of accretion disks
an azimuthal domain of at least $\pi$ $(180^\circ)$ is needed to present the most 
realistic turbulent and mean field evolution as the full $2\pi$ model.
%his is mainly visible in the results from the accretion stress, the $\alpha \Omega$
%dynamo and the two-point correlation function.
Here, the $\alpha \Omega$ dynamo plays a key role in determining the
saturation level of MRI.
Restricted domains of $\pi/4$ and $\pi/2$ amplify the MRI
turbulence due to a stronger axisymmetric magnetic fields.\\
%This effect becomes visible in $\pi/4$ models
%but can also be observed for $\pi/2$.
%even they are
%reduced.\\
%These results should help setting up future global simulations.
%while
%the $\pi/2$ model 

We thank Andrea Mignone for providing us with the newest code version and
the discussion on the numerical configuration.
We thank Sebastien Fromang for the helpful comments on the global models.
We thank also G\"unther R\"udiger and Rainer Arlt for their comments 
on the manuscript.
We thank Geoffroy Lesur for the discussion about the dynamo effect.
%We also thank Willy Kley for the comments on the viscous model.
%We thank Frederic A. Rasio and an anonymous referee for the fast and very
%professional processing of
%this work.
H. Klahr, N. Dzyurkevich and M. Flock have been supported in part by the
Deutsche Forschungsgemeinschaft DFG through grant DFG Forschergruppe 759
"The Formation of Planets. The Critical First Growth Phase". 
Neal Turner was supported by a
NASA Solar Systems Origins grant through the Jet Propulsion
Laboratory, California Institute of Technology, and by an Alexander
von Humboldt Foundation Fellowship for Experienced Researchers.
Parallel
computations have been performed on the Theo cluster of the MaxPlanck
Institute for Astronomy Heidelberg as well as the GENIUS Blue Gene/P cluster
both located at the computing center of the MaxPlanck Society in Garching.
%
%The increases artificially the accretion stress. 
%We find the first $\alpha \Omega$ dynamo
%in stratified 3D global MHD simulations.
%The dynamo is present for all azimuthal extents. The mean field and the
%turbulent EMF component are oscillating with the period of the butterfly.
%With shortening the azimuthal domain the influence of the mean field
%becomes visible in a higher accretion stress.\\
%III. The present two-point correlations for the velocity and the magnetic
%field and confirm the locality and anisotropy of the MRI which was already presented in
%local box simulations \citep{gua09}.
%First with a azimuthal domain of $\pi$ we can present very similar results as the full $2\pi$.
%We present the velocity and magnetic field correlation which show locality and anisotropy of the 
%MRI.
%\section{TO DO}
%I. Further correlation study: Does the angle change with different heights ?\\
%II. Insert Table with $\lambda_{maj}$; $\lambda_{min}$ ; $\lambda_z$ for all azimuthal extents.\\
%III. Plot the effect of the mean field (radial and azimuthal ) over time, is 
%there an increasing mean field after each butterfly loop ?

\bibliographystyle{apj}
\bibliography{dynamo}

\begin{thebibliography}{59}
\expandafter\ifx\csname natexlab\endcsname\relax\def\natexlab#1{#1}\fi

\bibitem[{{Arlt} \& {Brandenburg}(2001)}]{arl01b}
{Arlt}, R. \& {Brandenburg}, A. 2001, \aap, 380, 359

\bibitem[{{Arlt} \& {R{\"u}diger}(2001)}]{arl01}
{Arlt}, R. \& {R{\"u}diger}, G. 2001, \aap, 374, 1035

\bibitem[{{Armitage}(1998)}]{arm98}
{Armitage}, P.~J. 1998, \apjl, 501, L189

\bibitem[{{Balbus} \& {Hawley}(1991)}]{bal91}
{Balbus}, S.~A. \& {Hawley}, J.~F. 1991, \apj, 376, 214

\bibitem[{{Balbus} \& {Hawley}(1998)}]{bal98}
---. 1998, Reviews of Modern Physics, 70, 1

\bibitem[{{Beckwith} {et~al.}(2011){Beckwith}, {Armitage}, \& {Simon}}]{bec11}
{Beckwith}, K., {Armitage}, P.~J., \& {Simon}, J.~B. 2011, ArXiv e-prints

\bibitem[{{Blackman}(2010)}]{bla10}
{Blackman}, E.~G. 2010, Astronomische Nachrichten, 331, 101

\bibitem[{{Brandenburg} \& {Donner}(1997)}]{bra97}
{Brandenburg}, A. \& {Donner}, K.~J. 1997, \mnras, 288, L29

\bibitem[{{Brandenburg} {et~al.}(1995){Brandenburg}, {Nordlund}, {Stein}, \&
  {Torkelsson}}]{bra95}
{Brandenburg}, A., {Nordlund}, A., {Stein}, R.~F., \& {Torkelsson}, U. 1995,
  \apj, 446, 741

\bibitem[{{Brandenburg} \& {Subramanian}(2005)}]{bra05}
{Brandenburg}, A. \& {Subramanian}, K. 2005, \physrep, 417, 1

\bibitem[{{Brandenburg} \& {von Rekowski}(2007)}]{bra07}
{Brandenburg}, A. \& {von Rekowski}, B. 2007, Memorie della Societa Astronomica
  Italiana, 78, 374

\bibitem[{{Davis} {et~al.}(2010){Davis}, {Stone}, \& {Pessah}}]{dav10}
{Davis}, S.~W., {Stone}, J.~M., \& {Pessah}, M.~E. 2010, \apj, 713, 52

\bibitem[{{Dzyurkevich} {et~al.}(2010){Dzyurkevich}, {Flock}, {Turner},
  {Klahr}, \& {Henning}}]{dzy10}
{Dzyurkevich}, N., {Flock}, M., {Turner}, N.~J., {Klahr}, H., \& {Henning}, T.
  2010, \aap, 515, A70

\bibitem[{{Elstner} {et~al.}(1996){Elstner}, {Ruediger}, \& {Schultz}}]{els96}
{Elstner}, D., {Ruediger}, G., \& {Schultz}, M. 1996, \aap, 306, 740

\bibitem[{{Fleming} \& {Stone}(2003)}]{fle03}
{Fleming}, T. \& {Stone}, J.~M. 2003, \apj, 585, 908

\bibitem[{{Flock} {et~al.}(2010){Flock}, {Dzyurkevich}, {Klahr}, \&
  {Mignone}}]{flo10}
{Flock}, M., {Dzyurkevich}, N., {Klahr}, H., \& {Mignone}, A. 2010, \aap, 516,
  A26

\bibitem[{{Flock} {et~al.}(2011){Flock}, {Dzyurkevich}, {Klahr}, {Turner}, \&
  {Henning}}]{flo11}
{Flock}, M., {Dzyurkevich}, N., {Klahr}, H., {Turner}, N.~J., \& {Henning}, T.
  2011, ArXiv e-prints

\bibitem[{{Foglizzo} \& {Tagger}(1995)}]{fog95}
{Foglizzo}, T. \& {Tagger}, M. 1995, \aap, 301, 293

\bibitem[{{Fromang}(2010)}]{fro10}
{Fromang}, S. 2010, \aap, 514, L5

\bibitem[{{Fromang} \& {Nelson}(2006)}]{fro06}
{Fromang}, S. \& {Nelson}, R.~P. 2006, \aap, 457, 343

\bibitem[{{Fromang} \& {Nelson}(2009)}]{fro09}
---. 2009, \aap, 496, 597

\bibitem[{{Gardiner} \& {Stone}(2005)}]{gar05}
{Gardiner}, T.~A. \& {Stone}, J.~M. 2005, Journal of Computational Physics,
  205, 509

\bibitem[{{Gellert} {et~al.}(2007){Gellert}, {R{\"u}diger}, \&
  {Fournier}}]{gel07}
{Gellert}, M., {R{\"u}diger}, G., \& {Fournier}, A. 2007, Astronomische
  Nachrichten, 328, 1162

\bibitem[{{Gressel}(2010)}]{gre10}
{Gressel}, O. 2010, \mnras, 404

\bibitem[{{Guan} \& {Gammie}(2011)}]{gua11}
{Guan}, X. \& {Gammie}, C.~F. 2011, \apj, 728, 130

\bibitem[{{Guan} {et~al.}(2009){Guan}, {Gammie}, {Simon}, \& {Johnson}}]{gua09}
{Guan}, X., {Gammie}, C.~F., {Simon}, J.~B., \& {Johnson}, B.~M. 2009, \apj,
  694, 1010

\bibitem[{{Hawley}(2000)}]{haw00}
{Hawley}, J.~F. 2000, \apj, 528, 462

\bibitem[{{Hawley} \& {Balbus}(1991)}]{haw91}
{Hawley}, J.~F. \& {Balbus}, S.~A. 1991, \apj, 376, 223

\bibitem[{{Hawley} \& {Balbus}(1992)}]{haw92}
---. 1992, \apj, 400, 595

\bibitem[{{Hawley} {et~al.}(1995){Hawley}, {Gammie}, \& {Balbus}}]{haw95}
{Hawley}, J.~F., {Gammie}, C.~F., \& {Balbus}, S.~A. 1995, \apj, 440, 742

\bibitem[{{Hawley} {et~al.}(1996){Hawley}, {Gammie}, \& {Balbus}}]{haw96}
---. 1996, \apj, 464, 690

\bibitem[{{Hawley} {et~al.}(2011){Hawley}, {Guan}, \& {Krolik}}]{haw11}
{Hawley}, J.~F., {Guan}, X., \& {Krolik}, J.~H. 2011, ArXiv e-prints

\bibitem[{{Heinemann} \& {Papaloizou}(2009)}]{hei09}
{Heinemann}, T. \& {Papaloizou}, J.~C.~B. 2009, \mnras, 397, 64

\bibitem[{{Inutsuka} \& {Sano}(2005)}]{inu05}
{Inutsuka}, S. \& {Sano}, T. 2005, \apjl, 628, L155

\bibitem[{{Johansen} {et~al.}(2009){Johansen}, {Youdin}, \& {Klahr}}]{joh09}
{Johansen}, A., {Youdin}, A., \& {Klahr}, H. 2009, \apj, 697, 1269

\bibitem[{{Krause} \& {Raedler}(1980)}]{kra80}
{Krause}, F. \& {Raedler}, K.-H. 1980, {Mean-field magnetohydrodynamics and
  dynamo theory}, ed. {Williams, L.~O.}

\bibitem[{{Lesur} \& {Ogilvie}(2008{\natexlab{a}})}]{les08b}
{Lesur}, G. \& {Ogilvie}, G.~I. 2008{\natexlab{a}}, \mnras, 391, 1437

\bibitem[{{Lesur} \& {Ogilvie}(2008{\natexlab{b}})}]{les08}
---. 2008{\natexlab{b}}, \aap, 488, 451

\bibitem[{{Matsumoto} \& {Tajima}(1995)}]{mat95}
{Matsumoto}, R. \& {Tajima}, T. 1995, \apj, 445, 767

\bibitem[{{Miller} \& {Stone}(2000)}]{mil00}
{Miller}, K.~A. \& {Stone}, J.~M. 2000, \apj, 534, 398

\bibitem[{{Miyoshi} \& {Kusano}(2005)}]{miy05}
{Miyoshi}, T. \& {Kusano}, K. 2005, Journal of Computational Physics, 208, 315

\bibitem[{{Noble} {et~al.}(2010){Noble}, {Krolik}, \& {Hawley}}]{nob10}
{Noble}, S.~C., {Krolik}, J.~H., \& {Hawley}, J.~F. 2010, \apj, 711, 959

\bibitem[{{Papaloizou} \& {Terquem}(1997)}]{pap97}
{Papaloizou}, J.~C.~B. \& {Terquem}, C. 1997, \mnras, 287, 771

\bibitem[{{Rekowski} {et~al.}(2000){Rekowski}, {R{\"u}diger}, \&
  {Elstner}}]{rek00}
{Rekowski}, M.~v., {R{\"u}diger}, G., \& {Elstner}, D. 2000, \aap, 353, 813

\bibitem[{{R{\"u}diger} {et~al.}(2007){R{\"u}diger}, {Hollerbach}, {Gellert},
  \& {Schultz}}]{gue07b}
{R{\"u}diger}, G., {Hollerbach}, R., {Gellert}, M., \& {Schultz}, M. 2007,
  Astronomische Nachrichten, 328, 1158

\bibitem[{{R{\"u}diger} \& {Pipin}(2000)}]{rue00}
{R{\"u}diger}, G. \& {Pipin}, V.~V. 2000, \aap, 362, 756

\bibitem[{{Ruediger} \& {Kichatinov}(1993)}]{rue93}
{Ruediger}, G. \& {Kichatinov}, L.~L. 1993, \aap, 269, 581

\bibitem[{{Sano} {et~al.}(2000){Sano}, {Miyama}, {Umebayashi}, \&
  {Nakano}}]{san00}
{Sano}, T., {Miyama}, S.~M., {Umebayashi}, T., \& {Nakano}, T. 2000, \apj, 543,
  486

\bibitem[{{Simon} {et~al.}(2011){Simon}, {Hawley}, \& {Beckwith}}]{sim11}
{Simon}, J.~B., {Hawley}, J.~F., \& {Beckwith}, K. 2011, \apj, 730, 94

\bibitem[{{Sorathia} {et~al.}(2010){Sorathia}, {Reynolds}, \&
  {Armitage}}]{sor10}
{Sorathia}, K.~A., {Reynolds}, C.~S., \& {Armitage}, P.~J. 2010, \apj, 712,
  1241

\bibitem[{{Sorathia} {et~al.}(2011){Sorathia}, {Reynolds}, {Stone}, \&
  {Beckwith}}]{sor11}
{Sorathia}, K.~A., {Reynolds}, C.~S., {Stone}, J.~M., \& {Beckwith}, K. 2011,
  ArXiv e-prints

\bibitem[{{Stone} {et~al.}(1996){Stone}, {Hawley}, {Gammie}, \&
  {Balbus}}]{sto96}
{Stone}, J.~M., {Hawley}, J.~F., {Gammie}, C.~F., \& {Balbus}, S.~A. 1996,
  \apj, 463, 656

\bibitem[{{Terquem} \& {Papaloizou}(1996)}]{ter96}
{Terquem}, C. \& {Papaloizou}, J.~C.~B. 1996, \mnras, 279, 767

\bibitem[{{Turner} {et~al.}(2010){Turner}, {Carballido}, \& {Sano}}]{tur10}
{Turner}, N.~J., {Carballido}, A., \& {Sano}, T. 2010, \apj, 708, 188

\bibitem[{{Umebayashi}(1983)}]{ume83}
{Umebayashi}, T. 1983, Progress of Theoretical Physics, 69, 480

\bibitem[{{Umebayashi} \& {Nakano}(2009)}]{ume09}
{Umebayashi}, T. \& {Nakano}, T. 2009, \apj, 690, 69

\bibitem[{{Uribe} {et~al.}(2011){Uribe}, {Klahr}, {Flock}, \&
  {Henning}}]{uri11}
{Uribe}, A., {Klahr}, H., {Flock}, M., \& {Henning}, T. 2011, ArXiv e-prints

\bibitem[{{Wardle}(2007)}]{war07}
{Wardle}, M. 2007, \apss, 311, 35

\bibitem[{{Ziegler} \& {R{\"u}diger}(2000)}]{zie00}
{Ziegler}, U. \& {R{\"u}diger}, G. 2000, \aap, 356, 1141

\end{thebibliography}
\end{document}